\documentclass[
	oldfontcommands,
	openany,
	oneside,	
	12pt,				
	a4paper,			
	english,			
	brazil				
	]{{abntex2_unifor}}

\usepackage[brazilian,hyperpageref]{backref}
\usepackage[alf]{{abntex2_unifor_cite}}
\usepackage{calc}
\usepackage{lmodern}		
\usepackage[T1]{fontenc}
\usepackage[utf8]{inputenc}
\usepackage{lastpage}
\usepackage{indentfirst}
\usepackage{color}
\usepackage{graphicx}
\usepackage[space]{grffile}
\usepackage{microtype}
\usepackage{amsmath}
\usepackage{amssymb}
\usepackage{booktabs}
\usepackage{caption}
\usepackage{subcaption}
\usepackage{pdfpages}
\usepackage{hyperref}
\usepackage[brazil]{babel}
\usepackage{hyphenat}
\usepackage{rotating}
\usepackage{lscape}
\usepackage{pdflscape}
\usepackage{algorithm}
\usepackage{algpseudocode}

\graphicspath{{Figuras/}}
\include{0b_macros}

\renewcommand*{\backrefalt}[4]{
	\ifcase #1 %
		Nenhuma citação no texto.%
	\or
		Citado na página #2.%
	\else
		Citado #1 vezes nas páginas #2.%
	\fi}

\titulo{Descoberta  de relações alométricas entre população e crime dentro de uma grande metrópole}
\autor{Carlos de Oliveira Caminha Neto}
\local{Fortaleza-CE, Brasil}
\data{2017}
\orientador{Prof. João José Vasco Peixoto Furtado}
\instituicao{%
  Programa de Pós-Graduação em Informática Aplicada (PPGIA)	
  \par
  Universidade de Fortaleza (UNIFOR)}
  
\tipotrabalho{Tese}

\preambulo{Tese apresentada ao Programa de Pós-Graduação em Informática
Aplicada da Universidade de Fortaleza, como parte dos requisitos para
obtenção do título de Doutor em Informática Aplicada.}

\definecolor{blue}{RGB}{41,5,195}

\makeatletter
\hypersetup{
		pdftitle={\@title}, 
		pdfauthor={\@author},
    	pdfsubject={\imprimirpreambulo},
	    pdfcreator={LaTeX com uma versão modificada do abnTeX2},
		pdfkeywords={abnt}{latex}{abntex}{abntex2}{trabalho acadêmico}, 
		colorlinks=true,       		
    	linkcolor=blue,          	
    	citecolor=blue,        		
    	filecolor=magenta,          
		urlcolor=blue,
		bookmarksdepth=4
}
\makeatother

\makeindex

\begin{document}

\begin{figure}[h]
\center
\includegraphics[width=0.1\textwidth]{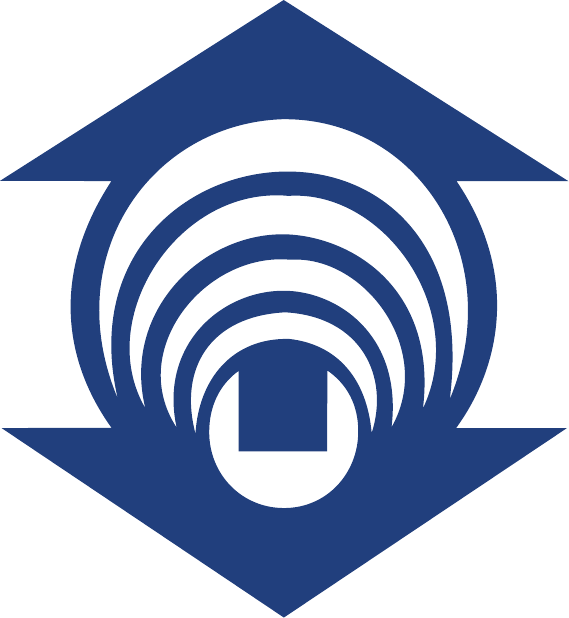}
\end{figure}

\selectlanguage{brazil}
\frenchspacing 


\imprimircapa
\imprimirfolhaderosto*

\includepdf{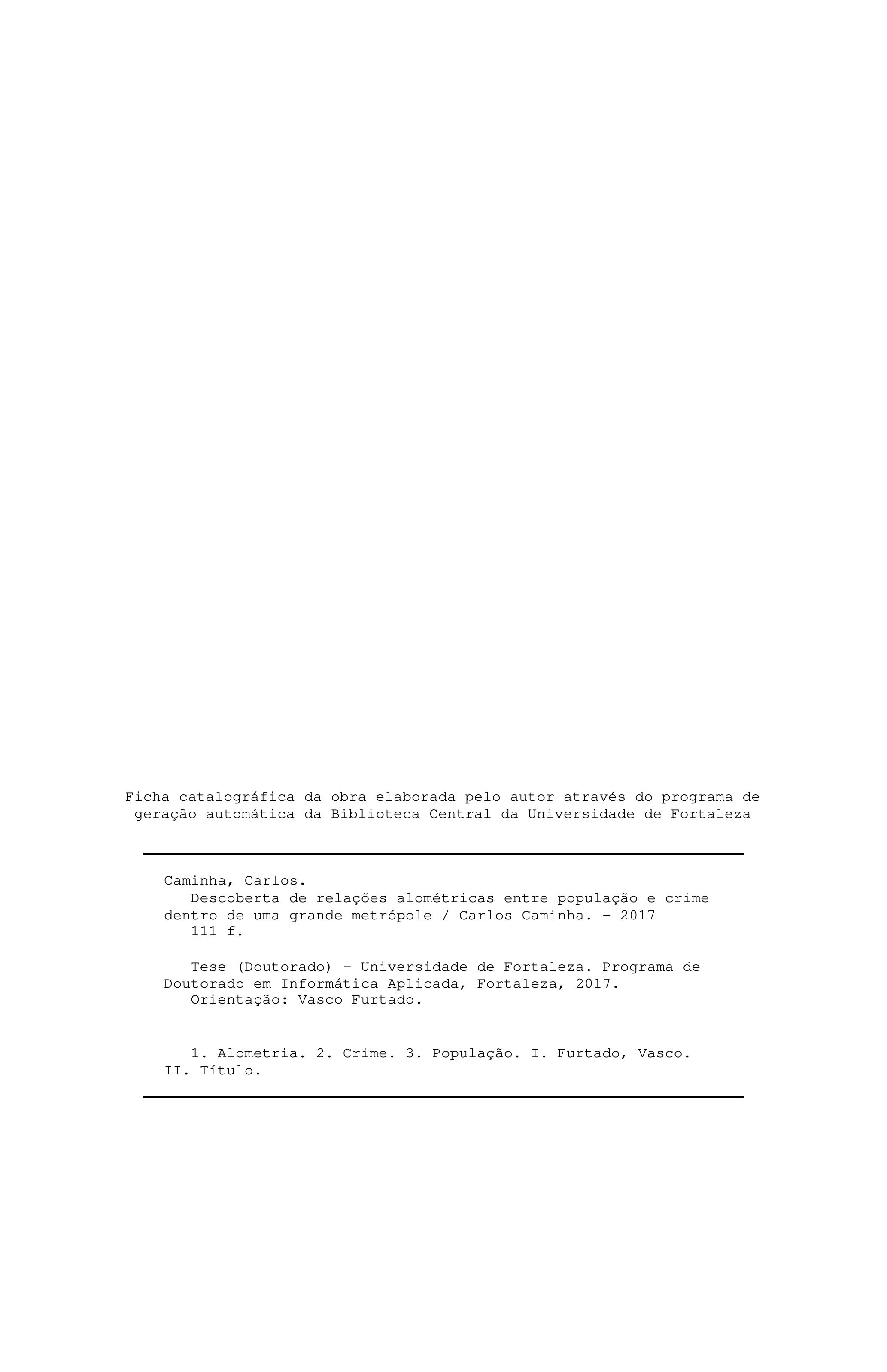}

\begin{folhadeaprovacao}

\includepdf{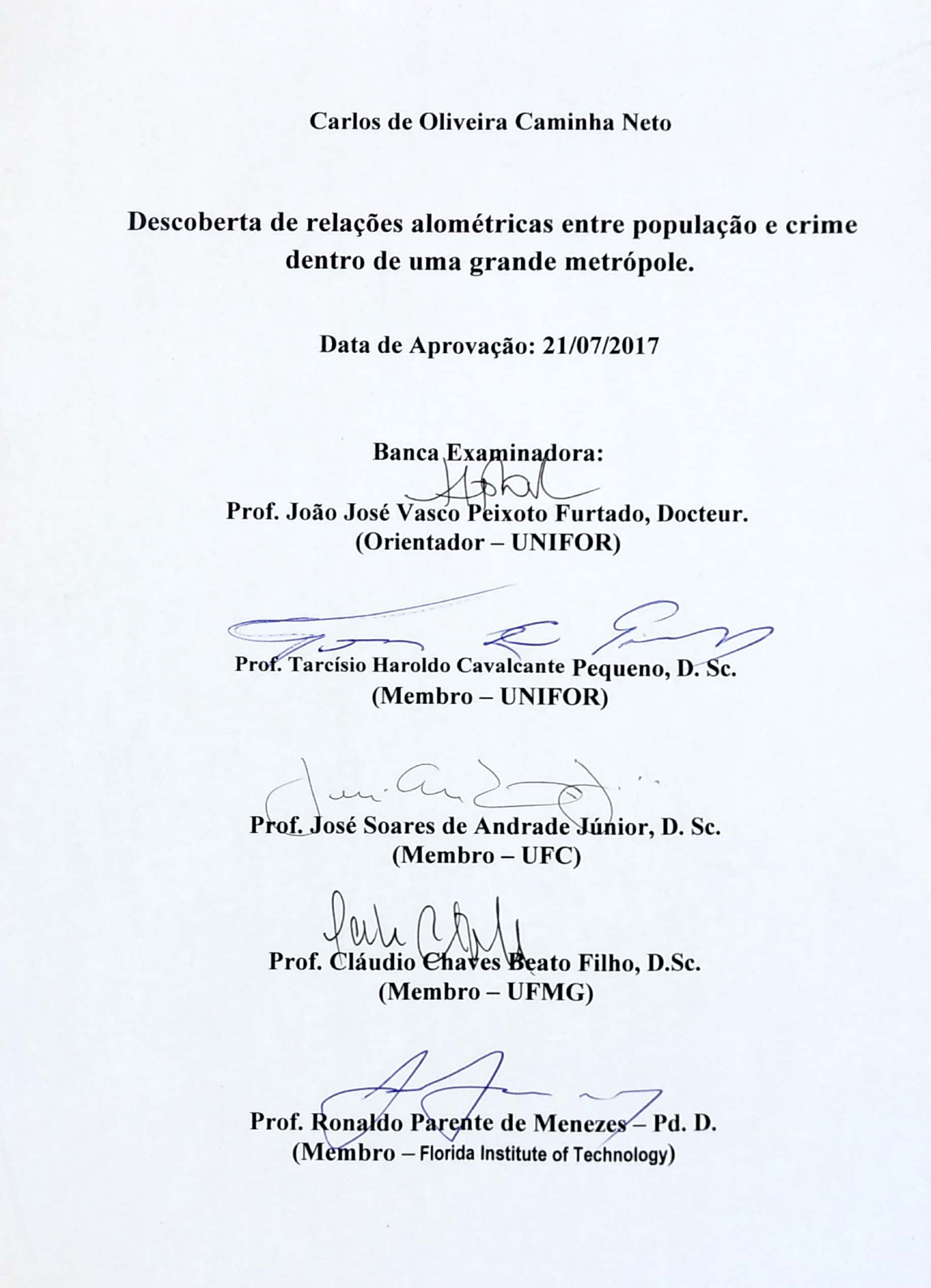}

\end{folhadeaprovacao}

\newpage

\begin{epigrafe}
    \vspace*{\fill}
	\begin{flushright}
	    
		\textit{
		     Aos meus pais, esposa, \\ %
		     irmãs e sobrinha,\\ %
       		     que estiveram comigo \\ %
 		     em todos os momentos. \\ %
       		     Amo vocês! \\ %
		}
	\end{flushright}
\end{epigrafe}

\begin{agradecimentos}

Em primeiro lugar, agradesço aos meus pais, Carlos e Nágela, pelo apoio
incondicional em todos os momentos, dando-me liberdade e acreditando em minhas
escolhas. A eles dedico esta tese de doutorado.

À minha esposa Sarah, pela paciência, compreensão e apoio nos vários momentos em
que estive ausente por estar desenvolvendo esta tese. Obrigado por tudo, meu
amor.

Às minhas irmãs, Naira, Natália e Maria Clara, meus sobrinhos Larinha e Davi e
meus sogros Hosanildo e Amparo, por todo amor e respeito que sempre tiveram por
mim. 

Aos amigos do Laboratório Engenharia do Conhecimento da UNIFOR que conviveram e
compartilharam comigo quase que diariamente o esforço dispensado nesta
tese.

Ao departamento de Física da Universidade Federal do Ceará, especialmente o
Professor Soares, Dr. Erneson Oliveira e Professor Hygor Piaget, por me
receberem em seu ambiente de trabalho e por permitirem que eu aprendesse com
suas ideias e experiências.

Ao Professor Vasco Furtado, meu orientador, por todo o seu tempo e dedicação não
somente a esta pesquisa, mas também a toda a minha formação acadêmica e
profissional, sempre atuando com sua competência e experiência. Extendo esse
agradesciento a todos os professores e funcionários do PPGIA e da UNIFOR que
também contribuíram para o meu crescimento pessoal.

\end{agradecimentos}

\begin{epigrafe}
    \vspace*{\fill}
	\begin{flushright}
	    
		\textit{
		    "Descobrir consiste em \\ %
                    olhar para o que todo \\ %
                    mundo está vendo e pensar \\ %
                    uma coisa diferente.'' \\ %
		    Roger Von Oech \\ %
		}
	\end{flushright}
\end{epigrafe}

\cleardoublepage

\setlength{\absparsep}{18pt} 
\begin{resumo}
    
Recentemente a humanidade atravessou um marco importante em sua história com a
constatação de que a maioria da sua população agora se concentra nas grandes
cidades. Essa concentração populacional é capaz de potencializar o crescimento
de indicadores positivos como inovação, produção de novas patentes e empregos
supercriativos, porém potencializa a propagação de doenças e a ocorrência de
crimes. Diante da constatação de que as taxas de crimes crescem ano após ano em
muitos desses grandes centros urbanos, buscou-se compreender a dinâmica do crime
dentro de uma cidade. Foi investigada a incidência de crimes contra o
patrimônio, tanto em função de população residente quanto em função de população
flutuante, em aglomerados de população dentro de uma grande metrópole
brasileira. Foi encontrada uma relevante relação alométrica, mas que só pôde ser
observada entre o crimes contra o patrimônio e população flutuante. Mais
precisamente, a evidência de um comportamento alométrico superlinear indica que
um número desproporcional de crimes ocorre em regiões onde o fluxo de pessoas é
maior. Também foi descoberto que o número de chamadas à polícia por perturbação
de sossego alheio é correlacionado, também de forma superlinear, com a população
residente. Esta tese levanta a interessante possibilidade de que a
superlinearidade observada em estudos anteriores [Bettencourt {\it et al.},
Proc. Natl. Acad. Sci. USA {\bf 104}, 7301 (2007) e Melo {\it et al.}, Sci. Rep.
{\bf 4}, 6239 (2014)] para crimes sérios em função de população, em escala de
cidade, pode ter origem na mobilidade humana, e não na presença de residentes
como se pensava. Esse achado foi a motivação para a codificação de um {\it
framework}, que apoia a análise de dados de população e crime para que sejam
propostas divisões de cidade que permitam a alocação de policiais por
estatísticas de população flutuante e residente.

    \textbf{Palavras-chave}: Alometria, Crime, População
\end{resumo}

\begin{resumo}[Abstract]
    \begin{otherlanguage*}{english}

Recently humanity has just crossed an important landmark in its history with the
majority of people now living in large cities. This population concentration is
capable of boosting the growth of positive indicators such as innovation, the
production of new patents and supercreative employment, but increases the spread
of diseases and the occurrence of crimes. Faced with the realization that crime
rates grow year after year in these large urban centers, we sought to understand
the dynamics of crime within cities. We investigate at the subscale of the
neighborhoods of a highly populated city the incidence of property crimes in
terms of both the resident and the floating population. Our results show that a
relevant allometric relation could only be observed between property crimes and
floating population. More precisely, the evidence of a superlinear behavior
indicates that a disproportional number of property crimes occurs in regions
where an increased flow of people takes place in the city. For comparison, we
also found that the number of crimes of peace disturbance only correlates well,
and in a superlinear fashion too, with the resident population. Our study raises
the interesting possibility that the superlinearity observed in previous studies
[Bettencourt {\it et al.}, Proc. Natl. Acad. Sci. USA {\bf 104}, 7301 (2007) and
Melo {\it et al.}, Sci. Rep. {\bf 4}, 6239 (2014)] for homicides versus
population at the city scale could have its origin in the fact that the floating
population, and not the resident one, should be taken as the relevant variable
determining the intrinsic microdynamical behavior of the system. This finding
was the motivation for the codification of a framework that supports the
analysis of population and crime data to propose city divisions that allow the
allocation of police by floating population and resident population statistics.     

    \vspace{\onelineskip}
    \end{otherlanguage*}
\end{resumo}

\chapter*{Lista de Abreviaturas e Siglas}

\begin{table}[!ht]
 
    \begin{tabular}{ p{4em} p{23em}}
    
	APF & Alocação por População Flutuante \\	

	APR & Alocação por População Residente \\	
	
	CCA & {\it City Clustering Algorithm} \\

	CIOPS & Coordenadoria Integrada de Operações de Segurança \\
	
	CCP & Crimes Contra o Patrimônio \\

	GPS & {\it Global Positioning System} \\

	IBGE & Instituto Brasileiro de Geografia e Estatística	\\

	OLS & {\it Ordinary Least Square} \\

	PSA & Perturbação de Sossego Alheio \\

	PF & População Flutuante \\

	PR & População Residente \\
     
    \end{tabular}

\end{table}

\newpage

\pdfbookmark[0]{\listfigurename}{lof}
\listoffigures*
\cleardoublepage

\pdfbookmark[0]{\listtablename}{lof}
\listoftables*
\cleardoublepage

\pdfbookmark[0]{\contentsname}{toc}
\tableofcontents*
\cleardoublepage

\textual

\chapter{Introdução}\label{cap1}

\section{Contexto}

Relações sociais diretas, como interações entre conhecidos, ou indiretas, como o
uso compartilhado de recursos naturais ({\it e.g.} visitar uma praia ou um
parque ecológico) ou artificiais ({\it e.g.} usar um transporte coletivo) são
importantes elementos que potencializam o que abstratamente é conhecido como a
dinâmica de uma cidade ou, mais genericamente, de um aglomerado urbano ({\it
e.g.} regiões metropolitanas). Essas relações têm especial impacto no
comportamento do crime, sendo objeto de estudo de áreas diversas.
Majoritariamente, as ciências ditas sociais têm buscado explicar fenômenos
relacionados ao comportamento violento e delituoso por meio de pesquisas
qualitativas \cite{michael1933, cohen1979, felson2002, clarke1993}. A era
digital, em que dados sobre o comportamento humano passaram a ser registrados em
um nível de granularidade nunca antes imaginado, trouxe novos atores à cena da
pesquisa em ciências sociais com novos instrumentos para apoiar a compreensão da
dinâmica das cidades e do crime em particular baseados em modelos quantitativos
\cite{beato2001,guedes2014, wang2013, kiani2015, caminha2012, Furtado2012,
wang2005, melo2014, alves2013a, alves2013b, orsognad2015, alves2015b, short2008,
short2010}.

Um exemplo de como esses modelos quantitativos podem ser aplicados refere-se à
forma como o conceito de alometria - um termo originalmente concebido na
Biologia para descrever relações de causa e efeito com característica livre de
escala, {\it e.g.}, a relação entre a massa e a taxa metabólica de organismos
\cite{kleiber1932} – foi adaptado ao contexto social para compreensão da
dinâmica de cidades \cite{naroll1956}.

Na Biologia, mostrou-se que a taxa metabólica dos organismos escala
sublinearmente em relação à massa dos mesmos. Ou seja, o gasto de energia dos
animais é desproporcional a sua massa, indicando uma certa eficiência dos
organismos vivos, pois, por exemplo, o dobro da massa acarretaria um aumento
menor do que duas vezes no gasto metabólico. De maneira análoga aos estudos
biológicos, a população de uma cidade equivaleria a massa dos organismos vivos,
tendo o papel de dimensionar o objeto de estudo. A ideia é de que sirva para
compreender o efeito que o seu crescimento tem sobre indicadores
correlacionados, como indicadores sociais (crimes) e ambientais (emissão de
$CO_2$). Uma relação alométrica, em um contexto social, pode ser descrita por
uma função de lei de potência,
\begin{equation}
Y=aX^\beta,
\end{equation}
\noindent onde usualmente $X$ representa a população, $Y$ quantifica um
indicador social, $a$ é uma constante e $\beta \neq 1$ é o expoente alométrico.

A descoberta de relações alométricas em fenômenos sociais tornou-se uma
promissora linha de pesquisa, sendo capaz de estimar o impacto que as relações
sociais podem causar em indicadores relacionados à dinâmica das cidades \cite{
gomez2012, ignazzi2014, hanley2016, alves2015a, arcaute2016, cottineau2016,
leitao2016, bettencourt2013, arbesman2009}. Digno de menção é o trabalho de
Bettencourt {\it et al.} \cite{bettencourt2007}, que revelou que relações
alométricas estavam presentes em muitos aspectos infraestruturais e dinâmicos de
cidades americanas. Os autores utilizaram população residente como medida de
dimensão de cidades e estudaram o efeito que o crescimento das mesmas exerce
sobre indicadores urbanos.

Dentre os seus principais resultados, os autores observaram uma relação
alométrica superlinear ($\beta > 1$) característica entre a quantidade de
população residente e o número de crimes sérios, que indica que o crescimento
populacional de uma cidade ocorre a custo de um crescimento maior que o
proporcional da sua quantidade de crimes. Posteriormente, Melo {\it et al.}\cite
{melo2014} confirmaram esse comportamento para as cidades brasileiras, no
entanto encontraram forte evidência quantitativa de que suicídios escalam de
forma alométrica sublinear ($\beta < 1$) com população residente.

\section{Motivação e Questões de Pesquisa}

Grande parte dos trabalhos que buscam compreender a dinâmica das cidades e o
impacto que a população exerce sobre indicadores sociais e/ou ambientais
relacionados à vida urbana, em especial os que buscam compreender a dinâmica do
crime em aglomerados urbanos, fazem-no considerando a existência de uma relação
alométrica intercidades explicada pela presença de população residente. Apesar
da importância dos resultados apresentados nesses trabalhos, seus impactos no
planejamento urbano, mais especificamente no desenvolvimento de políticas
públicas, são limitados devido à sua macroescala, diferente de um estudo
intracidade que objetiva um entendimento mais profundo das causas que levam a
tal comportamento desproporcional. Desta forma, uma abordagem em microescala,
baseada no estudo da relação entre população e indicadores urbanos em
sub-regiões dentro das cidades, tornou-se a principal motivação desta pesquisa
visando ampliar o entendimento acerca dos resultados obtidos. Isso nos levou à
formulação da primeira questão de pesquisa desta tese:

{\it $QP_1$ - Há relações alométricas entre população e crime dentro de uma
cidade?}

No percurso de exploração dessa primeira questão de pesquisa, verificou-se que a
literatura científica, tanto na criminologia, como nas ciências sociais,
enfatizava o papel determinante da movimentação das pessoas no espaço urbano e
das relações sociais que nele ocorrem \cite{cohen1979, clarke1993}. Um exemplo
marcante é a Teoria das Atividades Rotineiras, proposta por Cohen e Felson
\cite{cohen1979}, que afirma que crimes, como roubos e furtos, ocorrem pela
convergência das rotinas de um agressor motivado a cometer um crime, com uma
vítima vulnerável, somada a ausência de um guardião capaz de proteger essa
vítima. Essa convergência de rotinas é evidentemente determinada pelo padrão de
movimentação das pessoas na cidade. Na verdade, vários outros estudos
criminológicos levam em conta a movimentação das pessoas dentro das cidades para
explicar a Jornada para Crime ({\it Journey to Crime}) \cite{rengert2004} e a
Vitimização Repetida ({\it Repeat Victimization}) \cite{lauritsen1995}, para
mencionar alguns. Dependendo do tipo de crime, a movimentação das pessoas pode
ter maior ou menor relevância em gerar relações sociais que possam provocar
crimes. Em certos crimes ou delitos, as relações de vizinhança seriam mais
determinantes o que justificaria considerar os dados da população residente. Em
outros, como roubos e furtos, os dados relativos aos deslocamentos das pessoas
seriam mais determinantes de relações sociais que levariam a crimes.

Com base nisso, construiu-se a questão de pesquisa $QP_2$, uma especificação da
primeira. Visa-se explorar se há relações alométricas em regiões de uma
cidade considerando tanto população flutuante como população residente.

{\it $QP_2$ - A quantidade de população residente e flutuante é igualmente
importante para explicar a quantidade de crimes em diferentes regiões de uma
cidade?}

\section{Desenvolvimento e Resultados}

A busca por respostas para as questões de pesquisa elaboradas nesta tese
trouxeram desafios técnicos à parte, quais sejam:

\begin{enumerate} [label=(\roman*)]

\item Estimar a população flutuante de uma cidade com um nível de granularidade
que permitisse verificar a relação entre população e crime em sub-aglomerados
dentro da cidade. Para tal efeito, desenvolveu-se um método que permite estimar
a movimentação das pessoas baseado em dados de bilhetagem de ônibus e do GPS dos
ônibus.

\item Estimar o tamanho e a forma dos sub-aglomerados urbanos adequados para a
identificação de relações alométricas. Desenvolveu-se uma metodologia fazendo
uso do algoritmo de aglomeração {\it City Clustering Algorithm} (CCA)
\cite{makse1998, rozenfeld2008, giesen2010, rozenfeld2011, duranton2013,
gallos2012, duranton2015, eeckhout2004}.

\end{enumerate}

A aplicação da metodologia desenvolvida se deu a partir de dados
georreferenciados de chamadas à polícia para crimes contra o patrimônio e
perturbação de sossego alheio, registrados na cidade brasileira de Fortaleza,
entre 2014 e 2016. Também foram utilizados dados de população residente,
fornecidos pelo Instituto Brasileiro de Geografia e Estatística (IBGE)
\footnote{disponível em http://www.ibge.gov.br/}. A estimativa de população
flutuante foi feita com base em dados de bilhetagem e GPS ({\it Global
Positioning System}) de ônibus do ano de 2015.

Dentre os resultados obtidos verificou-se que a incidência de crimes contra o
patrimônio tem relação alométrica superlinear com a população flutuante em
diversas áreas de Fortaleza. Este resultado implica que o aumento do fluxo de
pessoas em uma determinada área da cidade acontecerá a custo de uma taxa
proporcionalmente maior de crimes na mesma região. Esse comportamento
superlinear dentro de aglomerados na cidade amplia o entendimento acerca dos
resultados intercidades encontrados em \cite{bettencourt2007, melo2014}, pois,
em microescala, é a população flutuante a responsável por criar a
relação superlinear evidenciada em macroescala. Também foi encontrada
uma relação alométrica superlinear entre população residente e o número de
chamadas de perturbação de sossego alheio. Esse resultado mostra que o efeito da
influência social deve estar adequadamente correlacionado com a população
residente ou flutuante, dependendo do tipo de crime considerado.

Diante do maior entendimento acerca das relações entre população e crime dentro
de uma cidade, buscou-se avaliar como esse entendimento pode alterar políticas
de segurança pública. Foi desenvolvido um {\it framework} com módulos de
mineração de dados (de população e crime) e alocação policial. Esse {\it
framework} permitiu o estudo do impacto que a população, especialmente a
flutuante, deveria exercer na atividade de alocação policial, essencialmente
devido a constatação de que uma lei de potência conduz a relação entre população
flutuante e crimes contra o patrimônio dentro de uma cidade. Foram comparadas
duas estratégias de alocação policial, ambas baseadas no modelo de alocação
por alta densidade de crimes, que aloca a quantidade de recursos proporcionalmente
em função da quantidade de crimes na região. A primeira é uma estratégia
convencional de alocação, onde os recursos policiais são distribuídos por
unidades administrativas de terreno, normalmente definidas a partir de
estatísticas de população residente. A segunda estratégia aloca os mesmos
recursos por aglomerados de população flutuante estimados pelo CCA. Observou-se
uma diferença substancial entre as quantidades de policiais alocados de acordo
com as duas estratégias.

\section{Estrutura da tese}

Esta tese de doutorado é dividida em seis capítulos, incluindo esta introdução.
No capítulo \ref{cap2} é apresentada a fundamentação teórica que cerca este
trabalho. No capítulo \ref{proc-pf} será detalhado como foram estimados os
padrões de mobilidade dos cidadãos de Fortaleza e como esses padrões quantificam
o volume de população flutuante em diversas regiões da cidade. No capítulo
\ref{cap3} será realizada uma discussão a respeito de como uma cidade pode ser
dividida para compreensão da dinâmica do crime. Adicionalmente serão
apresentados os conjuntos de dados utilizados nesta pesquisa. No capítulo
\ref{cap4} será demonstrado como foram definidas as fronteiras da influência
social na cidade, também será explicitado como foram constatadas as relações
alométricas entre população e crime. No capítulo \ref{cap5} será apresentado um
exercício de aplicação motivado pela existência de uma relação superlinear entre
população flutuante e crimes contra o patrimônio. Especificamente será discutido
a respeito de como esse conhecimento poderia ser aplicado à alocação de recursos
policiais. Por fim, no capítulo \ref{cap6}, serão feitas considerações finais
acerca deste trabalho e apresentada uma discussão a respeito dos trabalhos
futuros identificados.

\chapter{Fundamentação teórica}\label{cap2}

Este capítulo tem como objetivo descrever os principais estudos que investigaram
aspectos reacionados a este trabalho. Na seção \ref{alometria-sec}, serão
explicitados trabalhos que fazem uso de técnicas quantitativas para compreender
fenômenos sociais, com foco principal em identificação de Leis de Escala em
indicadores urbanos. Na seção \ref{criminologia}, serão apresentadas algumas das
principais teorias da sociologia e da criminologia que, de alguma maneira,
relacionam a ocorrência de crimes com as rotinas diárias e mobilidade das
pessoas. Na seção \ref{mobilidade} serão descritos trabalhos que utilizaram
dados de mobilidade humana para compreender fenômenos sociais e urbanos.


\section{Análise quantitativa para compreensão de fenômenos sociais e urbanos}
\label{alometria-sec}

Métodos quantitativos têm sido aplicados amplamente no contexto da compreensão
de fenômenos sociais e urbanos. Enquanto métodos de Regressão Logística têm sido
utilizados para compreender o crescimento de cidades \cite{hu2007} e Modelos de
Regressão para detectar padrões de mobilidade e de expansão urbana
\cite{sampson1999}, o método de Regressão Linear (descrito no Apêndice I) tem
sido utilizado para estimar correlações entre inúmeros indicadores, provendo
explicação quantitativa para diversos fenômenos a nível de escala \cite{
bettencourt2007, melo2014, oliveira2014}.

\begin{figure}[!h]
\includegraphics[width=1\textwidth]{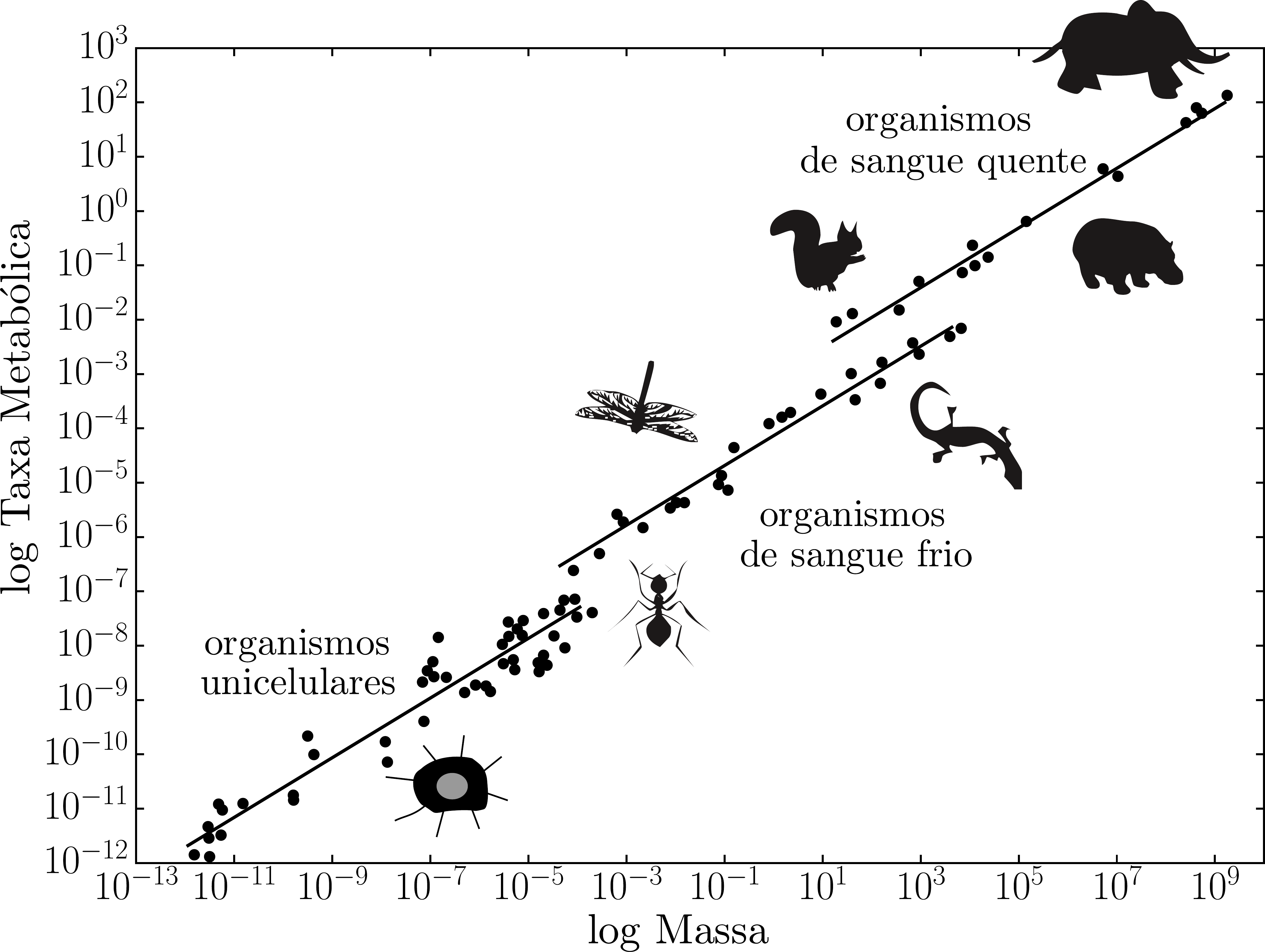}
\caption{{\bf Lei de escala observada por Kleiber.} Mostrou-se que a taxa
metabólica basal e a massa corporal de uma ampla gama de seres vivos estavam
relacionadas por uma potência de 3/4.}
\label{alom-bio}
\end{figure}

Em um contexto mais amplo que apenas a compreensão de fenômenos sociais e
urbanos, é observado que Leis de Escala estão presentes na ciência há séculos.
Em 1838, \cite{rameaux1838} propuseram que a produção de calor em mamíferos era
proporcional à potência de 2/3 da sua respectiva massa corporal. Essa hipótese
ficou conhecida como a Lei de Superfície e sua verificação experimental foi
observada alguns anos depois por \cite{rubner1883}, que mostrou que a taxa
metabólica de cães escalava com a massa corporal elevada a uma potência de 2/3.
Em 1932, \cite{kleiber1932} mostrou que a taxa metabólica basal e a massa
corporal de uma ampla gama de mamíferos estavam relacionadas por uma potência de
3/4 ao invés de 2/3 .Tal resultado ficou conhecido como a Lei de Kleiber e pôs
fim à hipótese defendida pela Lei de Superfície, uma vez que os resultados
obtidos por Kleiber eram irrefutáveis. Vale ressaltar que a Lei de Kleiber é uma
das leis de escala mais famosas da Biologia, uma vez que ela é satisfeita para
vários outros seres vivos, de organismos unicelulares até grandes mamíferos
\cite{kleiber1961,brown2000,simini2010}. A Figura \ref{alom-bio} ilustra a lei
de escala observada por Kleiber: o autor descobriu que que a taxa metabólica de
diversos organismos escalava com a sua massa corporal elevada a uma potência de
3/4.

O termo ``alometria'' foi usado para descrever leis de escala biológicas gerais
durante décadas. Formalmente, uma equação alométrica descreve o
crescimento correlacionado de uma variável $Y$ em função de outra variável $X$,
através de uma função $Y=aX^\beta$, onde $a$ é o coeficiente alométrico e
$\beta$ é o expoente alométrico. O coeficiente alométrico, $a$, é
responsável por dimensionar a diferença de escala entre $X$ e $Y$. Por exemplo,
se $X$ e $Y$ têm crescimento correlacionado, e os valores iniciais de $X$ estão
na ordem de $10^3$ e $Y$ na ordem de $10^2$, $a$ assume valor aproximado de
$0.1$. Caso os valores iniciais de $X$ e $Y$ estejam na mesma ordem de grandeza,
$a \approx 1$. Caso os valores iniciais de $X$ estejam na ordem de $10^3$ e $Y$
na ordem de $10^4$, por exemplo, $a \approx 10$.

\begin{figure}[!h]
\includegraphics[width=1\textwidth]{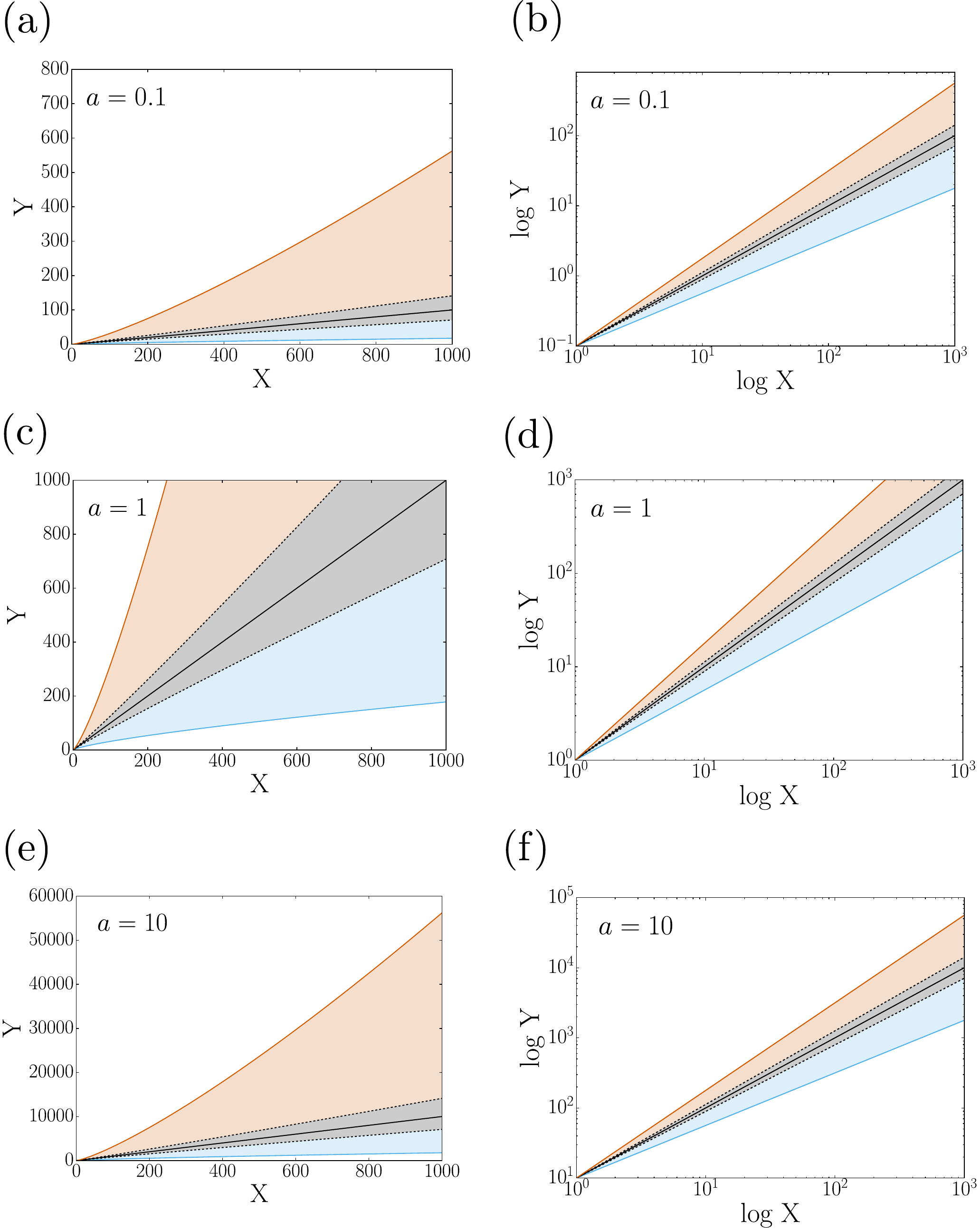}
\caption{{\bf Variação do coeficiente alométrico e expoente alométrico.} Em (a)
é ilustrada a variação do expoente alométrico ao fixar o coeficiente alométrico
com valor $a=0.1$. Em (c) o processo é repetido, no entanto, para $a=1$. Por
fim, em (e), a variação do expoente alométrico é ilustrada para um valor fixo de
$a=10$. E (b), (d) e (f), são ilustradas as mesmas distribuições de (a),(c) e
(d) respectivamente, no entanto os valores das distribuições são ilustrados em
escala logarítmica.}
\label{alometria}
\end{figure}

O expoente alométrico, $\beta$, é responsável por explicar o crescimento
de $Y$ em função de $X$ a nível de proporcionalidade. Quando $\beta \approx 1$ é
observado que o crescimento de $Y$ ocorre de maneira proporcional a $X$,
evidenciando uma relação isométrica entre $X$ e $Y$. $\beta>1$ indica que
$Y$ tem crescimento maior que o proporcional em função do crescimento de $X$,
sendo evidenciada, nesse caso, uma relação alométrica superlinear. Por
fim, quando $\beta<1$, é observado que $X$ cresce a custo de um crescimento
menor que o proporcional de $Y$, indicando uma relação alométrica
sublinear. A Figura \ref{alometria} ilustra a variação de $a$ e $\beta$ para
valores de $X$ e $Y$ em escala linear e logarítmica. Nos exemplos gerados com
valores de $X$, $a$ e $\beta$ sintéticos, o coeficiente alométrico assumiu
os valores 0.1 (a-b), 1 (c-d) e 10 (e-f), já o expoente alométrico assumiu
valores sublineares, isométricos e superlineares, respectivamente, 0.75 (linhas
azuis), 1 (linhas contínuas pretas) e 1.25 (linhas vermelhas). As linhas
pontilhadas pretas representam limiares aceitáveis \cite{bettencourt2007} para
classificar uma relação como isométrica, respectivamente, as linhas pontilhadas
superiores e inferiores foram geradas para $\beta = 1.05$ e $\beta = 0.95$.

Nos anos seguintes surgiram os primeiros trabalhos de alometria em estudos não
biológicos. Nesses trabalhos foi observado o crescimento de indicadores sociais
e urbanos em função de medidas de dimensão de cidades. Uma medida de dimensão de
cidade é um atributo que quantifica o tamanho de uma cidade, {\it e.g.},
população ou características espaciais. Alguns trabalhos investigaram o tamanho
da população e a organização interna das cidades, exemplos incluem: o trabalho
de Zipf, em 1950, sobre o tamanho das cidades e o número de estabelecimentos de
manufatura \cite{zipf1950}; o trabalho de Naroll e Bertalanffy, em 1956, sobre o
tamanho do maior assentamento em tribos primitivas, o número de especialidades
ocupacionais e o número de tipos organizacionais \cite{naroll1956}; o trabalho
de Woldenberg, no ano de 1973, para áreas residenciais, instalações urbanas e
uso da terra \cite{woldenberg1973}; e o trabalho de Newling que, em 1966,
observou uma relação alométrica a partir de dados de densidade populacional
urbana \cite{newling1966}. 

No entanto, em 2007, \cite{bettencourt2007} trouxeram a alometria em estudos
urbanos de volta à discussão acadêmica após extenso período. Em um trabalho
realizado com diversas cidades americanas, chinesas e alemãs, os autores
descobriram relações alométricas e isométricas entre população residente e
diversos indicadores urbanos. Ao utilizar população residente como medida de
dimensão de cidades, os autores acabaram por mensurar indiretamente o volume de
influência social exercido sobre as cidades. Já as relações alométricas e
isométricas encontradas dimensionaram o papel que essa influência social tem
sobre indicadores urbanos a nível de escala. De forma análoga aos trabalhos
realizados na biologia, os autores mensuraram essa relação por uma função do
tipo $Y=aX^\beta$, no entanto, no trabalho de Bettencourt {\it et al.}, $X$ é a
quantidade de população residente e $Y$ corresponde a quantidade de um indicador
urbano. Cada ponto plotado nos gráficos de dispersão de Bettencourt {\it et al.}
correspondia a uma cidade. Os autores delimitaram as cidades pelas suas
fronteiras administrativas.

Dentre os principais resultados obtidos por \cite{bettencourt2007}, destacam-se
as observações de que as grandes cidades potencializam a produção de inovação e
de novas patentes. Observou-se uma relação alométrica superlinear entre
população residente e quantidade de novas patentes e de invenções, em outras
palavras, quando a população de uma cidade dobra, a quantidade de novas patentes
e invenções mais do que dobra. Os autores mostraram também que existe uma
relação alométrica sublinear entre população residente e, por exemplo, a
quantidade de postos de gasolina. Nesse ponto os autores apontam que as grandes
cidades alcançam eficiência similar a de sistemas biológicos (Lei de Kleiber),
devido a constatação de que se uma cidade dobra a sua população é necessário
menos que o dobro de postos de gasolinas para atender a sua demanda. Por fim,
ainda foi observado que alguns indicadores se relacionavam de forma isométrica
com a população, o aumento dos indicadores urbanos lineares são proporcionais ao
tamanho da população das cidades refletindo as necessidades individuais comuns
de cada ser humano, como o consumo de água, o número de empregos e o número de
residências. A Tabela \ref{res-bett} ilustra as relações estatísticas observadas
por Bettencourt {\it et al.}

\begin{table}[!ht]
  \begin{center}
  \caption{Relações alométricas e isométricas observadas por Bettencourt e
  coautores. Retirado de \cite{bettencourt2007}. }
  \label{res-bett}
    \begin{tabular}{ c c c c c c}
      \hline
      Y & $\beta$ & C.I. & $R^2$ & Observações & País Ano \\
      \hline
      Novas patentes & $1.27$ & $[1.25,1.29]$ & $0.72$ & $331$ & US 2001 \\
      
      Inventores & $1.25$ & $[1.22,1.27]$ & $0.76$ & $331$ & US 2001 \\
      
      Empregos de P\&D & $1.34$ & $[1.29,1.39]$ & $0.92$ & $266$ & US 2002 \\
      
      Empregos supercriativos & $1.15$ & $[1.11,1.18]$ & $0.89$ & $287$ & US 2003 \\
      
      Estabelecimentos de P\&D & $1.19$ & $[1.14,1.22]$ & $0.88$ & $287$ & US 1997 \\
      
      Empregos de P\&D & $1.26$ & $[1.18,1.43]$ & $0.93$ & $295$ & China 2002 \\

      Salários & $1.12$ & $[1.09,1.13]$ & $0.96$ & $361$ & US 2002 \\

      Depósitos bancários & $1.08$ & $[1.03,1.11]$ & $0.91$ & $267$ & US 1996 \\

      PIB & $1.15$ & $[1.06,1.23]$ & $0.96$ & $295$ & China 2002 \\

      PIB & $1.26$ & $[1.09,1.46]$ & $0.64$ & $196$ & EU 1999-2003 \\

      PIB & $1.13$ & $[1.03,1.23]$ & $0.94$ & $37$ & Alemanha 2003 \\

      Consumo elétrico total & $1.07$ & $[1.03,1.11]$ & $0.88$ & $392$ & Alemanha 2002 \\

      Novos casos de AIDS & $1.23$ & $[1.18,1.29]$ & $0.76$ & $93$ & US 2002-2003 \\

      Crimes Sérios & $1.16$ & $[1.11,1.18]$ & $0.89$ & $287$ & US 2003 \\

       &  &  &  &  &  \\
      
      Casas & $1.00$ & $[0.99,1.01]$ & $0.99$ & $316$ & US 1990 \\

      Empregos & $1.01$ & $[0.99,1.02]$ & $0.98$ & $331$ & US 2001 \\

      Consumo elétrico doméstico & $1.00$ & $[0.94,1.06]$ & $0.88$ & $377$ & Alemanha 2002 \\

      Consumo elétrico doméstico & $1.05$ & $[0.89,1.22]$ & $0.91$ & $295$ & China 2002 \\

      Consumo de água doméstico & $1.01$ & $[0.89,1.11]$ & $0.96$ & $295$ & China 2002 \\

       &  &  &  &  &  \\

      Postos de gasolina & $0.77$ & $[0.74,0.81]$ & $0.93$ & $318$ & US 2001 \\

      Vendas de gasolina & $0.79$ & $[0.73,0.80]$ & $0.94$ & $318$ & US 2001 \\

      Comp. da fiação elétrica & $0.87$ & $[0.82,0.92]$ & $0.75$ & $380$ & Alemanha 2002 \\

      Superfície da estrada & $0.83$ & $[0.74,0.92]$ & $0.87$ & $29$ & Alemanha 2002 \\

      \hline
    \end{tabular}
  \end{center}  
\end{table}

O trabalho de \cite{bettencourt2007} teve ampla divulgação na comunidade
científica e meios de imprensa. Um dos fatos mais discutidos a respeito desse
trabalho foi a relação superlinear encontrada entre população residente e crimes
sérios, com expoente alométrico $\beta=1.16 \pm 0.04$, em cidades dos Estados
Unidos no ano de 2003. Posteriormente, uma relação alométrica do mesmo tipo,
entre população residente e crimes violentos, foi observada também para cidades
brasileiras, com expoente alométrico $\beta=1.15 \pm 0.03$
\cite{bettencourt2010}.

\begin{figure}[!h]
\includegraphics[width=1\textwidth]{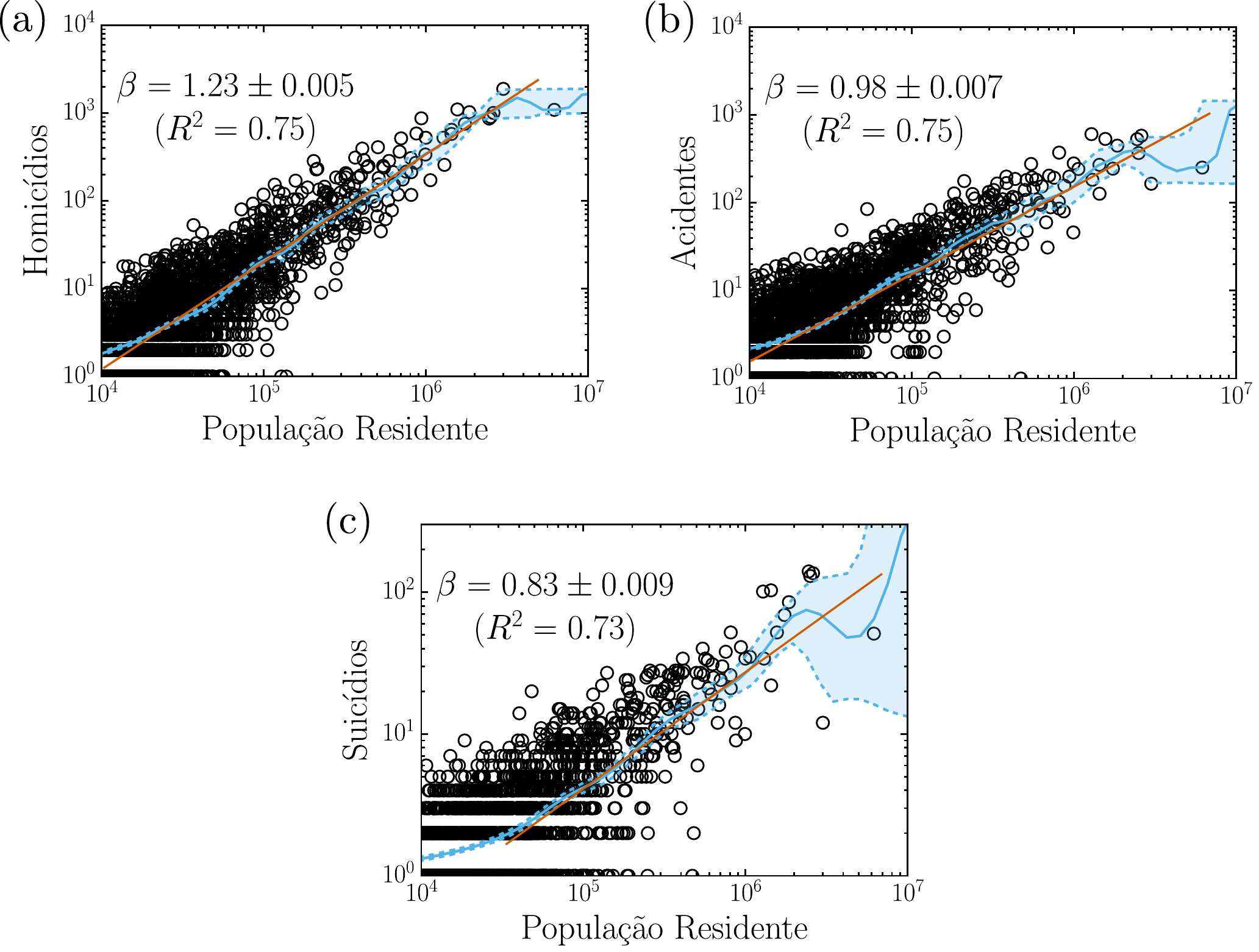}
\caption{{\bf Relações alométricas e isométricas observadas por Melo e
coautores.} (a) Relação alométrica superlinear entre população residente e
homicídios ($\beta=1.23 \pm 0.005$). (b) Relação isométrica entre população
residente e acidentes de trânsito ($\beta=0.98 \pm 0.007$). (c) Relação
alométrica sublinear entre população residente e suicídios ($\beta=0.83 \pm
0.009$). Em todas as figuras os círculos pretos representam cidades brasileiras
e as retas vermelhas representam regressões lineares aplicadas aos dados. As
linhas contínuas azuis foram estimadas pelo método Nadaraya-Watson
\cite{nadaraya1964,watson1964} e as linhas tracejadas azuis delimitam o
intervalo de confiança de 95\% estimado por {\it bootstrap}
\cite{racine2004,li2004}. Ambos os eixos estão em escala logarítmica.}
\label{homicidios-sup}
\end{figure}

Ainda sobre relações superlineares entre população e crime, \cite{melo2014}
encontram forte evidencia quantitativa de que homicídios em cidades brasileiras
escalam, com expoente $\beta=1.23 \pm 0.005$, de maneira superlinear com
população residente. Os autores observaram ainda a existência de uma relação
sublinear entre população residente e suicídios, com expoente alométrico
$\beta=0.83 \pm 0.009$, e relação isométrica entre população residente e
acidentes de trânsito ($\beta=0.98 \pm 0.007$). Foram utilizadas as fronteiras
administrativas definidas pelo governo para delimitar as cidades brasileiras. Na
discussão do trabalho os autores explicam que a influencia social potencializa a
propagação de crimes no espaço urbano, embora também apoie para que ocorram
menos suicídios. A Figura \ref{homicidios-sup} ilustra resultados reproduzidos a
partir do trabalho de \cite{melo2014}.

Trabalhos recentes demonstraram que unidades administrativas de terreno ({\it
e.g.} bairros, setores censitários ou fronteiras municipais) são imprecisas para
compreender fenômenos urbanos, principalmente porque estas são definidas com
restrições que buscam simplificar atividades de censo, como normalizações de
suas fronteiras por estatísticas de população residente \cite{oliveira2014,
arcaute2015, cottineau2016}. Existem diferentes métodos para reconstruir
sistemas urbanos, por exemplo através do crescimento populacional
\cite{rozenfeld2008, giesen2010, rozenfeld2011, duranton2013, gallos2012,
duranton2015, eeckhout2004}, ou métodos que utilizam a agregação por percolação
e limitação de difusão \cite{makse1998, makse1995, murcio2013, fluschnik2014}.

\begin{figure}[!h]
\includegraphics[width=1\textwidth]{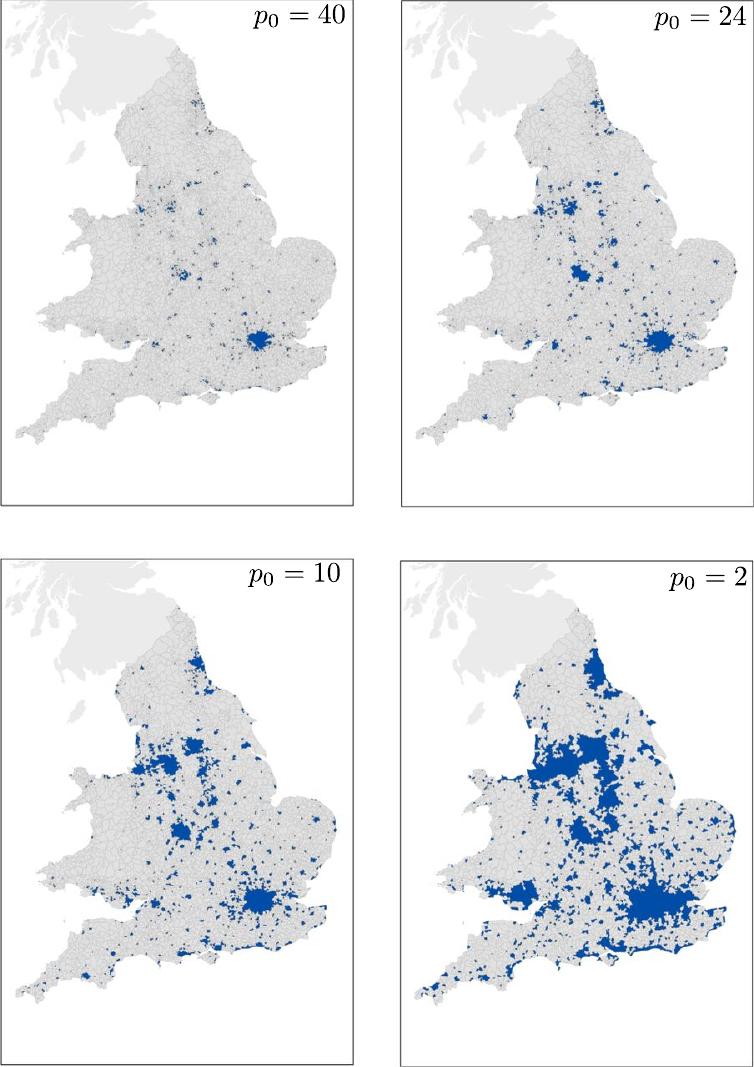}
\caption{{\bf Efeito da variação do parâmetro de densidade populacional na
formação de aglomerados urbanos.} Adaptado de \cite{arcaute2015}.}
\label{fig-paper-arcaute-2015}
\end{figure}

Recentemente, \cite{arcaute2015} desenvolveram um {\it framework} para definir
cidades além de suas fronteiras administrativas. A parte principal desse {\it
framework} é um algoritmo de aglomeração que utiliza a densidade populacional
como parâmetro principal, pois esta é uma propriedade intrínseca de espaços
urbanizados. Os autores definiram o parâmetro de densidade de população, $p_0$,
para situar-se dentro do intervalo [1-40], tendo sido esse parâmetro mensurado
em pessoas por hectare. Para cada limiar inteiro $p_0$ no intervalo, foram
agrupadas todas as unidades de aglomeração adjacentes com densidade $p_w$ tal
que $p_w \geq p_0$. Se uma unidade de aglomeração $k$ tiver uma densidade $p_k <
p_0$, mas é cercada por outra unidade de modo que, para cada unidade adjacente
($w$), $p_w \geq p_0$, a unidade $k$ também é incluída no aglomerado. Uma das
principais limitações desse algorítimo é que ele só produz aglomerados com
superfície contínua. Dentro do intervalo de variação do parâmetro, os autores
estimaram 40 divisões diferentes de sistemas de cidades para Inglaterra e País
de Gales. A Figura \ref{fig-paper-arcaute-2015} ilustra o efeito da variação de
$p_0$ na formação de aglomerados urbanos em um país inteiro.

\begin{landscape}

\begin{figure}[!h]
\includegraphics[width=1.55\textwidth]{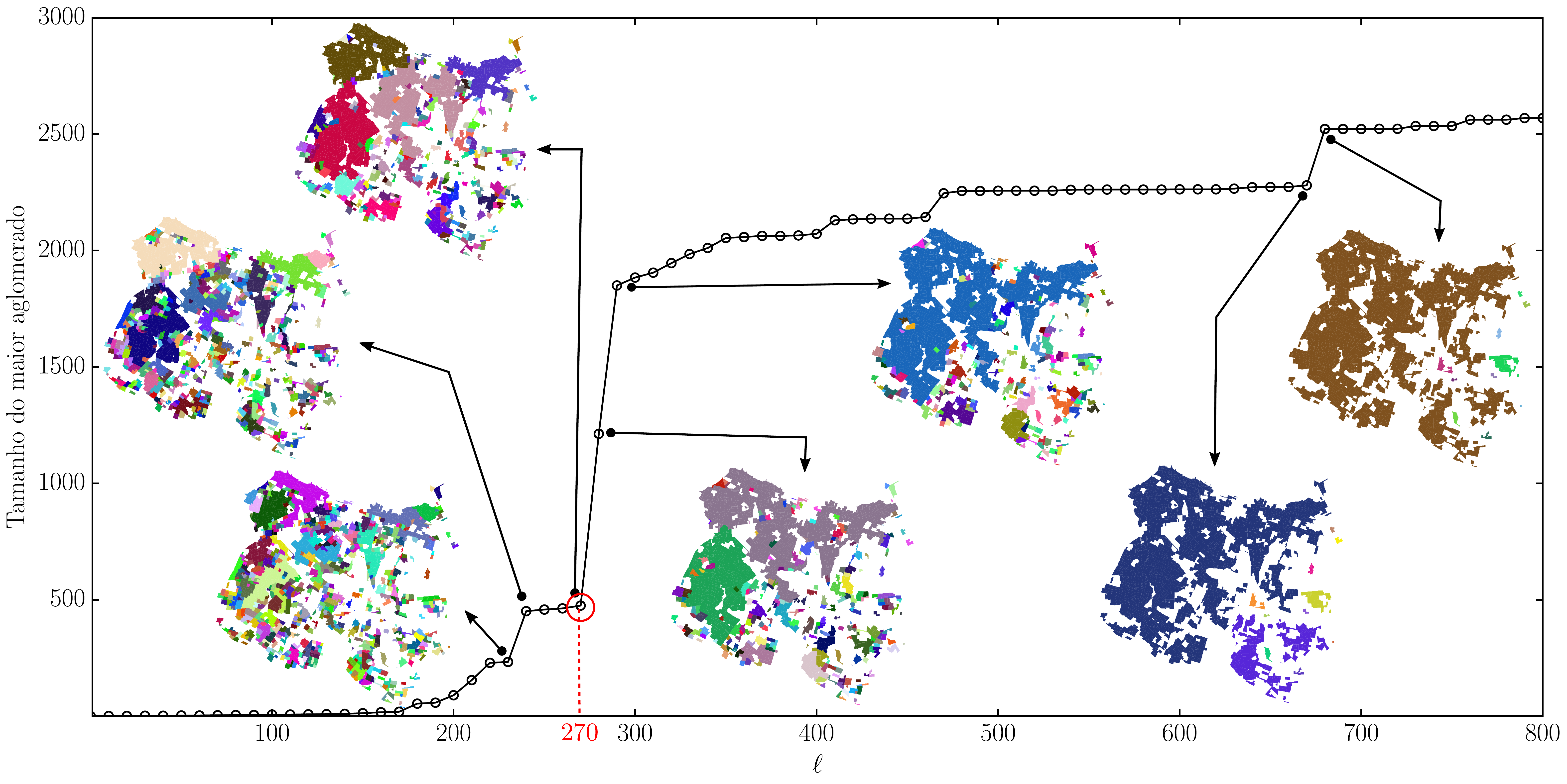}
\caption{{\bf Percolação em unidades administrativas de terreno da cidade de
Fortaleza.} É ilustrado o crescimento do maior aglomerado em função do
incremento de um limiar de distância ($\ell$). Nos mapas embutidos, cada cor
representa um aglomerado construído. Foi utilizado o algoritmo de aglomeração
{\it City Clustering Algorithm} (CCA) para agrupar setores censitários da cidade
de Fortaleza. É identificada uma transição de fase ao incrementar o valor de
$\ell$ de 270 para 280 metros, momento em que ocorre o maior salto no tamanho do
maior agregado. O tamanho do maior agregado foi mensurado a partir do total de
setores censitários agregados no mesmo. O CCA ainda possui um limiar de
densidade populacional ($D^*$), neste experimento foi utilizado $D^* = 6000$
pessoas por $km^2$.} 
\label{perc}
\end{figure}

\end{landscape}

Um ano depois, \cite{arcaute2016} também utilizaram percolação
\cite{stauffer1994} para escolher o melhor valor de $p_0$ na definição dos
aglomerados. Os autores mediram o tamanho do maior aglomero enquanto variavam
$p_0$ em busca de uma transição de fase \cite{stanley1971}. Mais precisamente,
os autores verificaram em que momento o maior aglomerado agregaria o segundo
maior aglomerado, momento esse dito como o ponto crítico do sistema, que ocorre
em uma transição de fase. Comumente é escolhido o valor do parâmetro de
agregação anterior a esse ponto crítico. A Figura \ref {perc} ilustra o processo
de percolação em uma agregação sistemática de unidades administrativas de
terreno da cidade de Fortaleza.

\begin{figure}[!h]
\includegraphics[width=1\textwidth]{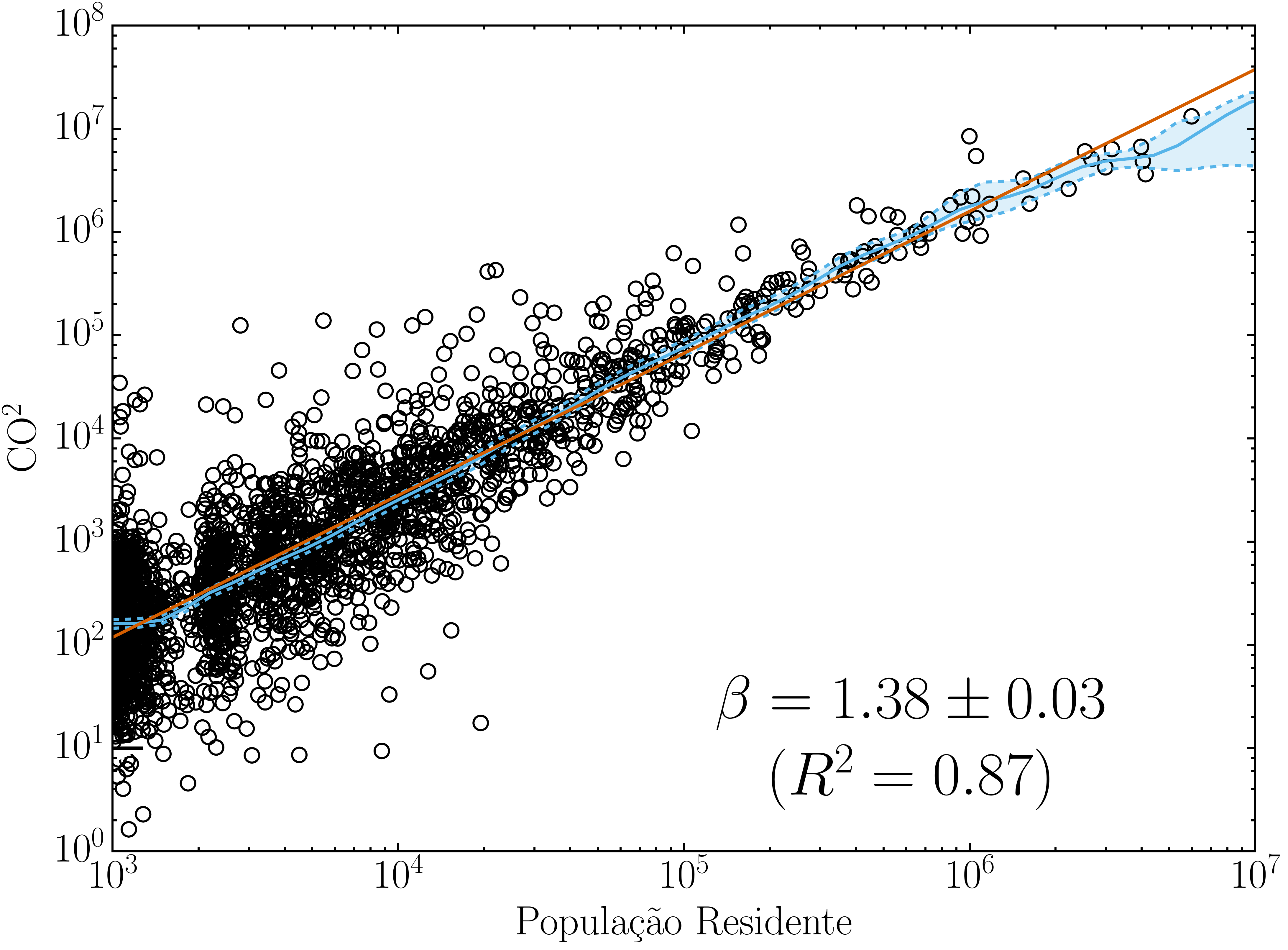}
\caption{{\bf Relação alométrica superlinear observada por Oliveira e
coautores.} Foi encontrado expoente alométrico $\beta=1.38 \pm 0.03$ entre
população residente e e emissão de gás carbônico em cidades americanas. Os
círculos pretos representam cidades americanas e a reta vermelha ilustra uma
regressão linear aplicada aos dados. A linha contínua azul foi estimada
utilizando o método Nadaraya-Watson \cite{nadaraya1964,watson1964} e as linhas
tracejadas azuis delimitam o intervalo de confiança de 95\% estimado por {\it
bootstrap} \cite{racine2004,li2004}. Ambos os eixos estão em escala
logarítmica.}
\label{co2-sup}
\end{figure}

Em 2014, \cite{oliveira2014} investigaram de forma mais profunda a construção de
aglomerados urbanos em um estudo com cidades americanas. Foi utilizado o {\it
City Clustering Algorithm} (CCA) para definir as fronteiras das cidades além dos
seus limites administrativos. Uma cidade do CCA é definida a partir de dois
parâmetros, um limiar de densidade populacional, $D^*$, e outro de distância,
$\ell$. Os autores estudaram o efeito que o crescimento populacional tinha sobre
os altos índices de emissão de gases, mais precisamente foram estudadas as
consequências do crescimento urbano e seus impactos no meio ambiente. É
intensamente debatido se grandes cidades podem ser consideradas ambientalmente
``verdes'', ou seja, se suas produtividades podem ser associadas a baixos
índices de emissão de poluentes \cite{bento2006, kahn2007, brownstone2009,
dodman2009, puga2010, glaeser2010}. Alguns estudos relatam que o nível de
comutação das pessoas, isto é, o processo que as pessoas fazem diariamente para
ir de casa para o trabalho e do trabalho para casa, tem uma grande contribuição
para a relação entre as emissões de poluentes e tamanho da cidade
\cite{bento2006, brownstone2009, glaeser2010}. Como consequência, as cidades
compactas seriam mais verdes, devido à atenuação da duração do percurso das
pessoas durante a comutação. Oliveira {\it et al.} \cite{oliveira2014} mostraram
que as relações alométricas entre as populações de cidades nos Estados Unidos e
suas respectivas emissões de dióxido de Carbono (CO$^2$) são superlineares,
concluindo que cidades grandes são menos verdes, isto é, poluem
proporcionalmente mais do que cidades pequenas. Os autores tiveram como
principal resultado a estimativa do expoente alométrico $\beta=1.38 \pm 0.03$. A
Figura \ref{co2-sup} ilustra um resultado reproduzido a partir do trabalho de
Oliveira {\it et al.} \cite{oliveira2014}.

\section{Crime e mobilidade humana: uma visão sociológica e criminológica}
\label{criminologia}

Ao longo dos anos, foram elaboradas uma série de teorias com objetivo de
explicar a dinâmica da criminalidade. A teoria da Escolha Racional, por exemplo,
parte do princípio de que as ações humanas são ações racionais, fruto de
decisões que envolvem a consideração de diversos fatores. Neste contexto, a
decisão de praticar um crime envolve a avaliação dos riscos e benefícios
decorrentes de diversos fatores, alguns deles presentes no ambiente onde o
indivíduo pretende consumar o ato criminoso, ainda que essa avaliação seja
limitada pelo não-conhecimento de todos os riscos envolvidos
\cite{espacialdistribuiccao, anselin2000a}.

\begin{figure}[!h]
\includegraphics[width=1\textwidth]{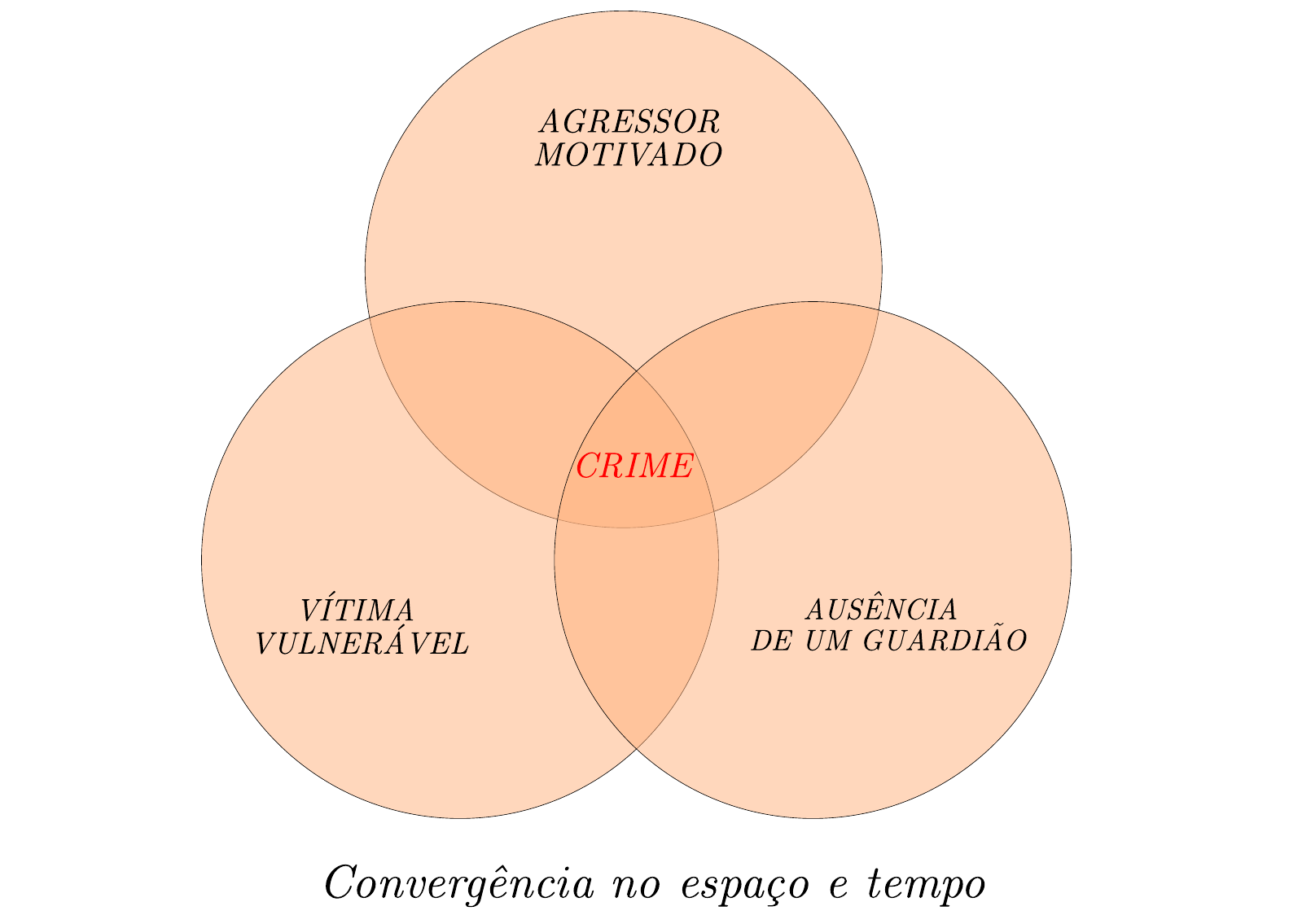}
\caption{{\bf Modelo gráfico da Teoria das Atividades Rotineiras.} São
ilustrados círculos que simbolizam as rotinas de três agentes. A teoria afirma
que um crime ocorre pela convergência espaçotemporal (interseção dos círculos)
das rotinas desses três agentes: Um agressor motivado; uma vítima desprotegida;
e a ausência de um guardião capaz de impedir a transgressão.}
\label{teoria-ativ-rot}
\end{figure}

Outra teoria de grande relevância, especialmente para este trabalho, denominada
teoria das Atividades Rotineiras, foi proposta por Cohen e Felson
\cite{cohen1979} em 1979. Os autores julgam que as atividades diárias criam a
convergência no tempo e espaço de três elementos necessários para que um crime
ocorra: um agressor motivado; um alvo vulnerável; e a ausência de um guardião
capaz. Um agressor motivado é fruto de uma desorganização social onde, a partir
de oportunidades, indivíduos cometem crimes com objetivo de maximizar seus
lucros. Alvos vulneráveis podem ser pessoas, locais e/ou produtos. Se o crime é
um arrombamento, por exemplo, o alvo adequado deve ser um local em que se
acredita haver um objeto de valor. Por outro lado, se o crime é um roubo a
pessoa, o alvo adequado será um indivíduo que é percebido carregando objetos de
valor, desprotegido e, possivelmente, sem condições de reagir. Por fim, um
guardião capaz é uma pessoa ou equipamento que desencoraje a prática do crime.
Esse guardião pode ser formal (policial ou sistemas de segurança) ou informal
(testemunhas). A Figura \ref{teoria-ativ-rot} ilustra o modelo gráfico clássico
dessa teoria.

Motivados pela grande dificuldade que os estudos contemporâneos apresentavam
para explicar as mudanças anuais nas tendências da taxa de criminalidade nos
Estados Unidos, essencialmente no período pós Segunda Guerra, os autores
iniciaram uma interessante discussão a respeito de como o incremento nas
oportunidades de desfrutar dos benefícios da vida também aumentam a oportunidade
de violações predatórias. Por exemplo, os automóveis proporcionam liberdade de
movimento aos infratores, bem como aos cidadãos comuns e oferecem alvos
vulneráveis para que roubos ocorram. A matrícula na faculdade, a participação da
força de trabalho feminina, a urbanização, a suburbanização, as férias e os
novos bens duráveis eletrônicos oferecem várias oportunidades de escapar dos
confins da família enquanto aumentam o risco de vitimização predatória.

A teoria ainda argumenta que as oportunidades disponíveis são um componente
importante na construção de um crime. Escolhas no estilo de vida por parte de
potenciais vítimas podem criar ou reduzir oportunidades de cometer crimes para o
criminoso motivado.

Alguns conceitos da teoria das Atividades Rotineiras foram aplicados por
Brahtingham e Brahtingham na elaboração da teoria dos Padrões Criminais
\cite{anselin2000b}. Os autores argumentam que a relação entre ambiente e crime
não deve ser avaliada sobre uma perspectiva espacial simples. Existem
características do ambiente que são potencialmente favoráveis à ocorrência de
eventos criminais e são percebidas por delinquentes e associadas à sua motivação
pessoal para cometer crimes, essencialmente, no contexto de seus deslocamentos
cotidianos. Através da interação entre percepção, cognição e comportamento, o
potencial criminoso constrói padrões criminais mentais para suas ações,
ajudando-o a avaliar um tipo particular de objeto, lugar ou situação, num
contexto que os autores denominam Mosaico Urbano. Por esse motivo, a
Criminologia Ambiental ({\it Environmental criminology}), como esse campo de
estudo é chamado, constitui uma abordagem que envolve conhecimentos de diversas
áreas {\cite{brantingham1981}}.

A teoria das Janelas Quebradas explica o efeito normativo e sinalizador da
desordem urbana e do vandalismo sobre o crime e o comportamento antissocial. A
teoria afirma que a manutenção e monitoramento de ambientes urbanos para
prevenir pequenos crimes, como vandalismo e uso excessivo de bebida alcoólica,
ajuda a criar uma atmosfera de ordem e legalidade, impedindo assim que pessoas
que trafegam pelo espaço urbano venham a cometer mais atos ilegais. A teoria foi
introduzida em 1982 pelos sociólogos James Q. Wilson e George L. Kelling
\cite{wilson1982, kelling1997}, desde então tem sido objeto de grande debate
tanto no âmbito das ciências sociais quanto no âmbito da esfera pública. A
teoria foi utilizada durante muitos anos como motivação para várias reformas na
política criminal norte americana.

Antes mesmo da introdução dessa teoria, o psicólogo Zimbardo organizou um
experimento testando o seu conceito principal em 1969 \cite{zimbardo1969}.
Zimbardo utilizou um automóvel sem placa de licença e o estacionou ocioso em um
bairro onde o espaço urbano estava deteriorado e um segundo automóvel, na mesma
condição, em outro bairro, onde não haviam sinais de deterioração. O primeiro
carro foi atacado poucos minutos depois de seu abandono. Zimbardo observou que
os primeiros praticantes de vandalismo a chegar eram uma família - um pai, uma
mãe e um filho - que removeram o radiador e a bateria do veículo. Dentro de
vinte e quatro horas de seu abandono, tudo de valor tinha sido retirado do
veículo. Depois disso, as janelas do carro foram quebradas, estofados rasgados,
e as crianças estavam usando o carro como um {\it playground}. Ao mesmo tempo, o
segundo veículo permaneceu intacto por mais de uma semana até que Zimbardo se
aproximou do veículo e o despedaçou deliberadamente com uma marreta. Logo depois
observou-se que outras pessoas passaram também e cometer atos de vandalismo com
o veículo. Zimbardo ainda observou que a maioria dos adultos vândalos, em ambos
os casos, eram principalmente pessoas bem vestidas e aparentemente indivíduos
respeitáveis, muitos vezes vindos de outras regiões da cidade. Acredita-se que,
em um bairro cujo o espaço urbano é deteriorado, onde é predominante a presença
de propriedades abandonadas, por exemplo, roubos são mais prevalentes e
vandalismos ocorrem muito mais rapidamente. Eventos semelhantes podem ocorrer em
qualquer comunidade civilizada quando as barreiras comunais - o sentido de
respeito mútuo e obrigações de civilidade - são apequenadas por ações que
sugerem apatia.

Após a criação da teoria das Janelas Quebradas e diante da constatação recente
de que os estilos de vida das pessoas influenciam na ocorrência de crimes
\cite{cohen1979}, em 1984 \cite{brantingham1984} foi fortalecida a necessidade
de investigar a movimentação das pessoas. Na criminologia é estabelecido que os
criminosos não se envolvem em crimes em áreas distantes do seu local de
residência \cite{brantingham1984}. É extensa a lista de estudos que têm,
consistentemente, demonstrado que, em média, os infratores residem a uma
distância de duas milhas das áreas onde cometem seus crimes \cite{groff2012}.
Esse fenômeno é descrito como Degradação na Distância do Crime, pois na medida
que a distância entre a residência de um agressor e um local - que propicie
oportunidades de cometer crimes - aumenta, o número de crimes cometidos pelo
infrator diminui. De acordo com a pesquisa, esse padrão é resultado da
relutância dos infratores em viajar para áreas que não conhecem
\cite{brantingham1981}, pois as mesmas proporcionariam um incremento nas chances
de apreensão. Embora destinado a explicar padrões de mobilidade dos criminosos,
a degradação na distância do crime encontrou também alguma validade na
explicação dos padrões de vitimização. Estudiosos que examinaram os padrões de
mobilidade das vítimas, de fato, descobriram que a vitimização (semelhante aos
infratores) tende a ocorrer em áreas próximas da residência do indivíduo vítima
\cite{bullock1955,caywood1998,messner1985}.

A partir da degradação na distância do crime, Pizarro {\it et al.}
\cite{pizarro2007} decidiram examinar os padrões de mobilidade das vítimas de
homicídios e dos infratores. A motivação para tal estudo veio a partir da
observação de que existiam algumas limitações na literatura até então, pois
poucos estudos examinavam os padrões de mobilidade das vítimas e dos infratores
\cite{bullock1955, caywood1998, groff2012, messner1985, pokorny1965, rand1986,
tita2005}. Basicamente, Pizzarro {\it et al.} estavam interessados em responder
três questões de pesquisa relacionadas a homicídios:
\begin{enumerate} [label=(\roman*)]
\item A distância percorrida por suspeitos e vítimas a partir de suas respectivas
residências para o local do incidente varia de acordo com o fato que motivou o
homicídio?
\item A razão pela qual as vítimas e suspeitos de homicídio viajam para
o local do incidente variam de acordo com o tipo de motivo de homicídio?

\item Que características da vítima e suspeito predizem significativamente a
distância percorrida da residência para a localização do incidente? 

\end{enumerate}

As análises dos autores sobre os dados da unidade de homicídios do Departamento
de Polícia de Newark, mostraram que existem diferenças estatisticamente
significativas entre os tipos de homicídios em termos de duração das viagens das
vítimas e dos suspeitos.

Ainda a respeito de teorias criminológicas, a teoria da Vitimização Repetida
({\it Repeat Victimization}) \cite{lauritsen1995} também é de grande relevância.
De acordo com muitas das suas definições, na repetição da vitimização, ou
revitimização, o mesmo tipo de incidente criminal é experienciado pela mesma
vítima, - ou virtualmente a mesma – ou alvo, dentro de um período de tempo
específico. Um alvo considerado "virtualmente" o mesmo, pode ser representado
por duas ou mais pessoas distintas, que estão conectadas por um mesmo contexto,
como por exemplo, trafegar por um determinado local, portando um mesmo objetivo
de valor, na mesma faixa de horário. Uma característica crítica e consistente da
repetição da vitimização é que as infrações repetidas ocorrem rapidamente -
muitas repetições ocorrem dentro de uma semana da ofensa inicial, e algumas até
ocorrem dentro de 24 horas. Um estudo inicial sobre essa teoria mostrou o maior
risco de um roubo repetido foi durante a primeira semana após um roubo inicial
\cite{wartell2004}.

Em 1995, Lauritsen e Quinet \cite{lauritsen1995} expandiram os conhecimentos
existentes sobre a vitimização, descrevendo os padrões temporais de risco e
desenvolvendo e testando modelos explicativos do vínculo entre os riscos
passados e futuros. As análises baseadas em dados de painel do {\it National
Youth Survey} suportam a dependência do estado e as interpretações de
heterogeneidade da correlação no risco ao longo do tempo. Em outras palavras, a
vitimização prévia prevê o risco futuro, em parte, porque altera algo sobre o
indivíduo, e porque indica uma propensão não mensurada para a vitimização que
persiste ao longo do tempo. As implicações teóricas desses achados, incluindo a
viabilidade de uma perspectiva de rotulagem da vítima, foram discutidas no
trabalho.

Mais recentemente, em 2008, Daigle {\it et al.} \cite{daigle2008} perceberam um
grau significativo de repetição em crimes de violência contra mulheres
americanas em uma análise sobre dois conjuntos de dados de nível nacional. Os
autores observaram que uma pequena proporção de mulheres da faculdade são
vítimas de uma grande proporção de violências sexuais. Também foi constatado que
as mulheres são mais propensas a sofrer uma repetição de vitimização sexual do
que repetir incidentes de violência. Observou-se que a repetição da vitimização
tende a acontecer no mesmo mês da vitimização inicial, e o próximo tipo de
vitimização mais provável é, de longe, do mesmo tipo. Os autores ainda mostram
que, entre as mulheres vitimadas, a repetição da vitimização é uma experiência
comum, atingindo entre 14\% e 26\% delas durante um ano letivo. De maneira
adicional é mostrado que a repetição da vitimização ocorre rapidamente, com o
risco de outro pico de vitimização no tempo imediatamente após a vitimização
inicial e, em seguida, diminuindo ao longo do tempo. Por fim, os autores
reforçam que a repetição da vitimização, é pouco compreendida e não é
sistematicamente abordada por programas de prevenção sexual ou de violência.

\section{Estimação de padrões de mobilidade} \label{mobilidade}

Recentemente se percebeu um considerável crescimento na quantidade de artigos
científicos que fazem uso de dados de mobilidade humana para estudar os mais
diversos fenômenos. Para prever, desde a disseminação de vírus humanos e
eletrônicos, até fazer planejamento urbano e gerenciamento de recursos em
comunicação móvel, mostrou-se ser fundamental a utilização de dados precisos de
mobilidade. Apesar de ainda ser limitado o acesso a esses dados, devido a,
essencialmente, questões de privacidade, o avanço tecnológico e, principalmente,
o considerável incremento no número de sensores digitais, permite que, nos dias
de hoje, alguns cientistas estudem a mobilidade das pessoas. Atualmente,
observa-se que as informações mais detalhadas a respeito desse tipo de dado são
coletadas por operadoras de telefonia móvel \cite{gonzalez2008, song2010,
hidalgo2008, eagle2006, gonzalez2006, eagle2009, lambiotte2008}, no entanto são
inúmeros os trabalhos que utilizam dados de {\it survey} \cite{liang2013} e
bilhetagem eletrônica \cite{ gordillo2006, caminha2016a, caminha2016b} para
compreender fenômenos relacionados à mobilidade das pessoas.

A respeito da possibilidade de se inferir padrões de mobilidade através de dados
de dispositivos móveis, Gonzalez {\it et al.} \cite{gonzalez2008}, em 2008,
estudaram a trajetória de 100000 usuários de telefones celulares anônimos, cuja
posição foi monitorada por um período de seis meses. Foi constatado que, em
contraste com as trajetórias aleatórias preditas por modelos como {\it Lévy
Flight} e {\it Random Walk} \cite{brockmann2006}, as trajetórias humanas mostram
um alto grau de regularidade temporal e espacial, sendo cada indivíduo
caracterizado por uma distância de viagem independente do tempo e uma
probabilidade significativa de retornar a alguns locais altamente frequentados.
Os padrões individuais de viagem convergem em uma única distribuição de
probabilidade espacial, indicando que, apesar da diversidade dos dados
históricos de viagem, os seres humanos seguem padrões reproduzíveis simples. Os
autores argumentam que esta semelhança inerente nos padrões de viagem poderia
impactar em todos os fenômenos impulsionados pela mobilidade humana, tais como
prevenção de epidemias e planejamento urbano.

O trabalho de Gonzalez {\it et al.} \cite{gonzalez2008} inspirou outras
produções científicas, sempre objetivando prever a mobilidade das pessoas. Nesse
aspecto, destaca-se o trabalho de Chaoming Song {\it et al.} \cite{song2010} que
exploraram os limites da previsibilidade na dinâmica do deslocamento humano. Os
autores estudaram padrões de mobilidade de usuários de telefones celulares
anônimos e ao medir a entropia da trajetória de cada indivíduo, foi encontrado
que é possível potencialmente prever 93\% da mobilidade de usuários utilizando
dados históricos de telefonia móvel.

Após se perceber que é possível identificar padrões claros de deslocamento a
partir de dados de telefonia móvel, uma série de pesquisadores buscaram
compreender as relações sociais de indivíduos a partir de padrões de mobilidade.
Nesse aspecto é relevante o trabalho de Hidalgo e Rodriguez (2008)
\cite{hidalgo2008}, que definiram e mediram a persistência de laços de
relacionamento em um período de um ano usando dados de atividade de todas as
chamadas de voz realizadas por uma operadora de telefonia móvel de um país
industrializado. Foi construída uma rede de relacionamentos e mostrado que a
persistência dos laços depende de características topológicas dessa rede. Mais
precisamente, os autores mostraram que os laços persistentes tendem a ser
recíprocos e são mais comuns para pessoas com baixo grau \cite{albert2002} e
coeficiente de agrupamento alto \cite{holland1971}.

Ainda em 2008, Lambiotte {\it et al.} \cite{lambiotte2008}, detectaram um
padrão espacial de relacionamento entre pessoas a partir de dados de mobilidade.
Mais precisamente, os autores mostraram que a probabilidade de dois usuários
manterem uma relação, segue um modelo de gravidade \cite{anderson2003}, ou seja,
diminui como $d^{-2}$, onde $d$ é a distância entre esses usuários. Foram
utilizados dados de uma empresa de telefonia móvel, que possuía até o momento da
pesquisa 2.5 milhões de clientes e 810 milhões de comunicações (chamadas
telefônicas e mensagens de texto) durante um período de seis meses.

Em 2009, Eagle {\it et al.} \cite{eagle2009} formularam a seguinte hipótese: Os
dados coletados de telefones celulares têm o potencial de fornecer uma visão
sobre a dinâmica dos relacionamentos das pessoas, revelando organizações,
comunidades e, potencialmente, sociedades. Os autores validaram sua hipótese
mostrando que é possível inferir, com precisão, 95\% das amizades com base
apenas nos dados de operadoras de telefonia móvel. O autores validaram as
amizades inferidas comparando seus resultados com dados de relatórios auto
declarados dos usuários. Foi ainda mostrado que os pares de amigos demonstram
padrões temporais e espaciais independentes de sua proximidade física e padrões
de chamada.

Dados de telefonia móvel são possivelmente os mais precisos para estimar padrões
de mobilidade. Diante da dificuldade de obter esse tipo de dado, uma série de
pesquisadores buscaram estudar dados de GPS ({\it Global Positioning System}) e
bilhetagem (cobrança automatizada) de ônibus, com o objetivo, muitas vezes, de
estimar os mesmos padrões de movimentação das pessoas. Nesse contexto, Gordillo
\cite{gordillo2006} analisou dados de bilhetagem e GPS de ônibus em 2006. O
resultado da sua pesquisa é uma metodologia para descobrir as origens e destinos
específicas para diferentes períodos de tempo, dias da semana e estações do ano
de cidadãos de uma grande metrópole. Uma das grandes vantagens de se estimar
padrões de mobilidade a partir da análise de dados de cobrança automática é que,
essas análises, devem ser capazes de obter uma visão detalhada, contínua e
precisa do comportamento da mobilidade das pessoas, por uma fração do custo
normalmente obtido em pesquisas baseadas em coleta de dados via {\it survey}
\cite{lam1991,remya2013}.

Ainda a respeito do problema de estimativa de origens e destinos, enquanto Remya
e Mathew (2013) \cite{remya2013} utilizaram redes neurais artificiais
\cite{wang2003, park1991, hsu1995} para compreender a demanda de redes de ônibus
hipotéticas, Hua-ling (2007) \cite{hua2007} analisou o impacto de
congestionamentos de trânsito na estimativa da dinâmica das origens e destinos
das pessoas. Hua-ling fez uso de dados de ocupação de vias e avaliou os trechos
de uma cidade e o seu grau de ocupação para propor uma metodologia capaz de
estimar as origens e destinos mais frequentes da cidade.

Em 2016, Caminha {\it et al.} \cite{caminha2016a} utilizaram dados de bilhetagem
e GPS de ônibus para estimar origens e destinos de usuários do sistema de ônibus
de uma grande metrópole. Em um processo de caracterização da rede de ônibus
baseado no cálculo de métricas de redes complexas \cite{albert2000,
strogatz2001, albert2002}, os autores sugeriram a criação de linhas expressas
para levar passageiros entre partes de rede que aparentemente possuíam gargalos.
Ainda em 2016, os mesmos autores, em outro trabalho \cite{caminha2016b},
utilizaram as origens e destinos estimadas para reconstruir as rotas dos
usuários de ônibus. Em um estudo mais profundo a respeito da oferta e da demanda
de uma rede de ônibus, os autores apresentaram uma metodologia para encontrar
gargalos microscópicos e desequilíbrios entre oferta e demanda em uma rede de
transporte público. Na discussão final dos trabalhos, os autores ainda alertam
para o potencial que a reconstrução das rotas dos usuários tem na compreensão da
dinâmica de espalhamento de epidemias, como a dengue \cite{nogueira2001}.

A respeito disso, é digno de destaque que a mobilidade humana também tem sido
utilizada para explicar a propagação de vírus humanos e eletrônicos. Colizza
{\it et al.} \cite{colizza2007} mostraram que a avaliação do transporte aéreo é
crucial no cálculo da probabilidade de ocorrência de surtos globais de doenças.
Já Hufnagel {\it et al.} \cite{hufnagel2004} apresentaram um modelo
probabilístico que descreve a disseminação mundial de doenças infecciosas e
demonstram que uma previsão da propagação geográfica de epidemias é de fato
possível. Seu modelo combina uma dinâmica de infecção estocástica local entre
indivíduos com transporte estocástico em uma rede mundial. Os autores levaram em
conta o tráfego nacional e internacional de aviação civil. Wang {\it et al.}, em
2009, \cite{wang2009}, modelaram a mobilidade de usuários de telefones celulares
para estudar os padrões de propagação fundamentais que caracterizam um surto de
vírus eletrônico móvel. Os autores descobriram que, embora os vírus {\it
bluetooth} possam chegar a todos os aparelhos suscetíveis com o tempo, eles se
espalham lentamente devido à mobilidade humana, oferecendo amplas oportunidades
para implantar um {\it software} antivírus.


Em 2011, Andresen \cite{andresen2011} introduziu o conceito de população
ambiente, que é uma estimativa média, por período de 24 horas, da quantidade de
pessoas em uma determinada área. Andresen utilizou dados do Laboratorio Nacional
de Oak Ridge para estimar uma taxa denominada população em risco. A taxa
em questão mensurava o risco que vítimas corriam ao estar em um determinado
local de uma grande metrópole. Esse risco era maior ou menor dependendo do tipo
de crime que era correlacionado com essa população ambiente.

Em 2015, Malleson e Andresen \cite{malleson2015} publicaram um artigo de grande
relevância, especialmente para este trabalho. Os autores se interessaram por
estudar a taxa de criminalidade, uma estatística usada para resumir o risco de
eventos criminosos. Motivados por trabalhos anteriores que mostraram que a
escolha do denominador apropriado para medir essa taxa não é trivial
\cite{andresen2011} e pelo fato de que população residente é a medida
populacional mais comumente utilizada para mensurar risco, Malleson e Andresen
mostraram que população flutuante é mais adequada para mensurar o risco para
alguns tipos de crimes, {\it e.g.}, roubos e furtos. 

Malleson e Andresen utilizaram dados {\it crowdsourced} - especificamente, da
rede social {\it Twitter} - na cidade de Leeds (Inglaterra) para medir a
população em risco considerando o crimes violentos. Diante da dificuldade
constatada de se obter dados reais de mobilidade humana, os autores tiveram como
primeira contribuição mostrar que população flutuante pode ser estimada
indiretamente a partir de dados de redes sociais. A principal contribuição dos
autores foi mostrar que diferentes padrões espaciais de taxas de criminalidade
surgem quando se utilizam duas medidas de população em risco diferentes: a
população residente (medida pelo censo 2011 do Reino Unido); e a população
flutuante (medida pelo número de mensagens do {\it Twitter}). Possivelmente seu
resultado mais relevante esteja relacionado a análise do centro comercial da
cidade de Leeds. Embora esta área tenha um grande volume de crimes violentos, a
mesma não exibe uma taxa estatisticamente significativa quando a população
flutuante é usada para medir a população em risco. Consequentemente, apesar do
alto volume de crimes violentos, não há uma elevação estatisticamente
significativa no risco dessa região. Os autores ainda reforçam que nenhuma
conclusão desse tipo, em relação ao centro da cidade, teria sido alcançada
utilizando apenas a população residente.

\chapter{Estimação de população flutuante a partir de dados de GPS e
bilhetagem de ônibus} \label{proc-pf}

Neste capítulo será descrito o processo de estimativa do volume de população
flutuante de diversas sub-regiões da cidade de Fortaleza. Foram utilizados dados
gerados por sensores do sistema de ônibus da cidade para mensurar a quantidade
média de pessoas que passam por cada setor censitário da cidade no decorrer de
um dia útil. Com base em heurísticas encontradas no estado da arte de Engenharia
do Transporte, foram estimadas origens e destinos dos usuários do sistema e
reconstruídos seus caminhos completos pela rede de ônibus. A partir dessas 
estimativas foi mensurado o volume de população flutuante nos setores 
censitários de Fortaleza.

\section{Contexto}

O sistema de mobilidade urbana de uma grande cidade é composto por várias redes
interligadas, como redes de metrô, ônibus, bicicletas, táxis e veículos
particulares. Os ônibus são o principal meio de transporte para a maioria dos
habitantes da cidade de Fortaleza, sendo utilizados por $\sim 700\mathrm{k}$
pessoas diariamente. De acordo com o IBGE \footnotemark[1]\footnotetext[1]
{http://www.cidades.ibge.gov.br/v3/cidades/municipio/2304400}, Fortaleza tem
$\sim 556\mathrm{k}$ carros e $\sim 265\mathrm{k}$ motocicletas. Supondo que o
limite superior para o deslocamento diário utilizando carros e motocicletas seja
de ($\sim 821\mathrm{k}$), o total de deslocamentos em Fortaleza é $\sim
1.5\mathrm{M}$. Portanto, o sistema de ônibus representa pelo menos $\sim 46\%$
da mobilidade urbana dentro de Fortaleza e são conhecidas as rotas de pelo menos
$\sim 19\%$ do total de deslocamentos. Levando em conta esse fato, é assumido
que a mobilidade urbana dentro da cidade pode ser representada pelo uso do
sistema de ônibus. Assim, as trajetórias dos usuários de ônibus serão utilizadas
para inferir o volume de população flutuante nos diferentes pontos da cidade.

\section{Conjuntos de dados de mobilidade}

Para compreender o fluxo de pessoas em toda a cidade, foram utilizados quatro
conjuntos de dados espaço-temporais relacionados à rede de ônibus de Fortaleza,
que são: 
\begin{enumerate} [label=(\roman*)]
\item Paradas de ônibus.
\item Rotas de ônibus.
\item {\it Global Positioning System} (GPS).
\item Bilhetagem de ônibus (Programa do Bilhete Único).
\end{enumerate}

No total, Fortaleza possui 4783 paradas de ônibus atendidas por 2034 ônibus que
circulam em 359 rotas (linhas de ônibus) diferentes. Cada ônibus da cidade é
equipado com um sistema de GPS, que registra a posição de ônibus em intervalos
de aproximadamente trinta segundos. Em um dia útil, são registrados cerca de
quatro milhões de coordenadas geográficas para os ônibus de Fortaleza. 

O modelo de transporte integrado adotado pela prefeitura de Fortaleza, nomeado
Bilhete Único, permite que os usuários registrados façam uma transferência de
ônibus em qualquer lugar da cidade, contanto que seja dentro de duas horas desde
a última validação de seu cartão. O processo de validação do Bilhete Único é
entendido como o ato de o usuário deslizar o seu cartão no leitor eletrônico
localizado próximo a catraca do ônibus ou a catraca de um terminal de ônibus.
Geralmente, tal procedimento acontece no início da viagem, já que a catraca está
próxima à entrada do ônibus em Fortaleza. Em um dia útil ocorrem, em média, 1.2
milhões de validações do Bilhete Único.

Todos os conjuntos de dados de mobilidade utilizados referem-se a duas semanas
do mês de março de 2015, especificamente do dia 4 ao dia 18 de março. A seguir
serão detalhadas estimativas de mobilidade dos usuários, buscando-se estimar um
comportamento médio de mobilidade de um dia útil na cidade de Fortaleza. Nos
dias úteis a rede de ônibus tem seu pico de uso \cite{caminha2016a,
caminha2016b, furtado2017}.

\section{Estimativa e validação de uma matriz origem-destino}

Foi estimada uma matriz de origem-destino dos usuários de ônibus de Fortaleza
através de heurísticas retiradas do estado da arte de Engenharia do Transporte.
É assumido que o ponto de origem de um usuário é o local da sua primeira
validação diária \cite{gordillo2006,lam1991}. Utilizando dados do dia onze de
março, para cada usuário foi recuperado o local da sua primeira validação no dia
\cite{remya2013,hua2007}. Objetivando minimizar o problema de que um usuário
pode permanecer durante boa parte da viagem na parte de trás do veículo, fazendo
assim a validação do seu bilhete em um ponto distante da sua origem, foram
analisados dados do seu Bilhete Único durante duas semanas (uma semana antes e
uma semana depois), e foi assumindo como origem do mesmo o ponto de validação
mais próximo do início da linha de ônibus que ele utilizou. De forma similar, um
ponto de destino de um usuário pode ser assumido como o local da sua última
validação no dia, sendo verificado se o local se repete no intervalo de duas
semanas (também uma semana antes e uma semana depois) do dia onze
\cite{gordillo2006,lam1991}.

Ao todo, foram estimadas origens e destinos de aproximadamente 40\% dos usuários
do sistema de ônibus de Fortaleza. É digno de destaque que essa amostra é 
significativamente maior que amostras extraídas por técnicas de {\it survey},
onde normalmente são entrevistados entre 0.5\% e 1\% dos usuários. 

\begin{figure}[!h] \includegraphics[width=1\textwidth]{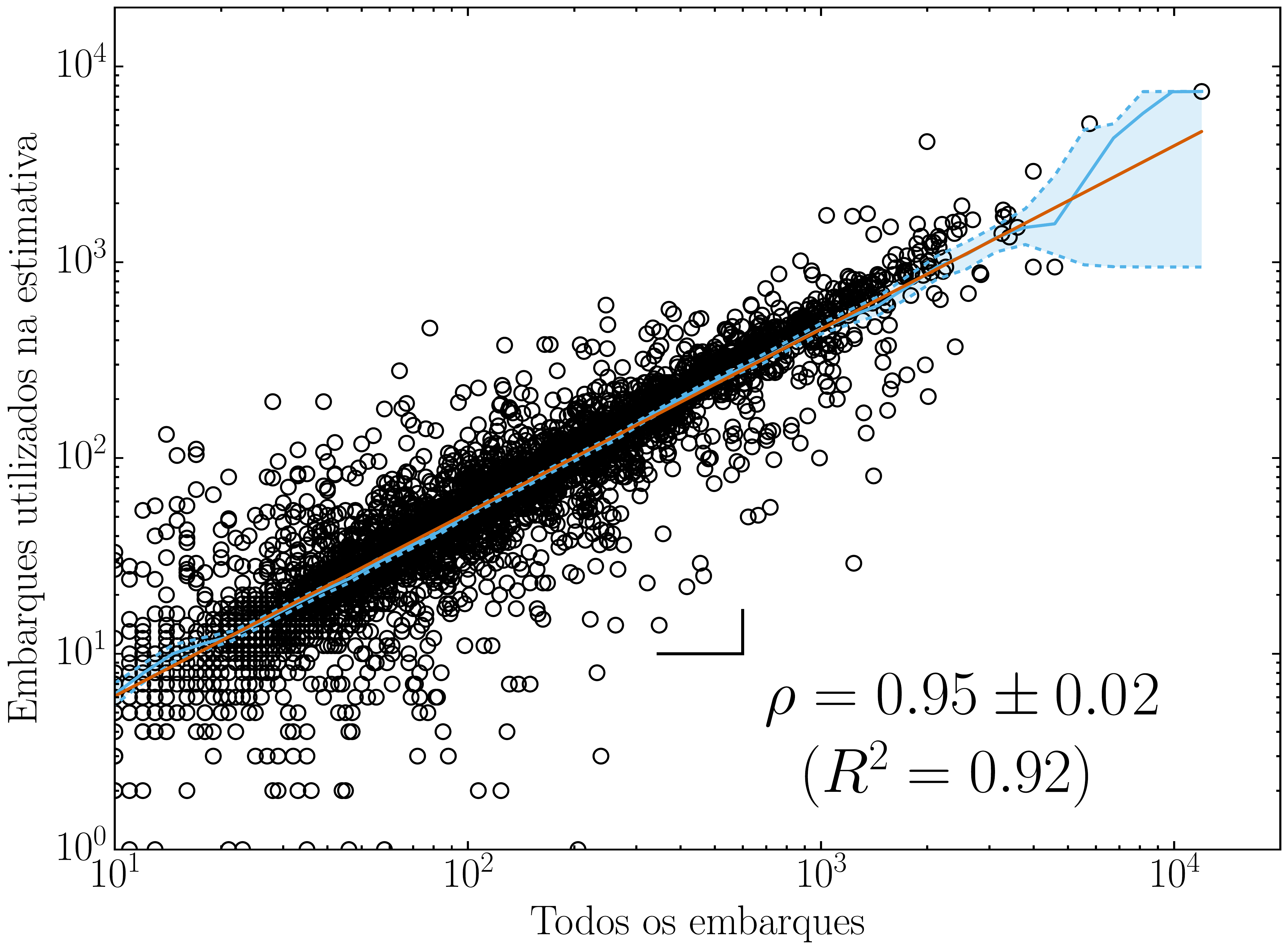}
\caption{{\bf Validação estatística das origens e destinos estimadas.} Cada
círculo preto representa uma parada de ônibus. A linha vermelha representa a
regressão linear aplicada aos dados, a linha contínua azul foi estimada pelo o
método Nadaraya-Watson \cite{nadaraya1964,watson1964} e as linhas tracejadas
azuis delimitam o intervalo de confiança de 95\% estimado por {\it bootstrap}
\cite{racine2004,li2004}. Uma relação isométrica foi encontrada, com expoente
$\rho = 0.95 \pm 0.02$ e $R^2=0.92$. O $R^2$ é o coeficiente de determinação da
regressão \cite{rawlings2001, montgomery2015}.}
\label{val-od}
\end{figure}

Foi ainda realizada uma validação espacial da amostra estimada. Espera-se que o
método de estimativa garanta, com certo grau de confiabilidade, que a amostra
não esteja polarizada espacialmente. No caso específico da técnica utilizada
nesta tese, é desejável que os embarques nos ônibus, usados nas estimativas,
sejam proporcionais (considerando cada parada de ônibus) ao número total de
embarques. Mesmo tendo em vista que as heurísticas utilizadas garantiriam mais
de 90\% de acerto nas origens e destinos estimadas \cite{gordillo2006, lam1991,
remya2013, hua2007}, não havia qualquer garantia de que a amostra extraída não
fosse polarizada. Para validar espacialmente a amostra, verificou-se a
correlação entre os embarques utilizados para estimar os pares origem-destino e
o número total de embarques em pontos de ônibus em Fortaleza. A partir da
equação:
\begin{equation}
Y=aX^\rho, 
\end{equation}
\noindent onde $X$ representa o número total de embarques nos pontos
de ônibus, $Y$ representa o número de embarques utilizados nas estimativas e $a$
é uma constante de normalização, uma relação linear ou isométrica é evidenciada
pelo valor o expoente $\rho \approx 1$, indicando que, proporcionalmente, os
embarques utilizados nas estimativas são equivalentes ao número total de
embarques em pontos de ônibus, constatando que as amostras foram extraídas de
forma aleatória. Foi encontrado $\rho = 0.95 \pm 0.02$, a Figura \ref{val-od}
ilustra a regressão aplicada aos dados.

\section{Reconstrução das trajetórias dos usuários}

As trajetórias dos usuários de ônibus são definidas pela composição das rotas
dos ônibus que os levam entre seus pares origem-destino. O Algoritmo
\ref{algo-rota} mostra o pseudocódigo de estimativa da rota de um usuário. Em
resumo, o algoritmo inicialmente busca um caminho direto entre o par
origem-destino (do ponto $n_o$ até o ponto $n_d$), baseando-se na rota da linha
de ônibus ($L_0$) que o usuário utilizou. Caso esse caminho direto exista, é
definido como caminho do usuário o trecho da rota da linha $L_0$ de $n_o$ a
$n_d$. Caso não exista um caminho direto, busca-se um terminal que tenha alguma
linha que o leve até seu destino estimado. 

No contexto desta tese, um caminho não direto, ou indireto, pode ser definido
como um caminho em que é necessário realizar um processo de baldeação, trocando
de veículo antes para chegar a um destino. Em Furtado {\it et al}
\cite{furtado2017} foi mostrado que a estrutura da rede de ônibus de Fortaleza
permite que aproximadamente 40\% dos seus usuários façam um caminho direto da
sua origem até seu destino, considerando que o mesmo aceite se deslocar por até
500 metros a pé, tanto na origem quanto no destino.

\algnewcommand\algorithmicforeach{\textbf{for each}}
\algdef{S}[FOR]{ForEach}[1]{\algorithmicforeach\ #1\ \algorithmicdo}
    
\begin{algorithm}
\caption{Algoritmo de reconstrução da rota de um usuário}
\label{array-sum}
\begin{algorithmic}[1]
    \Statex 	
    \Procedure{ConstroiRotaUsuario}{$n_o$, $n_d$, $L_o$}
	\Statex
	\State $pathList =  \{ \emptyset \}$        
	\Statex
        \State $pathList$ = todos os caminhos diretos entre $n_o$ e $n_d$ por $L_o$            
	\Statex
\If {$pathList$ está vazia }
	\Statex
    \ForEach {terminal $t$ que seja um ponto válido de baldeação entre $n_o$ e $n_d$}
	\Statex
        \State $path1 = getPath(n_o,t,L_o)$
        \State $path2 = getPath(t,n_d)$
	\Statex	
	\State $pathList.add(join(path1,path2))$
	\Statex
    \EndFor  
	\Statex          
\EndIf
\Statex
\If {$pathList$ está vazia }
    \Statex
    \State $hashTerminais$ = menores caminhos diretos entre cada par de terminais
    \Statex
    \ForEach {par $(t_1,t_2)$ de terminais em $hashTerminal$}  
	\Statex          
        \State $path1 = getPath(n_o,t_1,L_o)$
        \State $path2 = hashTerminal.get(t_1,t_2)$
        \State $path3 = getPath(t_2,n_d)$
	\Statex
        \State $pathList.add(join(path1,path2,path3))$ 
	\Statex   
    \EndFor 
    \Statex	           
\EndIf
    \Statex 	
\State \textbf{retorne} o caminho com menor distância em $pathList$
    \Statex        
                
\EndProcedure
\Statex
\end{algorithmic}
\label{algo-rota}
\end{algorithm}

\section{Estimação de população flutuante a partir de trajetórias de usuários de
ônibus}

Após a estimativa dos caminhos dos usuários dentro da rede de ônibus, foram
modeladas as suas trajetórias como um grafo direcionado $G(V,E)$, onde $V$ e $E$
são o conjunto de vértices, $v$, e arestas, $e$, respectivamente. Uma aresta,
$e$, entre dois vértices $v_i$ e $v_j$ é definida por um par ordenado
$(v_i,v_j)$. Na abordagem adotada aqui, os vértices representam paradas de
ônibus e as arestas representam a demanda de usuários de ônibus entre duas
paradas de ônibus consecutivas. 

\begin{figure}[!h] \includegraphics[width=1\textwidth]{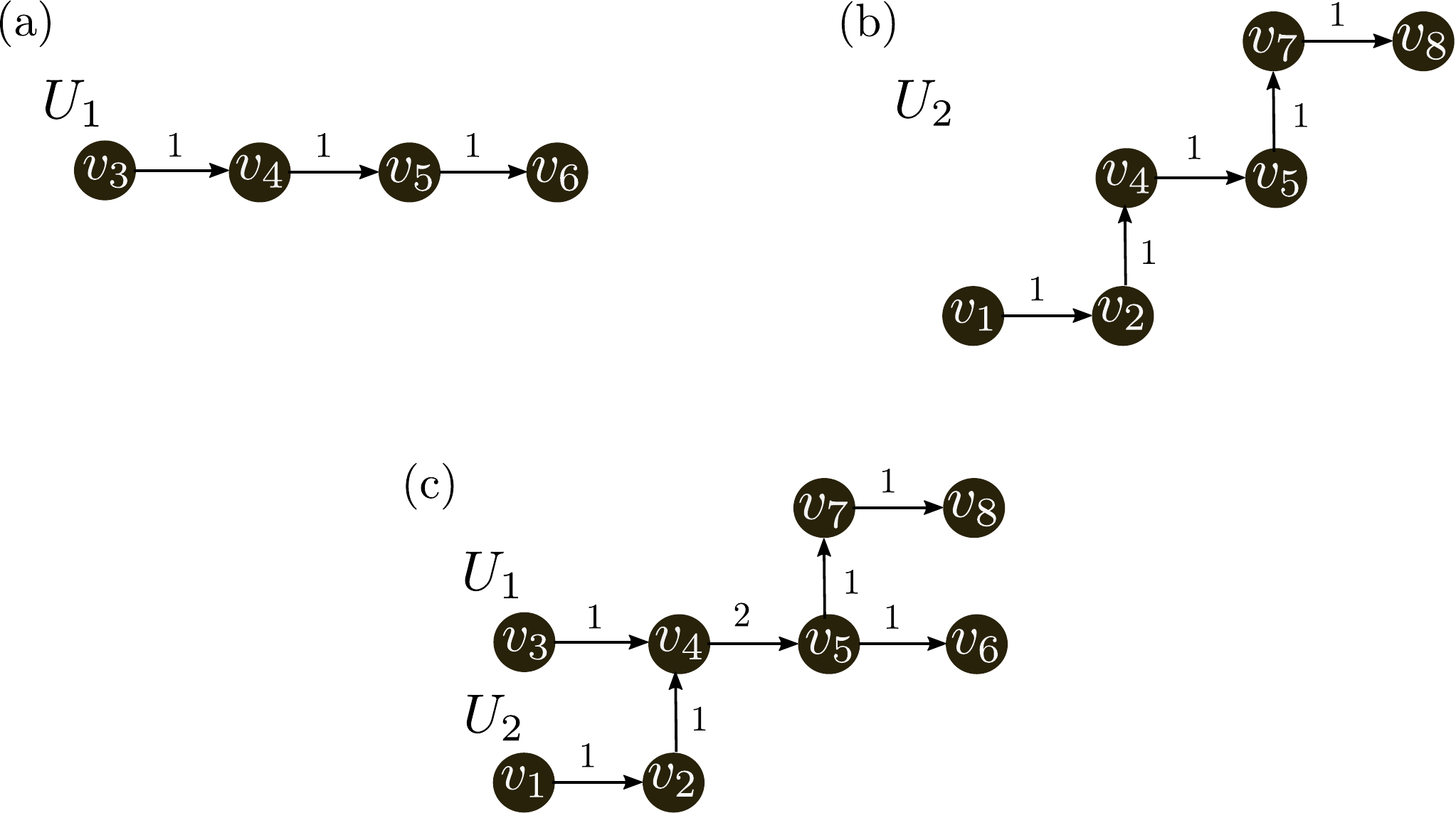}
\caption{{\bf Composição das rotas dos usuários na construção de uma rede de
fluxo.} Em (a) é ilustrada uma rota hipotética do usuário $U_1$. Em (b) é 
ilustrada a rota também hipotética de um usuário $U_2$. Em (c) é ilustrado como
é feita a composição das rotas para construir uma rede de fluxo.}
\label{demanda}
\end{figure}

A título de exemplo, a Figura \ref{demanda} ilustra uma representação hipotética
de duas rotas de usuários. No exemplo ilustrado essas duas rotas são combinadas
para formar uma pequena rede de fluxo. Em (a) é ilustrada a rota do primeiro
usuário ($U_1$), que visita em sequência as paradas de ônibus representadas
pelos vértices $v_3$, $v_4$, $v_5$ e $v_6$. Em (b), é ilustrada a rota de um
segundo usuário ($U_2$), que, dentro de um ônibus, visita em sequência as
paradas de ônibus representadas pelos vértices $v_1$, $v_2$, $v_4$, $v_5$, $v_7$
e $v_8$. Por fim, em (c), é ilustrada a rede resultante da combinação das duas
rotas. É possível observar que a aresta conectada por $v_4$ e $v_5$ teve seu
peso incrementado por ser um trecho em comum de passagem para $U_1$ e $U_2$.

\begin{figure}[!h] \includegraphics[width=1\textwidth]{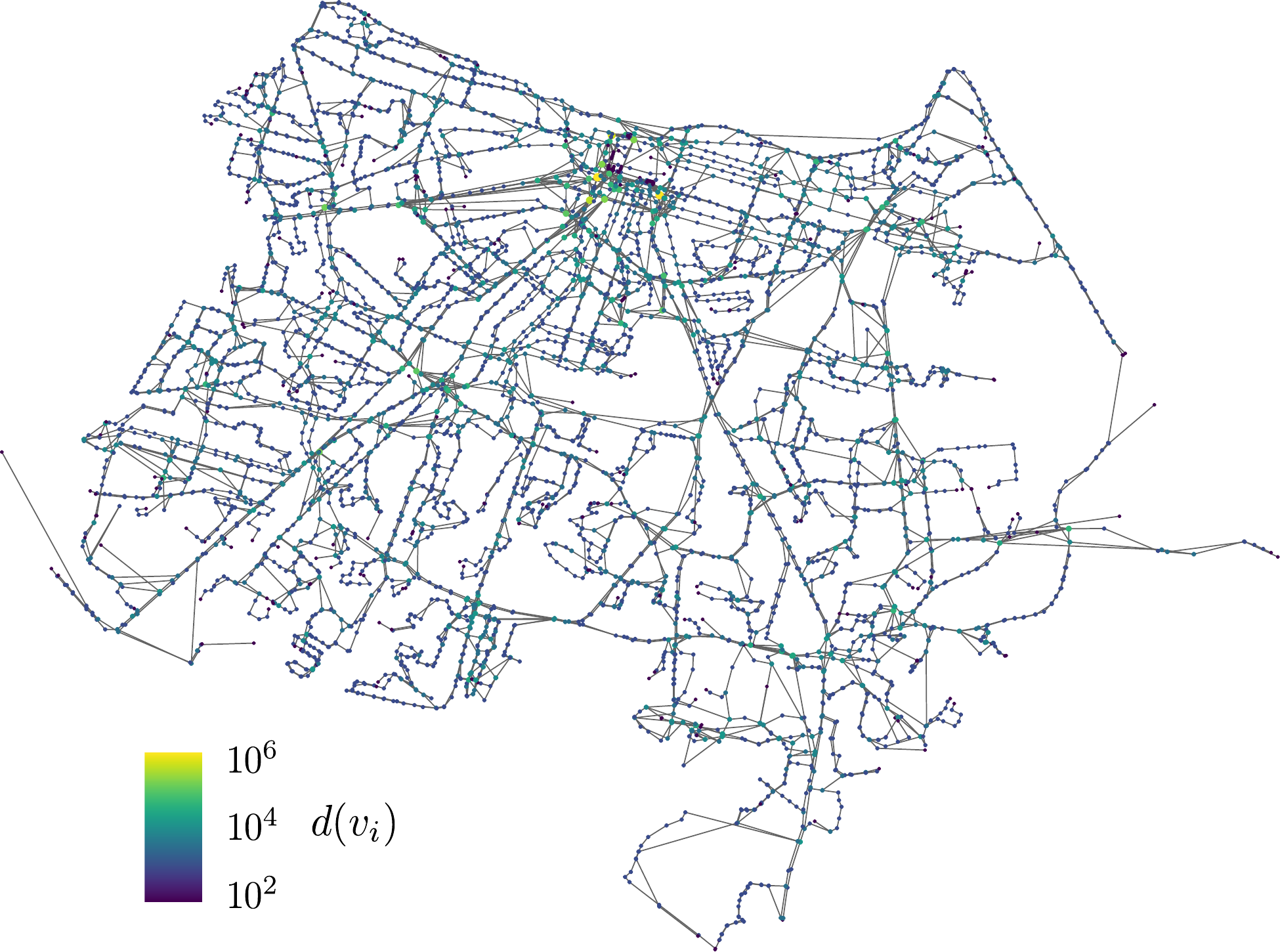}
\caption{{\bf Rede de fluxo estimada}. É ilustrada a distribuição espacial dos
vértices (que representam paradas de ônibus) dessa rede. Tamanho e cor dos nós
são definidos pelo seu grau. Quanto maior e mais claro é o nó, mais conexões ele
tem.}
\label{rede-fluxo} \end{figure}

Nesta tese, a rede de fluxo de pessoas foi construída a partir da combinação
das rotas de ônibus reais de usuários que utilizaram o sistema de ônibus de
Fortaleza no dia onze de março de 2015. A Figura \ref{rede-fluxo}, ilustra a
distribuição espacial dos vértices (paradas de ônibus) dessa rede, cada círculo
representa um vértice da mesma. Os vértices maiores e mais claros representam
paradas de ônibus que têm maior fluxo de pessoas. Ao todo, essa rede possui 4783
vértices e 5789 arestas. Seu maior componente conexo possui 4768 vértices e essa
rede só não é formada por um único componente conexo, porque existem linhas de
ônibus que apenas trafegam pelo Campus do Pici (passando por 15 paradas de
ônibus dentro do campus), da Universidade Federal do Ceará, e nenhuma outra
linha de ônibus que circula fora do campus, entra no mesmo, impedindo a conexão
dos dois componentes conexos.

\begin{figure}[!h] \includegraphics[width=1\textwidth]{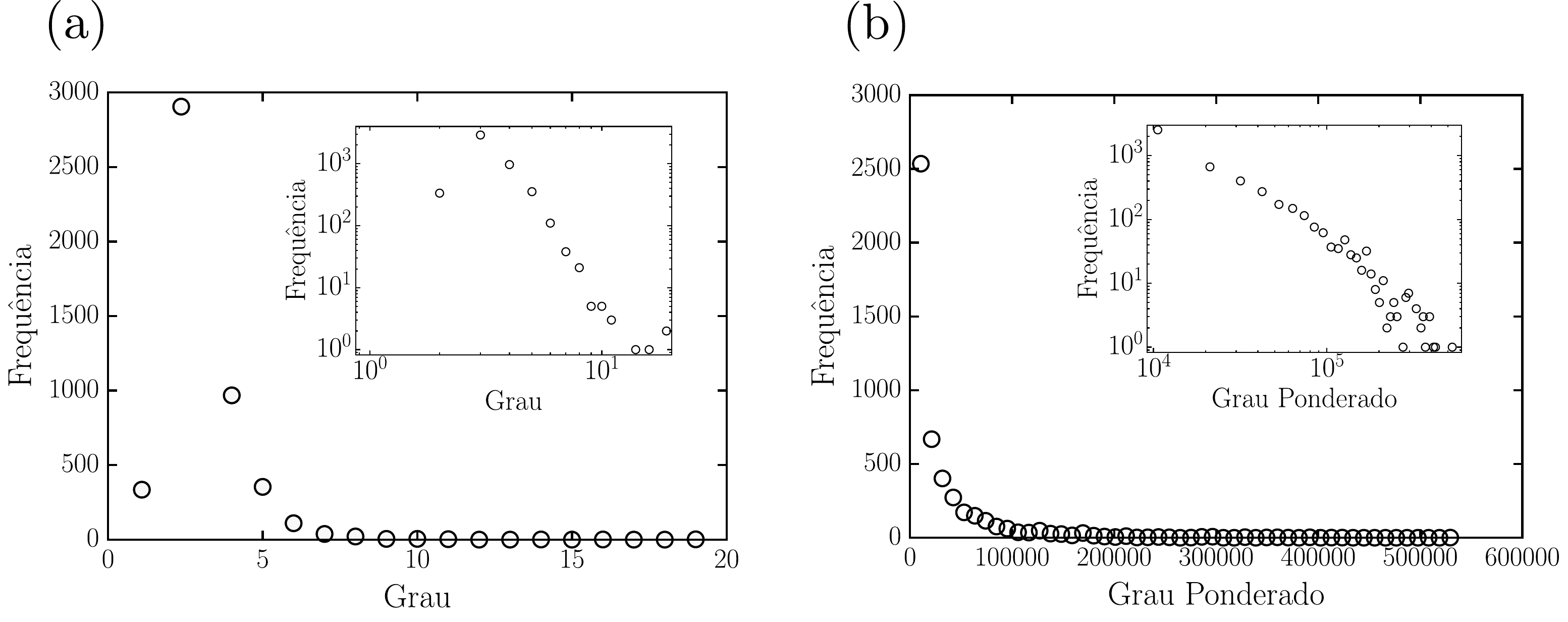}
\caption{{\bf Distribuições de graus da rede de fluxo estimada}. (a)
Distribuição de graus da rede. (b) Distribuição de graus ponderados: para melhor
visualização essa distribuição foi gerada em 50 {\it bins}. Em (a) e (b) ainda é
possível observar as suas respectivas distribuições em escala logarítmica
embutidas nos gráficos.}
\label{g-demanda} \end{figure}

Na Figura \ref{g-demanda} (a) é ilustrada a distribuição de graus dessa rede:
observa-se que vértices com grau dois são os mais frequentes, são quase 3000 ao
todo. Em (b) é ilustrada a distribuição de graus ponderados, sendo mais
frequentes vértices com grau ponderado entre 0 e 10000, ao todo, são mais de
2500 vértices com essa característica.

Por fim, para estimar o volume de população flutuante presente em cada setor
censitário, foi definida uma função, $d(v_i)$, para cada vértice $v_i$, como a
soma dos usuários que passam por $v_i$. Desta forma, o volume de população
flutuante, $vpf$, de um setor censitário, $sc$, é definido a partir do somatório
dos valores de $d(v_i)$ para cada vértice, $i$, dentro de $sc$. Dessa forma, é
possível definir formalmente $vpf_{sc}$ como:

\newpage

\begin{equation}
vpf_{sc} = \sum_{i=1}^{N} d(v_i),
\end{equation}

\noindent onde $N$ é a quantidade de vértices (paradas de ônibus) presentes
dentro do setor censitário $sc_i$.

\chapter{Divisões de cidade para compreensão da dinâmica do crime}\label{cap3}

\begin{figure}[!h]
\includegraphics[width=1\textwidth]{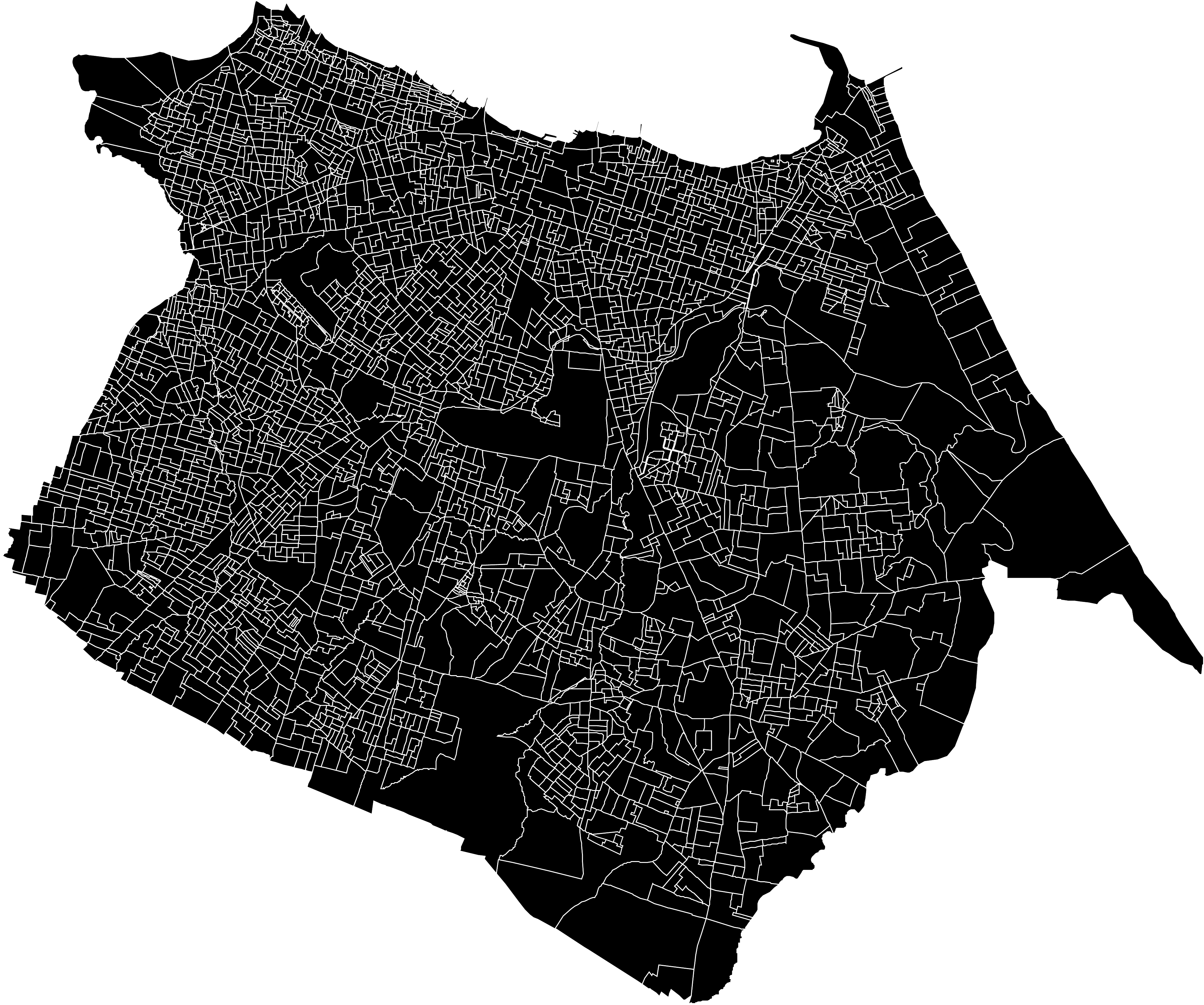}
\caption{{\bf Mapa de setores censitários de Fortaleza.} Esse tipo de divisão
administrativa de terreno segmenta Fortaleza em 3043 sub-regiões, que possuem em
média $\sim 800$ residentes. De uma maneira geral, os setores censitários são
muito pequenos, às vezes, menores que um quarteirão. O maior setor censitário da
cidade possui 7.92 $km^2$, o menor possui pouco mais de 0.001 $km^2$ (1000
$m^2$).} 
\label{mapa-setores}
\end{figure}

Neste capítulo busca-se verificar se as Leis de Escala encontradas por
Bettencourt {\it et al.} \cite{bettencourt2007} e Melo {\it et al.}
\cite{melo2014} em estudos intercidades podem ser constatadas também dentro de
uma cidade, principalmente, utilizando divisões administrativas de terreno. Será
utilizada a divisão por setores censitários, ilustrada na Figura
\ref{mapa-setores}. Um setor censitário é uma unidade administrativa de terreno
estabelecida para fins de controle cadastral, fornecida pelo Instituto
Brasileiro de Geografia e Estatística (IBGE) \footnote{disponível em
http://www.ibge.gov.br/}. Ao todo, Fortaleza possui 3043 setores censitários,
que somam aproximadamente 2,4 milhões de habitantes, segundo o censo 2010,
último realizado a nível de setores censitários no Brasil. De forma
complementar, neste capítulo, também serão descritos os conjuntos de dados e
variáveis utilizadas neste trabalho.

\section{Conjuntos de dados}

Neste capítulo serão estudadas quatro variáveis, que tiveram seus valores
extraídos de três conjuntos de dados. A partir do primeiro conjunto de dados são
obtidos os valores da variável {\it População Residente} (PR). Esse conjunto de
dados é fornecido pelo IBGE e quantifica o número de moradores por setor
censitário na cidade de Fortaleza.

A partir de um conjunto de dados de mobilidade do sistema de ônibus, detalhado
no capítulo \ref{proc-pf} desta tese, foi estimada a {\it População Flutuante}
(PF). Para cada setor censitário foi mensurado o volume de população flutuante a
partir da rede de fluxo de pessoas produzida através da análise de dados de
mobilidade do sistema de ônibus de Fortaleza. PF foi mensurado pelo número de
pessoas que trafegam por um setor censitário em um dia. A cidade de Fortaleza
possui 2034 ônibus, que circulam por mais de 359 rotas distintas, atendendo
aproximadamente 700 mil pessoas que usam diariamente o sistema de transporte
público da cidade. Vale ressaltar que, no caso de Fortaleza, os ônibus ainda
representam o principal meio de transporte da cidade.

Por fim, a partir do conjunto de dados de chamadas à polícia, disponibilizado
pela Coordenadoria Integrada de Operações de Segurança (CIOPS), que possui
172332 chamadas para o serviço 190 sobre {\it Crimes Contra o Patrimônio} (CCP)
e 97890 chamadas ao mesmo serviço sobre reclamação de {\it Perturbação de
Sossego Alheio} (PSA)\footnote{disponível em
http://wikicrimes.org/dados-oficiais/crimes.csv}. Todas as chamadas à polícia
foram feitas entre setembro de 2014 e março de 2016.

Aqui é assumido que mudanças maciças no ambiente urbano da cidade não podem ser
produzidas no período de investigação (intervalo entre 4 e 6 anos). Mudanças
relevantes nesse intervalo de tempo só existiriam se a sociedade estivesse
enfrentando uma severa ruptura social, devido, por exemplo, a desastres
naturais, graves crises econômicas ou guerras. O Apêndice II pode ser consultado
para obter informações adicionais sobre todos os conjuntos de dados utilizados
neste estudo, incluindo a URL para {\it download}.

\section{Relação entre população e crime a partir de análises sobre setores
censitários}

\begin{figure}[!h]
\includegraphics[width=1\textwidth]{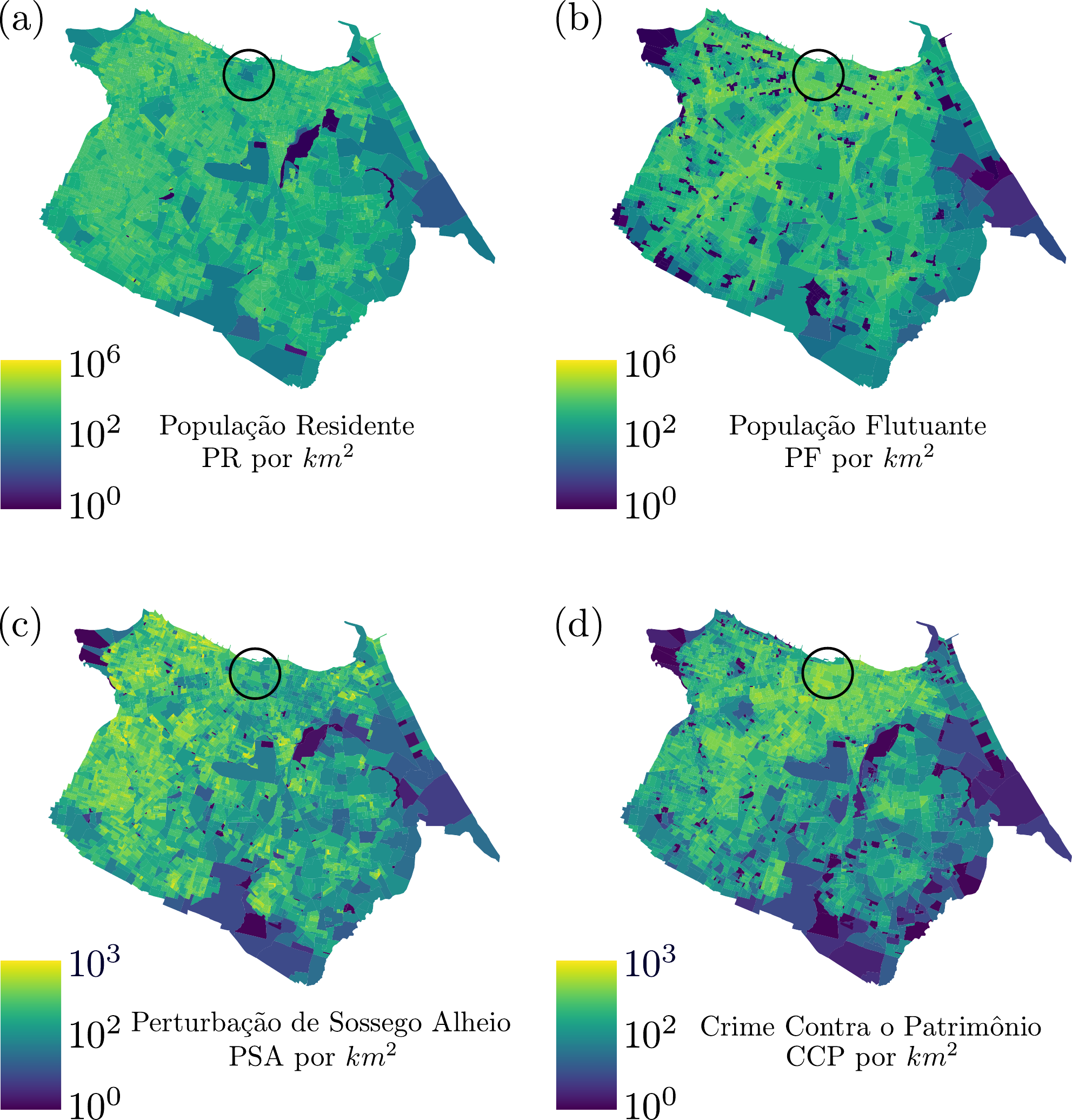}
\caption{{\bf Mapas de densidade das variáveis estudadas em Fortaleza.} (a)
População residente (PR) por ($\mathrm{km}^2$). (b) População flutuante (PF) por
$\mathrm{km}^2$. (c) Perturbação de sossego alheio (PSA) por $\mathrm{km}^2$.
(d) Crimes contra o patrimônio (CCP) por $\mathrm{km}^2$. O círculo preto
destaca o centro da cidade, essa região tem uma baixa densidade tanto de
residentes quanto de reclamações de perturbação de sossego alheio, em contra
partida é densa em fluxo de pessoas e crimes contra o patrimônio.} 
\label{mapas_pop_crime}
\end{figure}

Os mapas de calor nas Figuras \ref{mapas_pop_crime}(a)-\ref{mapas_pop_crime}(d)
fornecem a densidade local, em escala logarítmica, de PR, PF, PSA e CCP,
respectivamente, em Fortaleza. Observa-se alguma correlação espacial tanto entre
PR e PSA quanto em PF e CCP. Em especial, há uma evidente maior incidência de
{\it hot spots} em (b) e (d) se comparado com (a) e (c). Além disso, no centro
da cidade, destacado por círculos pretos em cada mapa, a alta densidade de PR é
compatível com as altas taxas de CCP, enquanto baixas densidades de PR parecem
explicar a baixa frequência de queixas PSA. Particularmente, o centro comercial
de Fortaleza é uma região de característica populacional única, enquanto
praticamente não há pessoas residindo no local, essa é a região da cidade com
maior fluxo de pessoas. A aparente correlação espacial entre PR e PSA sugere que
reclamação de perturbação de sossego é feita por residentes, possivelmente
quando estão em casa, momento que é intuitivo pensar que um indivíduo deseja
descansar e desta forma é mais sensível à perturbação. Vale ressaltar que os
dados confirmam essa intuição: quase 80\% das chamadas à polícia por PSA ocorrem
entre oito horas da noite e uma hora da manhã, intervalo que a maior parte da
população se encontra em casa, que representa pouco mais de 20\% do tempo de um
dia.

\begin{figure}[!h]
\includegraphics[width=1\textwidth]{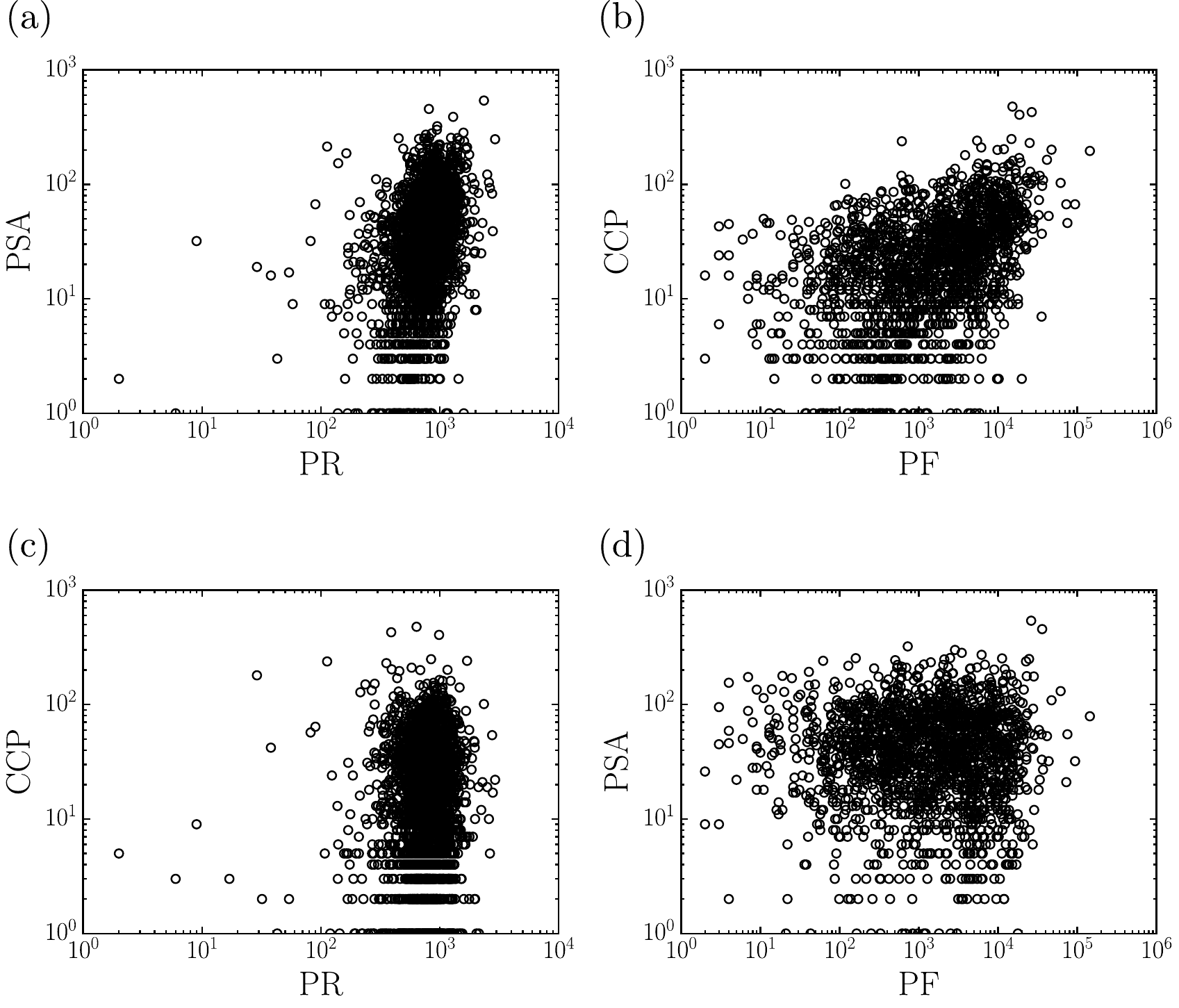}
\caption{{\bf Gráficos de dispersão aplicados às variáveis estudadas.} (a)-(d)
Revelam a ausência de correlação nas variáveis estudadas a nível de setor
censitário. Cada círculo preto representa um setor censitário e em
todos os gráficos foi estimado coeficiente de determinação \cite{rawlings2001,
montgomery2015} $R^2 < 0.15$.}
\label{correlacao_setores}
\end{figure}

Apesar de parecer trivial sugerir que há alguma correlação entre PR e PSA, assim
como entre PF e CCP a partir dos mapas de densidade ilustrados na
\ref{mapas_pop_crime}, os respectivos gráficos de dispersão visíveis na Figura
\ref{correlacao_setores} falham em capturar a aparente correlação existente,
devido à desagregação dos conjuntos dos dados de ambas populações. No presente
momento, conjetura-se que tais correlações existam de fato, mas estão escondidas
pelo nível de granularidade dos setores censitários. A maioria dos setores
censitários tem uma área pequena, às vezes do tamanho de um quarteirão, e é
provável que tal escala de granularidade seja insuficiente para capturar as
correlações e, portanto, revelar o impacto da influência social sobre PSA e CCP.
Adicionalmente, no caso de PR, a população tende a ser igualmente distribuída
entre os setores censitários e isto impossibilita a percepção do fenômeno a
nível de escala. De uma maneira geral, os setores censitários têm entre 600 e
800 residentes cada.

\section{Discussão}

No presente capítulo buscou-se verificar se as leis de escala observadas por
Bettencourt {\it et al.} \cite{bettencourt2007} e Melo {\it et al.}
\cite{melo2014}, em estudos intercidades, poderiam estar presentes dentro da
cidade de Fortaleza. Apesar da aparente correlação visual nos mapas de densidade
apresentados no painel da Figura \ref{mapas_pop_crime}, quando plotadas as
informações em gráficos de dispersão (Figura \ref{correlacao_setores}) não foi
possível verificar qualquer correlação estatística entre as variáveis
estudadas.

Apesar dos gráficos de dispersão (Figura \ref{correlacao_setores}) apresentados
neste capítulo sugerirem que não há qualquer correlação estatística entre
população e crime em Fortaleza, é precipitado descartar a hipótese de
que exista uma relação alométrica entre população e crime dentro da cidade. A
literatura mostra que unidades administrativas de terreno são imprecisas para
compreender fenômenos urbanos, essencialmente porque elas são definidas com
restrições que buscam simplificar atividades de censo e controle cadastral, como
normalizações por estatísticas de população residente \cite{oliveira2014,
arcaute2015, cottineau2016}.

Mesmo sabendo que setores censitários são definidos pela presença de população
residente é difícil capturar uma correlação estatística utilizando uma unidade
de terreno tão granular, mesmo para indicadores se são explicados pela presença
desses residentes. Setores censitários têm área reduzida, às vezes do tamanho de
uma quadra. Para uma possível correlação entre PR e PSA, pode-se imaginar um
cenário, por exemplo, em que ocorre uma grande festa em uma quadra da cidade.
Sabendo que grandes festas têm potencial de gerar reclamações de PSA muito além
de uma quadra de distância, para compreender a relação entre a ocorrência dessa
festa e suas reclamações de PSA, seria necessário agregar alguns setores
censitários próximos. Vale salientar que mesmo que fossem utilizadas unidades
administrativas de terreno menos granulares, como a divisão por bairros, ainda
assim haveria a questão de que essas unidades também tem restrições de
normalização por população residente na definição das suas fronteiras,
dificultando assim a compreensão de fenômenos a nível de escala. Em Fortaleza,
por exemplo, a maior parte dos bairros têm entre 30000 e 50000 residentes.

Por tudo que foi discutido neste capítulo, conjectura-se que a influência social
que a população de uma cidade exerce sobre indicadores urbanos não pode ser
explicada de maneira consistente utilizando divisões administrativas de terreno,
sendo necessário estimar as fronteiras dessa influência social utilizando
algorítimos de agrupamento isentos de qualquer tipo de restrição na definição
dessas fronteiras. Baseando-se na conjectura recém apresentada, foi considerado
fazer agrupamento das unidades territoriais de Fortaleza, utilizando a malha de
setores censitários como base, devido ser a unidade com máxima resolução
disponível, a fim de estimar as fronteiras de população flutuante e residente em
Fortaleza.

\chapter{Relações alométricas entre população e crime em agregados de população
 dentro de uma cidade }
\label{cap4}

Este capítulo tem como objetivo estimar estatisticamente o impacto que a
influência social exerce sobre o crime na cidade de Fortaleza. Para alcançar
tal objetivo, serão estudados tanto aglomerados de população residente, quanto
aglomerados de população flutuante, definidos além das fronteiras das unidades
administrativas de terreno dos setores censitários. Busca-se encontrar a
correlação espacial visualmente sugerida pela Figura \ref{mapas_pop_crime}, mas
estatisticamente negada pelo gráfico de dispersão ilustrado na Figura
\ref{correlacao_setores}. Conjectura-se que essa correlação exista, mas que a
mesma está sendo escondida pelo nível de desagregação dos setores censitários.

\section{Influência social mensurada a partir de agregados de população
residente e flutuante}\label{cap41}

\begin{figure}[!h]
\includegraphics[width=1\textwidth]{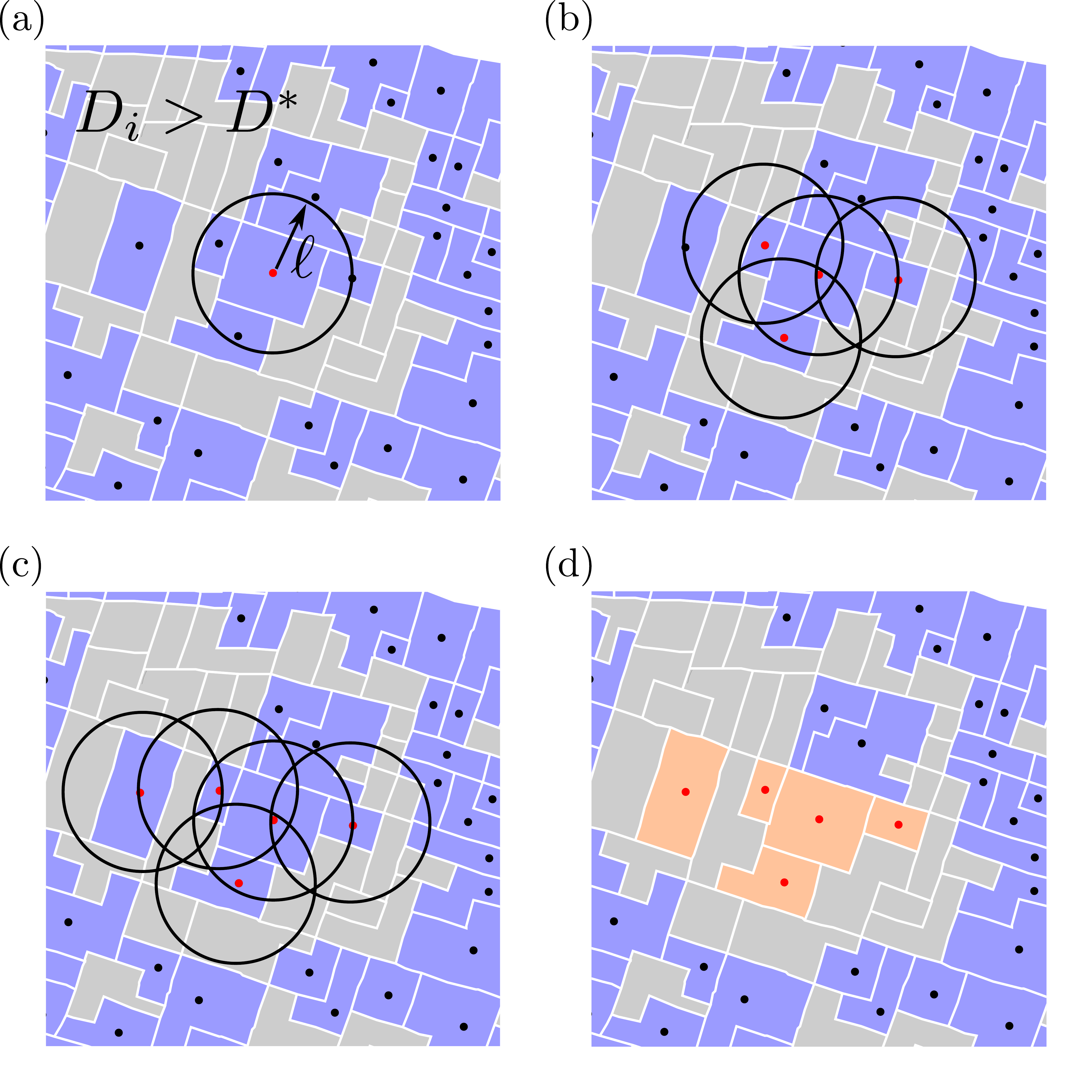}
\caption{{\bf O comportamento aglomerativo do (CCA).} Cada polígono representa
uma unidade de agrupamento, especificamente no caso deste trabalho, setores
censitários. Os polígonos na cor azul claro são candidatos a serem agrupados
$(D_i > D^*)$. Em contrapartida, os polígonos na cor cinza não podem ser
agrupados ($D_i \le D^*$). (a) O ponto vermelho representa o centro geométrico
do setor censitário $i$ e o círculo preto com raio $\ell$ procura vizinhos
pertencentes ao mesmo aglomerado. (b)-(c) A mesma operação de busca é feita para
os outros setores censitários até que não haja mais vizinhos dentro do raio de
operação. (d) O algoritmo termina a execução e o aglomerado é encontrado.}
\label{cca}
\end{figure}

Para definir limites espaciais além de fronteiras administrativas, foi
considerada a noção de continuidade espacial através da agregação de setores
censitários que estão próximos uns dos outros fazendo uso do {\it City
Clustering Algorithm} (CCA) \cite{makse1998, rozenfeld2008, giesen2010,
rozenfeld2011, duranton2013, gallos2012, duranton2015, eeckhout2004}. O CCA
encontra as fonteiras populacionais de uma área urbana considerando dois
parâmetros, o primeiro, $D^*$, um limiar de densidade populacional e o segundo,
$\ell$, um limiar de distância. Para um setor censitário $i$, a densidade
populacional $D_i$ é localizada em seu centro geométrico, se $D_i > D^*$, então
o setor censitário $i$ é considerado populado. O limiar de distância, $\ell$,
representa a distância de corte entre os setores censitários para considerá-los
como espacialmente contíguos, mais precisamente, todos os setores censitários
que estão em distâncias menores que $\ell$ são agrupados. Assim, um aglomerado
encontrado pelo CCA é definido por áreas povoadas dentro de uma distância
$\ell$, como pode ser observado esquematicamente na Figura \ref{cca}. Embora o
algoritmo inicie de um setor censitário semente arbitrário, ele não produz
aglomerados distintos ao variar esta semente. Os dois únicos fatores que são
responsáveis pelo comportamento aglomerativo do CCA são os parâmetros $\ell$ e
$D^*$. Estudos recentes \cite{oliveira2014,duranton2013,duranton2015}
demonstraram que os resultados produzidos pelo CCA são fracamente dependentes de
$D^*$ e $\ell$ para uma faixa dos valores dos parâmetros. Neste trabalho, $\ell$
será quantificado em metros e $D^*$ em população residente ou flutuante por
$\mathrm{km}^2$.

\begin{figure}[!h]
\includegraphics[width=1\textwidth]{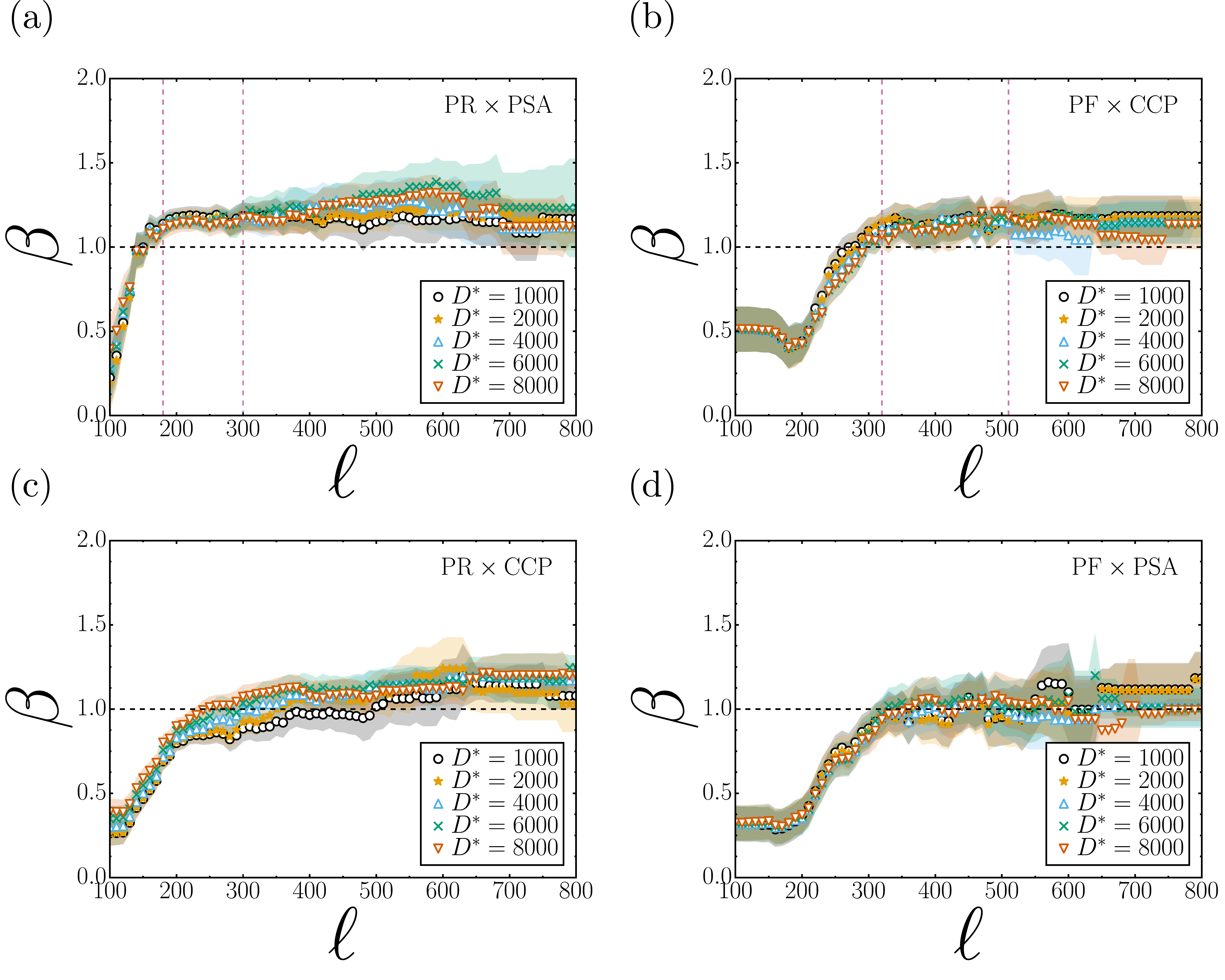}
\caption{{\bf Comportamento do expoente $\boldsymbol\beta$ ao variar os
parâmetros do CCA, $\boldsymbol\ell$ e $\boldsymbol D^*$.} Em (a) é ilustrada a
variação de $\beta$ na correlação entre PR e PSA. Em (b) A variação é ilustrada
para a correlação entre PF e CCP. Em (c) são ilustrados os valores de $\beta$ da
relação entre PR e CCP. Por fim, em (d), são ilustrados os valores do expoente
$\beta$ para a relação entre PF e PSA. Em todas as Figuras, (a-d), o eixo $x$
representa o valor de $\ell$, que foi variado a partir de 100 até 800 metros
(momento em que o maior aglomerado encontrado consome quase toda a cidade). O
valor do expoente $\beta$ é exibido no eixo $y$. As cores das linhas representam
a variação do parâmetro $D^*$, que correspondente à densidade de população
residente nos itens (a) e (c), e a densidade de população flutuante nos itens
(b) e (d). $D^*$ foi variado a partir de 1000 até 8000. As sombras representam o
erro padrão do coeficiente $\beta$. Os gráficos também mostram linhas
tracejadas vermelhas, entre essas linhas é destacado o intervalo onde,
independentemente da parametrização, o expoente $\beta$ tem intervalos menores
de variação. As linhas pretas pontilhadas destacam o expoente $\beta =
1$, situação em que a relação entre variáveis é isométrica. Em (a), a relação é
superlinear a partir de $\ell \ge 180$, já em (b), a superlinearidade aparece em
$\ell \ge 320$. Em todas as figuras, as sombras coloridas representam o Erro 
Padrão do Coeficiente, a cor da sombra indica a que valor de $D^*$ o erro padrão
está associado.}
\label{variacao_beta}
\end{figure}

Para determinar o efeito da parametrização sobre o valor do expoente $\beta$,
foi buscado um intervalo dentro dos parâmetros onde $\beta$ tem baixa
sensibilidade a essa variação. A Figura \ref{variacao_beta} ilustra o
comportamento do expoente $\beta$ em função da variação dos parâmetros CCA.
Conforme é ilustrado nas Figuras \ref{variacao_beta} (a) e (b), o valor do
expoente $\beta$, estimado utilizando o método de regressão linear (mais
detalhes no Anexo I), para PR contra PSA permanece praticamente insensível aos
parâmetros CCA na faixa $180 \ge \ell \ge 300$, independentemente dos valores de
$D^*$ adotados no processo de estimativa. Além disso, a média
$\beta=1.17\pm0.06$ fornece forte evidência de uma relação superlinear entre
essas variáveis. Um comportamento semelhante pode ser observado entre PF e CCP,
mas agora o expoente $\beta$ permanece praticamente invariável dentro do
intervalo $320 \geq \ell \geq 510$. O valor médio de $\beta=1.14\pm0.04$, também
nesse caso, indica a presença de uma relação alométrica superlinear. Nos itens
(c) e (d) não se pode sugerir nenhum tipo de relação entre as variáveis,
primeiro, devido a maior flutuação entre valores sublineares, isométricos e
superlineares, segundo, devido ao maior valor médio do Erro Padrão do
Coeficiente $\beta$, representado pelas sombras no gráfico. Este resultado
inclusive pode ser comparado com a variação do expoente alométrico para
diferentes processos de aglomeração urbana \cite{oliveira2014, arcaute2015,
cottineau2016}.

Neste ponto, é necessário determinar um critério para selecionar um valor
adequado do parâmetro $D^*$. São preferíveis pequenos valores de $\ell$, pois
estes conduzem a formação de muitos aglomerados no CCA, aumentando a quantidade
de observações a serem incluídas na estatística. Também são desejáveis
pequenos valores de $D^*$, pois desta forma poucos setores censitários são
eliminados do mapa, incluindo uma parcela maior da população na análise.

Adicionalmente, é proposto que a escolha de $D^*$ esteja associada a uma
distribuição espacial mais homogênea da população \cite{oliveira2014}. Em outras
palavras, deseja-se evitar que a divisão encontrada produza alguns aglomerados
com uma concentração muito maior (ou muito menor) de pessoas que outros. A
justificativa para evitar fazer análises mais profundas em divisões da cidade
com essa característica é que não se sabe ao certo o efeito que a concentração
de pessoas tem sobre o crime \cite{bettencourt2007, oliveira2014, melo2014}. Por
exemplo, locais com densidade populacional extremamente alta podem,
naturalmente, inibir que atividade criminosa ocorra, sendo esse possivelmente um
fator que traz insegurança a agressores motivados. Em contrapartida, alguns
locais com densidade populacional muito baixa podem fazer com que as rotinas de
criminosos, vítimas e guardiões jamais entrem em convergência \cite{cohen1979,
felson2002}. Para isolar esse fator, será escolhida uma divisão de agregados
que tenha a distribuição de densidade populacional mais próxima de uniforme
possível.

Uma configuração de divisão por aglomerados da cidade, que tem distribuição de
densidade populacional uniforme, pode ser obtida ao se buscar agrupamentos do
CCA cujas áreas se aproximem de forma mais isométrica possível com os dados
populacionais. Em outras palavras, é desejável escolher divisões que tenham a
seguinte característica: Quando a população cresce, a área do aglomerado cresce
proporcionalmente. Formalmente, uma relação isométrica entre população e área
dos aglomerados do CCA, pode ser definida a partir da seguinte função:
\begin{equation}
Y=aX^{\alpha}, 
\end{equation}
\noindent com $\alpha \approx 1$, onde $X$ é a população, $Y$ é a
área em $\mathrm{km}^2$ dos aglomerados, e $a$ é uma constante.

\begin{figure}[!h]
\includegraphics[width=1\textwidth]{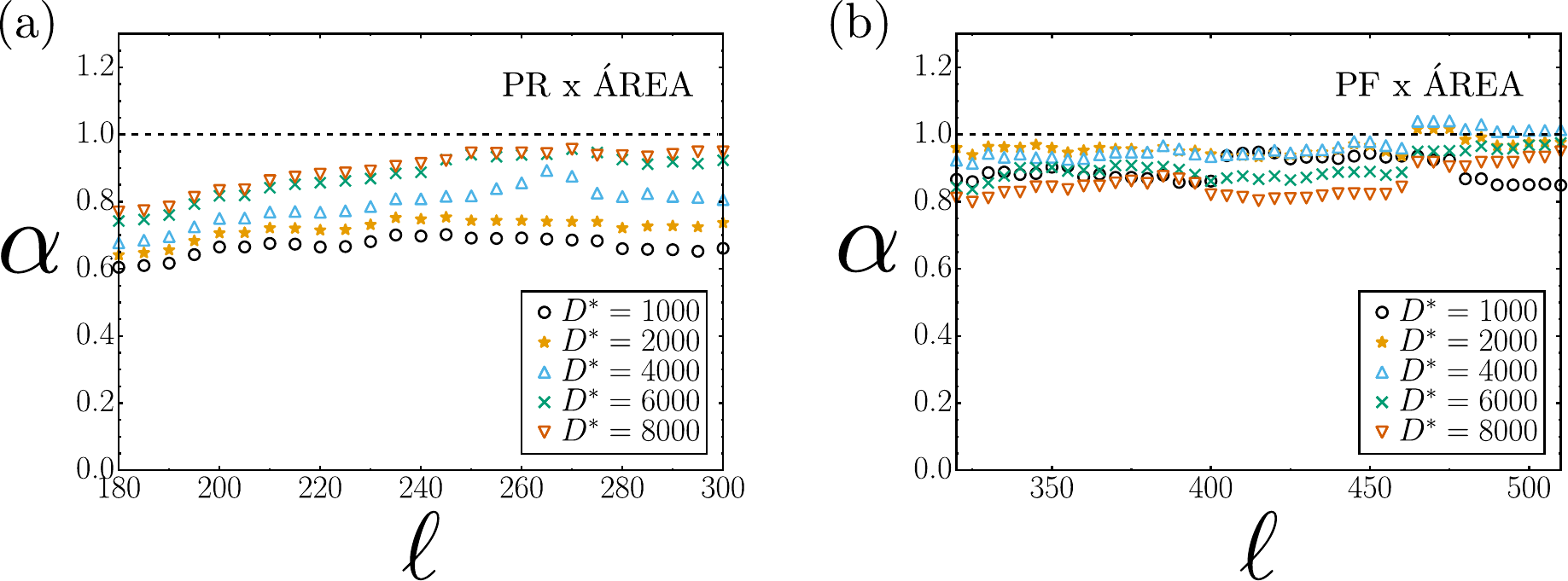}
\caption{{\bf Comportamento do expoente $\boldsymbol\alpha$ ao variar os
parâmetros do CCA, $\boldsymbol\ell$ e $\boldsymbol D^*$.} (a) Variação de
$\alpha$ nas correlações entre população residente (PR) e área em quilômetros
quadrados ($\mathrm{km}^2$) dos aglomerados descobertos pelo CCA. (b) Variação
dos expoentes das correlações entre a população flutuante (PF) e área. Em (a)
e (b), O eixo $x$ representa o parâmetro $\ell$, e o eixo $y$ representa o
expoente $\alpha$. As cores das linhas representam a variação do parâmetro
$D^*$. $D^*$ foi variado a partir de 1000 até 8000. As linhas pretas pontilhadas
destacam o expoente $\beta = 1$, situação em que a relação entre variáveis é
isométrica.}
\label{variacao_alpha}
\end{figure}

\begin{figure}[!h] 
\includegraphics[width=1.0\textwidth]{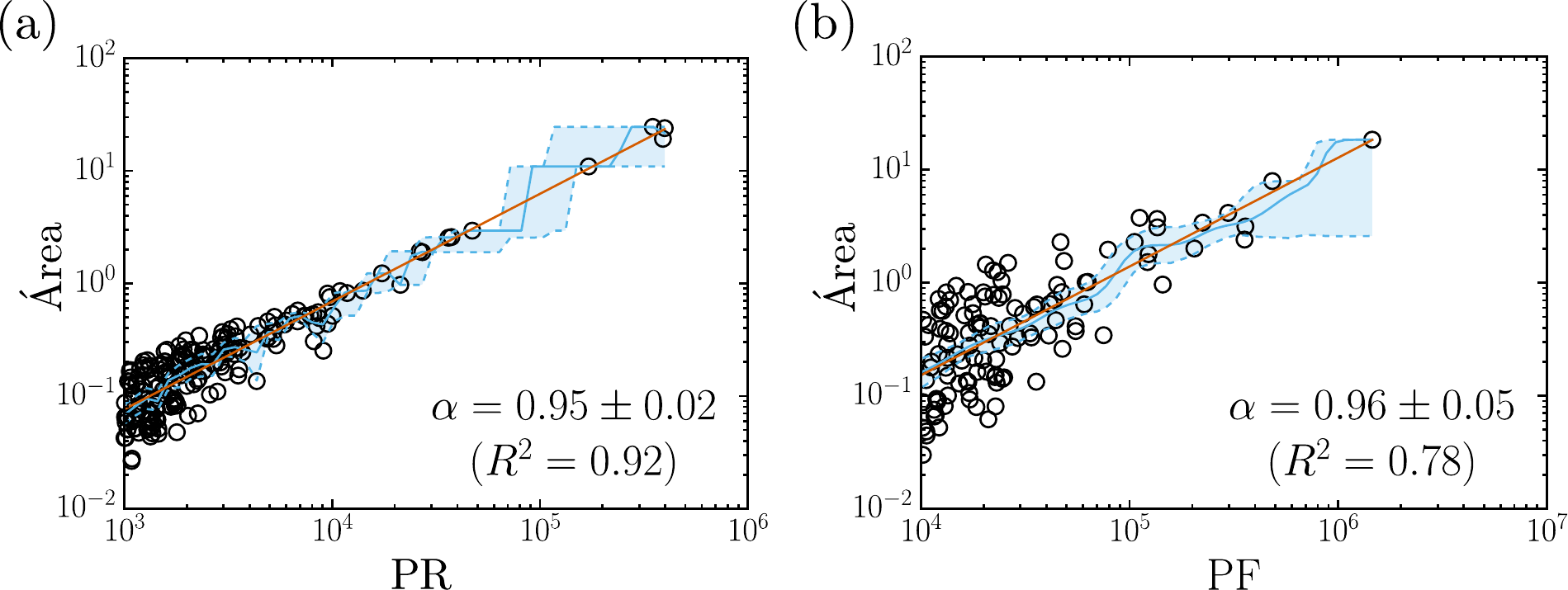}
\caption{{\bf Relação isométrica entre população residente (PR) e população
flutuante (PF) com a área de seus aglomerados em quilômetros quadrados.} (a)
Correlação entre PR e área dos aglomerados encontrados pelo {\it City Clustering
Algorithm} (CCA) para $D^*=6000$ residentes por $\mathrm{km}^2$ e $\ell=270$
metros. (b) Correlação entre PF e área para $D^*=2000$ pessoas passando por
$\mathrm{km}^2$ e $\ell = 320$ metros. A linha vermelha representa uma regressão
{\it Ordinary Least Square} (OLS) aplicada ao logaritmo dos dados
\cite{rawlings2001,montgomery2015}. A linha contínua azul indica a regressão de
kernel Nadaraya-Watson \cite{nadaraya1964,watson1964}. Finalmente, as linhas
tracejadas azuis delimitam 95\% de intervalo de confiança estimado a partir de
500 amostras de aleatórias com substituição. $R^2$ é o Coeficiente de
Determinação da regressão \cite{rawlings2001, montgomery2015}.}
\label{isome-arexpop}
\end{figure}

Seguindo o procedimento previamente descrito, a partir da Figura
\ref{variacao_alpha}a, obtém-se $\ell=270$ e $D^{*}=6000$ correspondendo ao par
de parâmetros do CCA que conduzem à relação isométrica mais próxima entre a área
dos aglomerados e PR. No caso de PF e área dos aglomerados, foram escolhidos os
valores $\ell=320$ e $D^{*}=2000$, a partir da Figura \ref{variacao_alpha}b. A
Figura \ref{isome-arexpop} ilustra as correlações encontradas entre as
populações estudadas e a área dos seus aglomerados. Em (a) foi estimado o valor
do expoente $\alpha=0.95 \pm 0.02$ ($R^2=0.92$) para PR contra a área dos seus
aglomerados. Já em (b) é ilustrada a correlação entre PF e a área dos seus
aglomerados, cujo foi estimado expoente $\alpha=0.96 \pm 0.05$ ($R^2=0.78$).

\begin{figure}[!h] 
\includegraphics[width=1.0\textwidth]{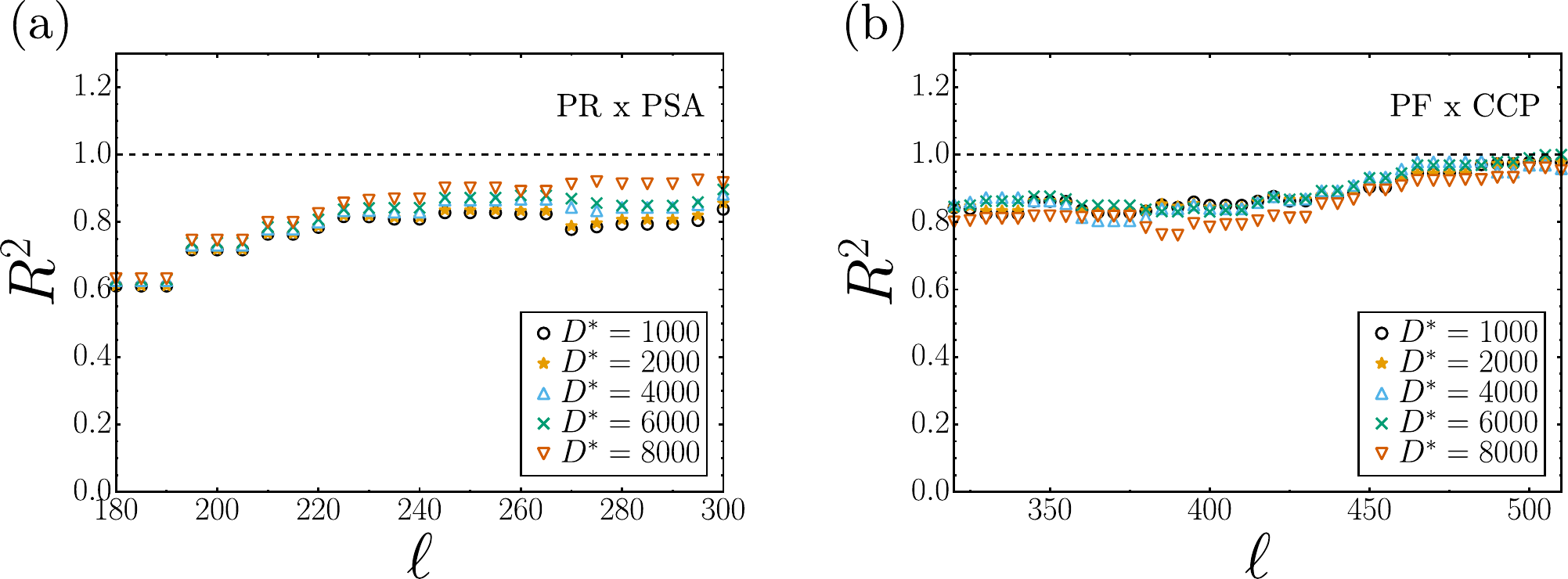}
\caption{{\bf Comportamento do Coeficiente de Determinação ao variar os
parâmetros do CCA, $\boldsymbol \ell$ e $\boldsymbol D^*$.} (a) A variação de
$R^2$ na correlação entre população residente (PR) e perturbação de sossego
alheio (PSA) é ilustrada. Em (b) é ilustrada a variação para correlações entre
população flutuante (PF) e crimes contra o patrimônio (CCP). $D^*$ foi variado a
partir de 1000 até 8000. As linhas pretas pontilhadas destacam o expoente $\beta
= 1$, situação em que a relação entre variáveis é isométrica.}
\label{variacao_r2}
\end{figure}

Outra abordagem para definir $D*$ poderia se basear na maximização $R^2$.
Infelizmente, no caso deste trabalho, $R^2$ não se mostra um bom indicador, pois
ele se mantém praticamente constante ao variar $\ell$ e $D^*$, como é possível
visualizar na Figura \ref{variacao_r2}. Outro problema de usar $R^2$ como
indicador é que o aumento dos valores de $\ell$ e $D^*$ pode conduzir a um
aumento artificial de $R^2$, devido à acentuada diminuição do número de
aglomerados, levando a uma configuração com apenas dois ou três pontos no
gráfico de correlação. Neste trabalho, buscou-se usar o menor valor de $D^*$
possível para se ter uma grande representação da população da cidade.

\section{Resultados} \label{cap42}

\begin{figure}[!h]
\includegraphics[width=1\textwidth]{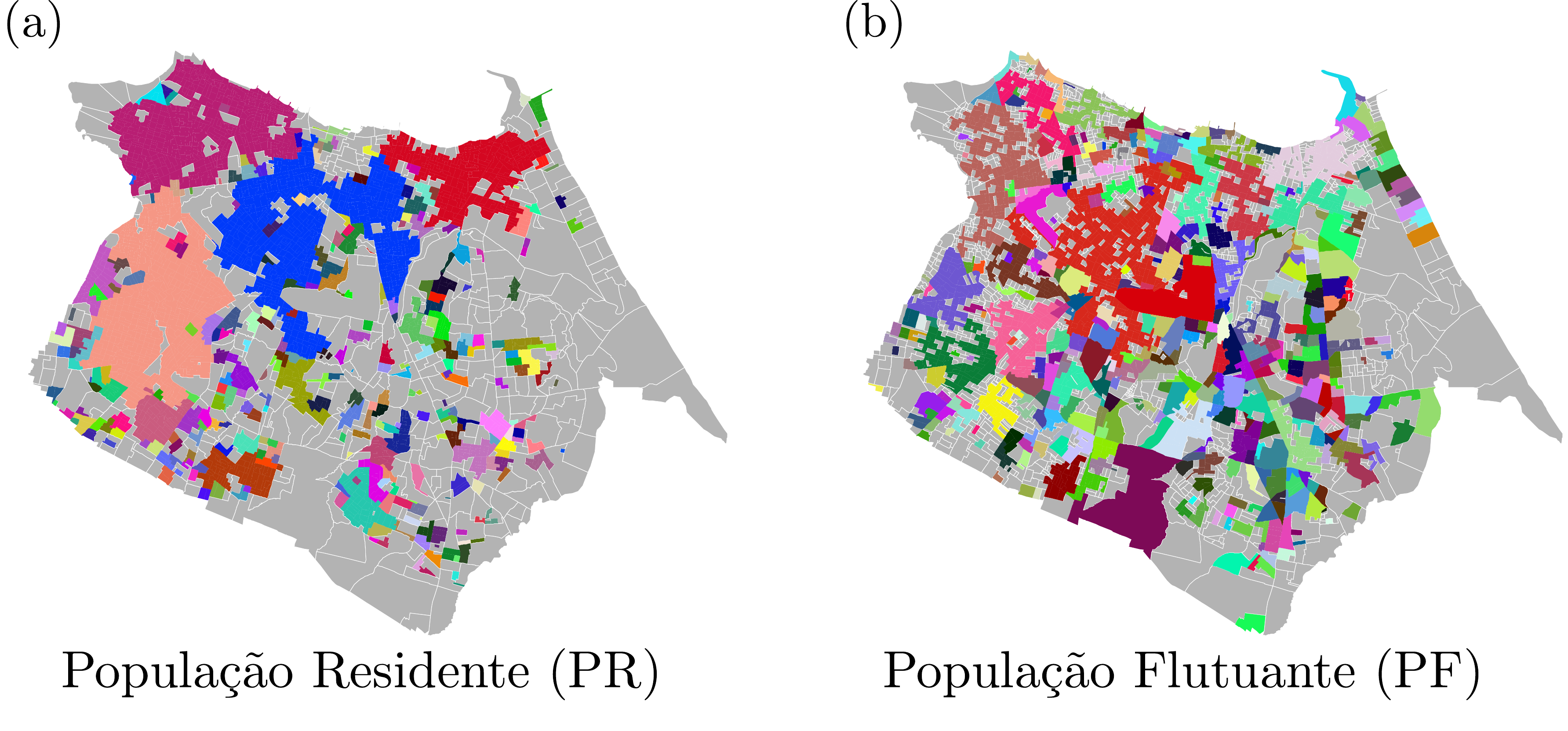}
\caption{{\bf CCA aplicado à população residente (PR) e população flutuante (PF)
em Fortaleza.} Cada cor representa um aglomerado, as áreas na cor cinza claro
correspondem a setores censitários que não foram agrupados por terem $D_i <
D^*$. (a) densidade populacional foi utilizada para encontrar os limites dos
aglomerados com $\ell=270$ metros e $D^*=6000$ pessoas residindo por
$\mathrm{km}^2$. (b) divisão encontrada considerando a mobilidade urbana é
ilustrada, o mapa foi gerado para $\ell=320$ metros e $D^*=2000$ pessoas
passando por $\mathrm{km}^2$ em um dia em Fortaleza.}
\label{clusters}
\end{figure}

Os setores censitários foram agrupados, usando o CCA, por PR e PF (Fig
\ref{clusters}). Na Figura \ref{clusters}a, a divisão alcançada por PR é
ilustrada, a cidade foi dividida usando $\ell=270$ metros e $D^*=6000$
residentes por $\mathrm{km}^2$. Na Figura \ref{clusters}b é ilustrada a divisão
encontrada para PF, usando $\ell=320$ metros e $D^*=2000$ pessoas passando por
$\mathrm{km}^2$ em um dia. Ressalta-se a existência de maiores lacunas no mapa
de PR (Fig \ref{clusters}a) do que ocorre no mapa de PF (Fig \ref{clusters}b).
Tal característica se dá pelo fato de Fortaleza ter regiões de natureza apenas
comercial, em outras palavras, regiões onde não existe uma grande presença de
residentes. Quanto a população flutuante, há pessoas se movendo praticamente por
toda a cidade, tanto em áreas comerciais, quanto em áreas residenciais.

\begin{figure}[!h]
\includegraphics[width=1\textwidth]{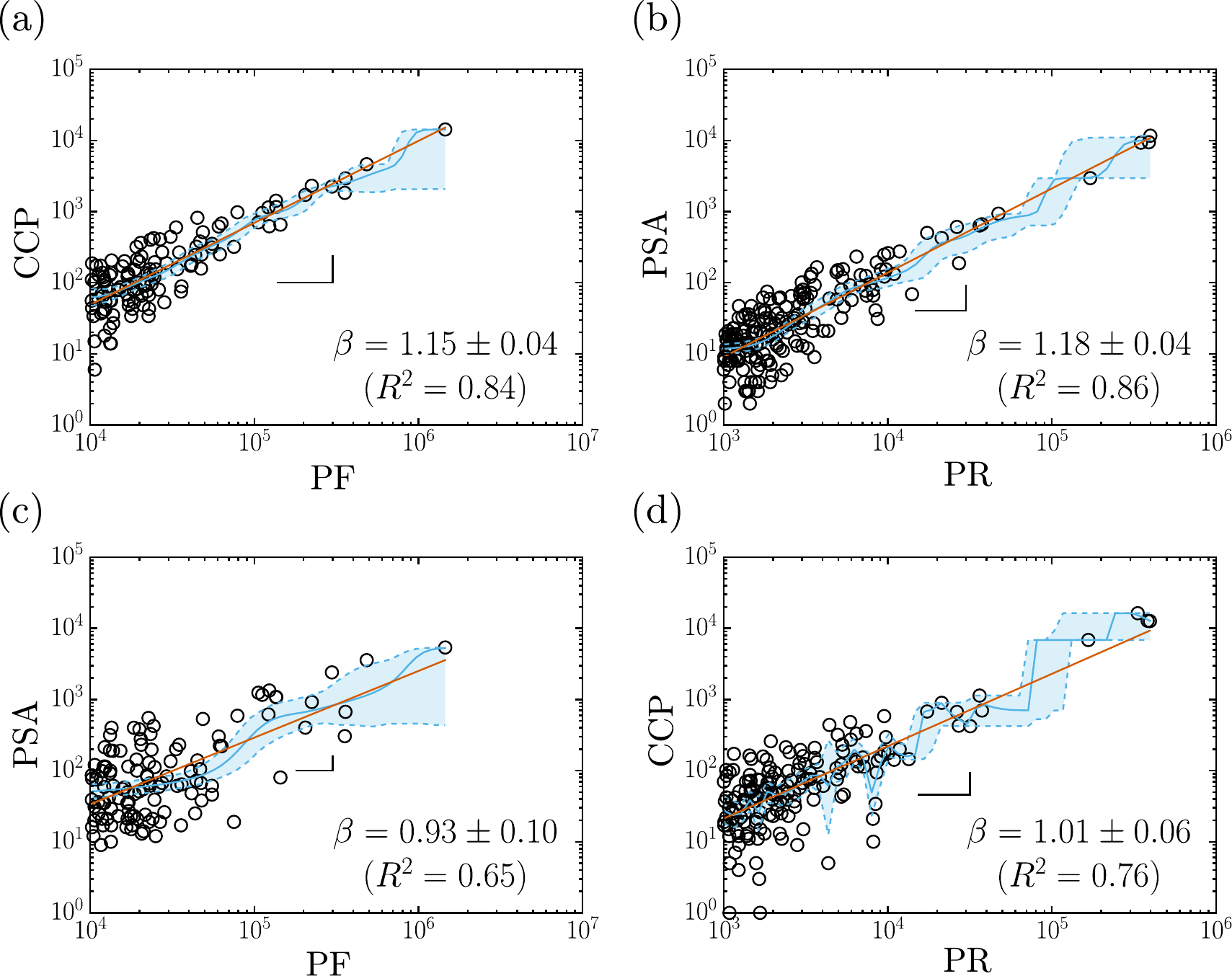}
\caption{{\bf Gráficos de dispersão calculados a partir dos aglomerados
encontrados pelo CCA.} As linhas vermelhas representam as regressões lineares
aplicadas aos dados, as linhas contínuas azuis representam o método
Nadaraya-Watson \cite{nadaraya1964,watson1964} e as linhas tracejadas azuis
delimitam o intervalo de confiança de 95\% estimado por {\it bootstrap}
\cite{racine2004,li2004}. (a) Uma relação superlinear foi encontrada, com
expoente $\beta = 1.15 \pm 0.04$, entre população flutuante (PF) e crimes contra
o patrimônio (CCP). (b) Uma relação superlinear também foi encontrada entre a
população residente (PR) e perturbação de sossego alheio (PSA), com expoente
$\beta = 1.18 \pm 0.04$. (c)-(d) Os gráficos de dispersão de PF com PSA e PR com
CCP mostraram uma relação isométrica entre as variáveis, mas com correlações
menores que (a) e (b). O $R^2$ é o Coeficiente de Determinação das regressões.}
\label{fig7}
\end{figure}

Ao comparar com os resultados apresentados na Figura \ref{correlacao_setores}, a
aplicação do CCA nos dados de população divulga um cenário diferente para as
correlações entre as variáveis investigadas neste trabalho. Primeiro, como
ilustrado na Figura \ref{fig7}a e \ref{fig7}b, relações superlineares em termos
de leis de potência, $Y = aX^\beta$, são reveladas entre PF e CCP, bem como entre
PR e PSA, com expoentes $\beta = 1.15 \pm 0.04$ e $\beta = 1.18 \pm 0.04$,
respectivamente. Em contrapartida, as relações obtidas entre PF e PSA, bem como
entre PR e CCP estão mais próximas de uma relação isométrica (linear), com
expoentes $\beta = 0.93 \pm 0.10$ e $\beta = 1.01 \pm 0.06$ respectivamente,
contudo, os valores baixos dos coeficientes de determinação correspondentes
indicam que estes resultados devem ser interpretados com cautela.

A relação alométrica superlinear entre PR e CCP, implica que o aumento do volume
de popuação flutuante em uma região da cidade acontecerá a custo de uma taxa
proporcionalmente maior de crimes na mesma região. Mais importante ainda, esse
comportamento superlinear ocorrendo em microescala (dentro de uma
cidade) fornece uma explicação plausível para a alometria de crimes sérios
encontrados por Bettencourt {\it et al.} \cite{bettencourt2010} e Melo {\it et
al.} \cite{melo2014} em estudos realizados com dados de cidades brasileiras e
estadunidenses, em escala macrodinâmica. Diferente do que se afirmava, não é a
população residente que explica o comportamento superlinear de crimes sérios,
mas a movimentação das pessoas dentro dos aglomerados urbanos.

É digno de menção ainda que, o resultado da correlação entre PR e CCP, está de
acordo com a Teoria das Atividades Rotineiras \cite{cohen1979}, que afirma que
um crime ocorre pela convergência das rotinas de um agressor motivado, de uma
vítima desprotegida, bem como a ausência de um guardião capaz de prevenir a
transgressão. Além de fornecer apoio quantitativo para essa teoria, mostrando
que a convergência das rotinas dos agentes aumenta o número de crimes, este
estudo também indica o fato notável de que esse aumento é não-linear. Em outras
palavras, este estudo provê a primeira explicação do fenômeno a nível de escala,
mostrando que esse fenômeno não-linear ocorre em aglomerados onde passam dezenas
de milhares, centenas de milhares e até milhões de pessoas em um dia.

\section{Discussão}

Os resultados obtidos ao fim deste capítulo respondem as questões de pesquisa
expostas no capítulo \ref{cap1} desta tese.

{\it $QP_1$ - Há relações alométricas entre população e crime dentro de uma
cidade?}

Sim. Foram encontradas relações alométricas entre população e crime na cidade de
Fortaleza. Utilizou-se o CCA para delimitar as fronteiras da influência social,
mensurada indiretamente pela densidade de PR e PF em diversas regiões da cidade.
Foi encontrado que PF escala de maneira alométrica com CCP, com expoente $\beta
= 1.15 \pm 0.04$ e PR escala com chamadas de PSA, também de forma superlinear,
com expoente $\beta = 1.18 \pm 0.04$. Variando os parâmetros $\ell$ e $D^*$ do
CCA, Fortaleza foi dividida de 3500 formas diferentes por estatísticas de
população residente e mais 3500 por estatísticas de população flutuante
(conforme ilustrado na Figura \ref{variacao_beta}). Em mais de 80\% dessas
divisões, foram constadas relações alométricas superlineares de PR contra PSA e
PF contra CCP, provendo forte evidência estatística de que existem tais
relações.

{\it $QP_2$ - A quantidade de população residente e flutuante é igualmente
importante para explicar a quantidade de crimes em diferentes regiões de uma
cidade?}

Não. Conforme já mencionado em $QP_1$, o aumento no número de chamadas de PSA é
correlacionado com o crescimento de PR e o mesmo ocorre entre CCP e PF. No
entanto as correlações substancialmente mais fracas encontradas entre PR contra
CCP e PF contra PSA, sugerem que PR e PF não são adequadas para explicar a
ocorrência de CCP e PSA, respectivamente. Esse resultado mostra que, diferente
do se pensava, não é a grande concentração de população residente que explica o
crescimento desproporcional de crimes violentos em muitas cidades, mas o
processo diário de comutação das pessoas, que em microescala, propicia
que haja convergência das rotinas de vítimas e criminosos, conforme é definido
pela Teoria das Atividades Rotineiras \cite{cohen1979}.

Os resultados descritos aqui trazem alternativas para a implementação de
práticas inovadoras para gestores públicos dentro de cidades. A mais óbvia delas
se refere ao fato de que, ao mostrar a correlação de diferentes tipos de crimes
não só com população residente, mas também com a população flutuante,
estratégias de alocação da força policial devem ser implementadas através da
análise de diferentes bases populacionais, dependendo do tipo de crime que se
deseja prevenir. Por exemplo, a alocação de policiamento comunitário, mais
apropriada para resolver conflitos que potencialmente podem emergir da
perturbação de sossego, deve ser planejada a partir de uma configuração de
aglomerados e uma análise de {\it hot spots} que foram produzidos da perspectiva
da densidade da população residente. Quando é necessário estabelecer uma
política de alocação de uma polícia uniformizada para mitigar os crimes contra o
patrimônio, a alocação da força policial deve ser realizada a partir de uma
análise do movimento das pessoas.

Além dessas estratégias de alocação policial, os resultados aqui descritos
constituem importantes indicadores para a formulação de políticas públicas de
uso do solo e de projeto ambiental em geral. Trabalhos nesta linha foram
desenvolvidos em \cite{taylor1996}, onde foi proposto um {\it framework} para
associar espaços físicos e o sentimento de segurança, bem como
\cite{brantingham1981}, que lançou a criminologia ambiental, focando seu estudo
criminológico sobre fatores ambientais ou de contexto que podem influenciar a
atividade criminosa. Esses fatores incluem espaço, tempo, lei, agressor e alvo
ou vítima. Estes cinco componentes são uma condição necessária e suficiente,
pois sem um, os outros quatro, mesmo juntos, não constituirão um incidente
criminal. A descoberta que existe uma relação superlinear entre crime e
população (residente ou flutuante) em aglomerados dentro das cidades, reforça a
alegação de que as mudanças no espaço urbano podem levar à redução da
criminalidade como discutido em \cite{kinney2008}. Neste contexto, acredita-se
que existam duas estratégias possíveis para reduzir os crimes nas cidades. Em um
curto período de tempo, pode-se tentar modificar a rota do transporte público, a
fim de evitar uma alta convergência espaço-temporal das pessoas. Em um longo
período de tempo, estimulando com isenções fiscais, por exemplo, poderiam ser
criados diversos centros autônomos na cidade, isso certamente reduziria o
deslocamento entre esses centros altamente convergentes.

\chapter{Uma ferramenta de apoio na tarefa de alocação policial} \label{cap5}

Diante da descoberta da existência de uma relação alométrica superlinear entre o
PF e CCP na cidade de Fortaleza, buscou-se avaliar o impacto que a mobilidade
urbana deveria exercer sobre a alocação de recursos policiais na cidade. Para
avaliar esse impacto, foi desenvolvido um {\it framework} de apoio a atividade
de alocação policial \cite{melo2005, di2004, gelman2007, agnew2007, ferrett2007,
guedes2014, kennedy1990, beato2014}. Esse {\it framework} possui dois módulos,
no primeiro, denominado módulo de análise de dados, é utilizada uma ferramenta
de {\it business intelligence} para fazer mineração de dados de crimes e
população em aglomerados urbanos. Essa ferramenta gera como saída uma divisão da
cidade, baseada em escolhas do usuário, que terá acesso a alguns dos gráficos
expostos no capítulo \ref{cap4} desta tese para dar apoio a sua decisão. O
segundo módulo, denominado módulo de alocação, é um {\it software} que recebe
como entrada um arquivo {\it Keyhole Markup Language} (KML), com a divisão da
cidade obtida a partir do primeiro módulo e a quantidade de recursos policiais
disponíveis para alocação. Esse {\it software} aloca os recursos policiais de
forma proporcional à quantidade de crimes que ocorrem nas regiões da cidade,
seguindo o modelo de alocação baseado em alta densidade de crimes
\cite{sherman1989, sherman1995, weisburd2006, ratcliffe2006, wortley2016,
groff2002, berk2011, kennedy2011}.

Além de descrever o {\it framework} desenvolvido, este capítulo tem como
objetivo realizar um exercício de aplicação para a descoberta de que a
mobilidade humana é a chave para entender a dinâmica de crimes contra o
patrimônio em uma grande metrópole. Para realizar tal exercício, será utilizado
o {\it framework} desenvolvido para gerar os aglomerados de população flutuante
estimados na seção \ref{cap41} desta tese. Esses aglomerados serão
utilizados em uma estratégia de alocação policial onde os recursos policiais
serão distribuídos por divisões de terreno definidas por estatísticas de
população flutuante. Essa estratégia será comparada com uma estratégia de
alocação em que os agentes da polícia são distribuídos por divisões
administrativas de terreno, comumente definidas a partir de estatísticas de
população residente.

\section{Descrição do framework}

\subsection{Módulo de análise de dados}

O módulo de análise de dados foi desenvolvido utilizando a plataforma de {\it
business intelligence} Qlik Sense
\footnote{http://www.qlik.com/us/products/qlik-sense}, que possibilita o
tratamento e visualização de grandes volumes de dados. Foi desenvolvido um {\it
dashboard} protótipo, que possui três painéis de visualização. O primeiro painel
apoia a análise de dados brutos de crimes e população, o principal propósito
desse painel é permitir que o usuário se familiarize com os dados, sendo
possível verificar inconsistências ou buscar faixas dos dados que merecem passar
por algum processo de higienização. O segundo painel dá suporte à análise de
agregações de unidades terreno realizadas pelo CCA, onde é possível variar os
parâmetros do algoritmo, avaliar a sensibilidade do expoente alométrico em
função dessa variação e ainda analisar a relação entre a área dos aglomerados e
sua população (tal qual foi apresentado na seção \ref{cap41} desta tese). Por
fim, o terceiro painel ilustra as correlações entre população e crime com seus
respectivos expoentes alométricos encontrados. Foi incluído nesse módulo de
análise de dados uma funcionalidade de exportação de delimitações definidas pela
parametrização escolhida pelo usuário. As Figuras \ref{dash1}, \ref{dash2} e
\ref{dash3} ilustram os painéis de visualização desenvolvidos. Todos os painéis
são interativos, sendo possível realizar filtros por áreas de mapas e barras ou
pontos de gráficos.

\begin{landscape}

\begin{figure}[!h] 
\includegraphics[width=1.5\textwidth]{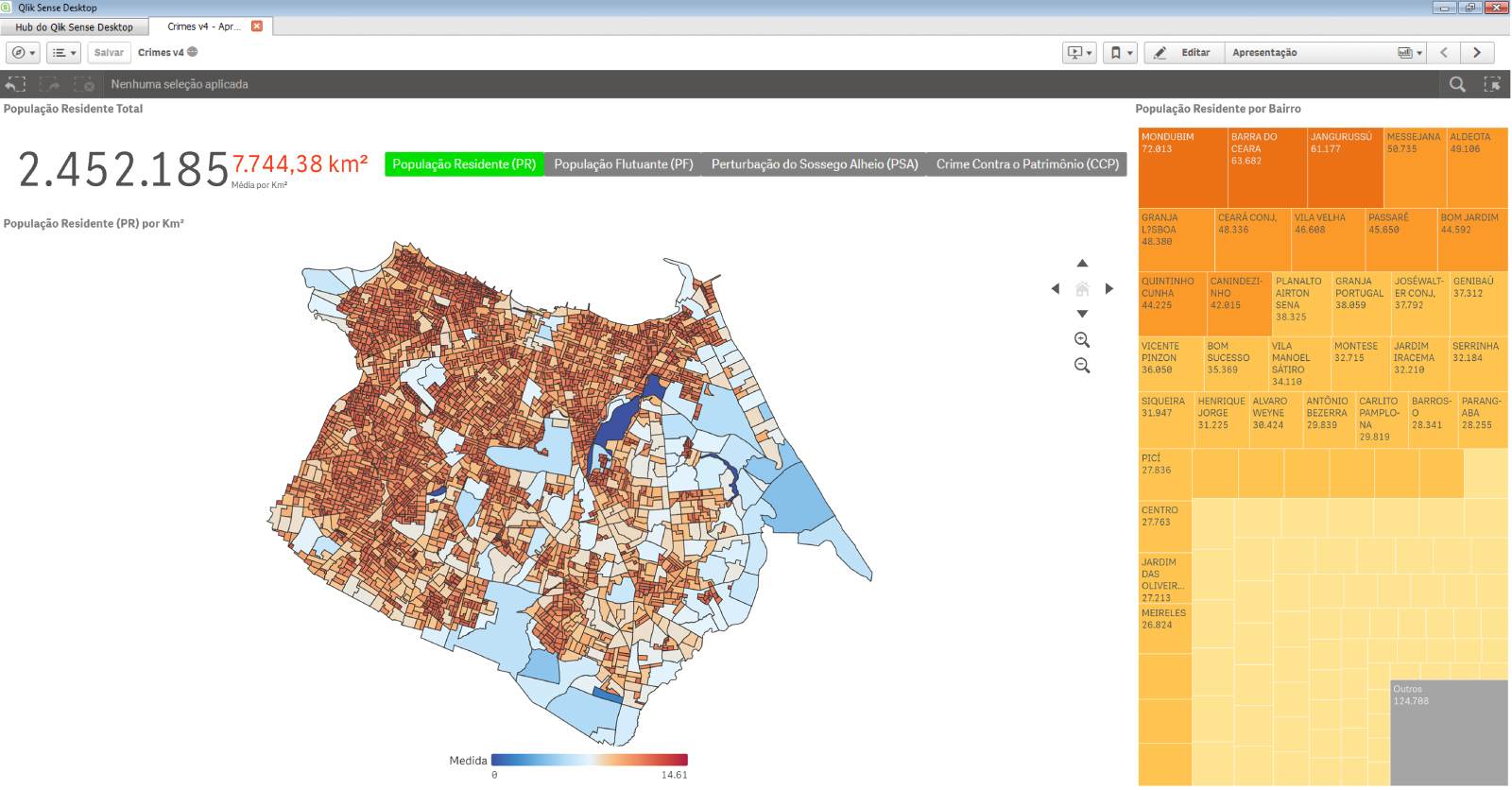}
\caption{{\bf Painel de análise de de dados brutos.} O mapa central ilustra a
densidade de crimes ou população, dependendo da escolha do usuário. É possível
observar ainda a distribuição de crimes por tipo, por bairro e por agência ({\it
e.g.} PM, SAMU e DETRAN). Esse painel tem grande utilidade na verificação de
inconsistências nos dados. A título de exemplo, foram observados erros de
georreferenciamento nos dados de crime. Ao selecionar um bairro (no gráficos em
tons de laranja à direita), observou-se que cerca de 5\% das latitudes e
longitudes apontavam para regiões fora do bairro. Devido a tal observação, esses
dados foram removidos das análises.}
\label{dash1}
\end{figure}

\begin{figure}[!h] 
\includegraphics[width=1.5\textwidth]{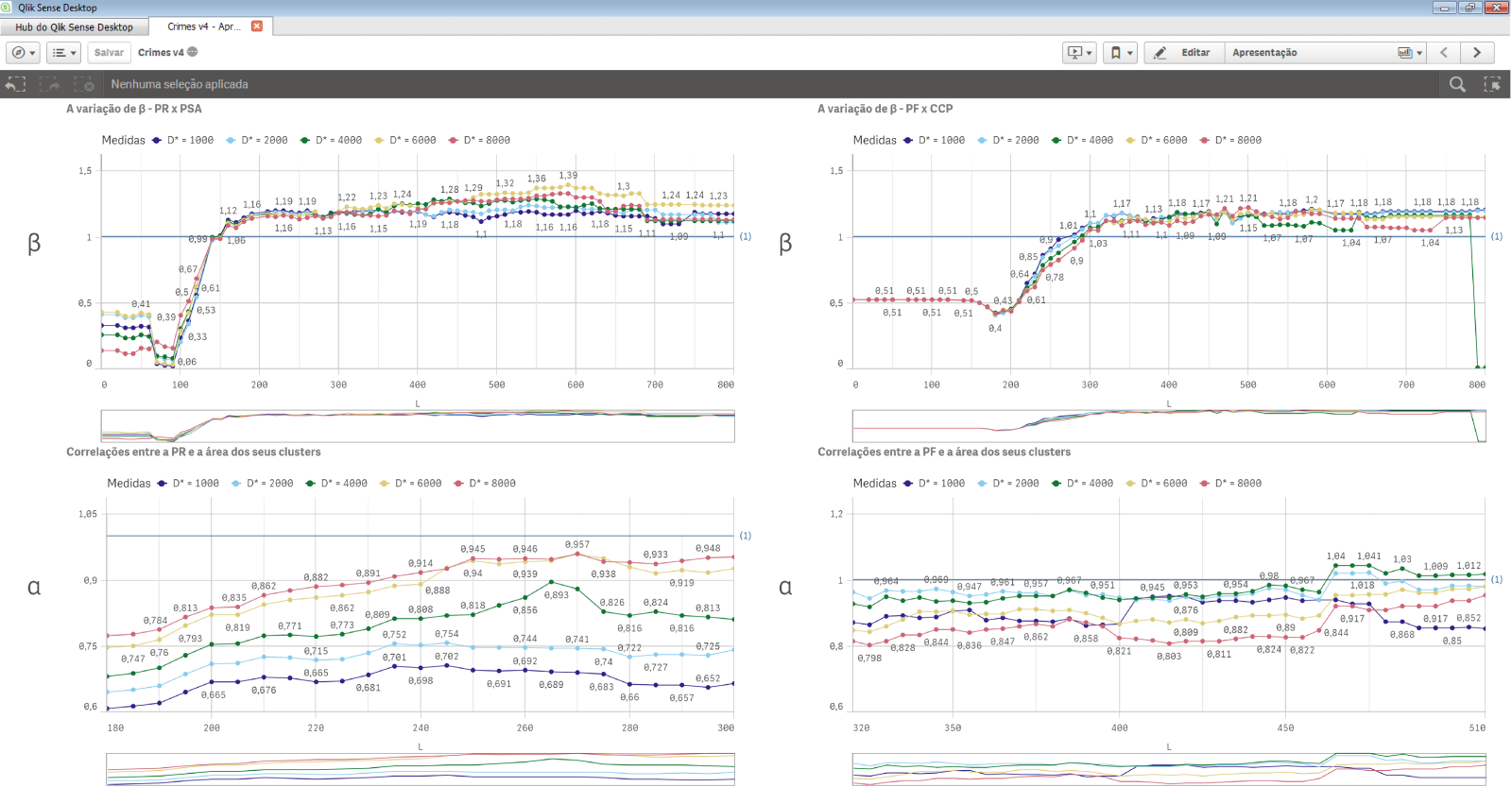}
\caption{{\bf Painel de análise de agregações de unidades terreno realizadas
pelo o CCA.} Esse painel permite avaliar o efeito da parametrização do CCA sobre
o expoente alométrico nas relações entre população e crime. Os gráficos
ilustrados são inspirados nas Figuras \ref{variacao_beta} e \ref{variacao_alpha}
desta tese.}
\label{dash2}
\end{figure}

\begin{figure}[!h] 
\includegraphics[width=1.5\textwidth]{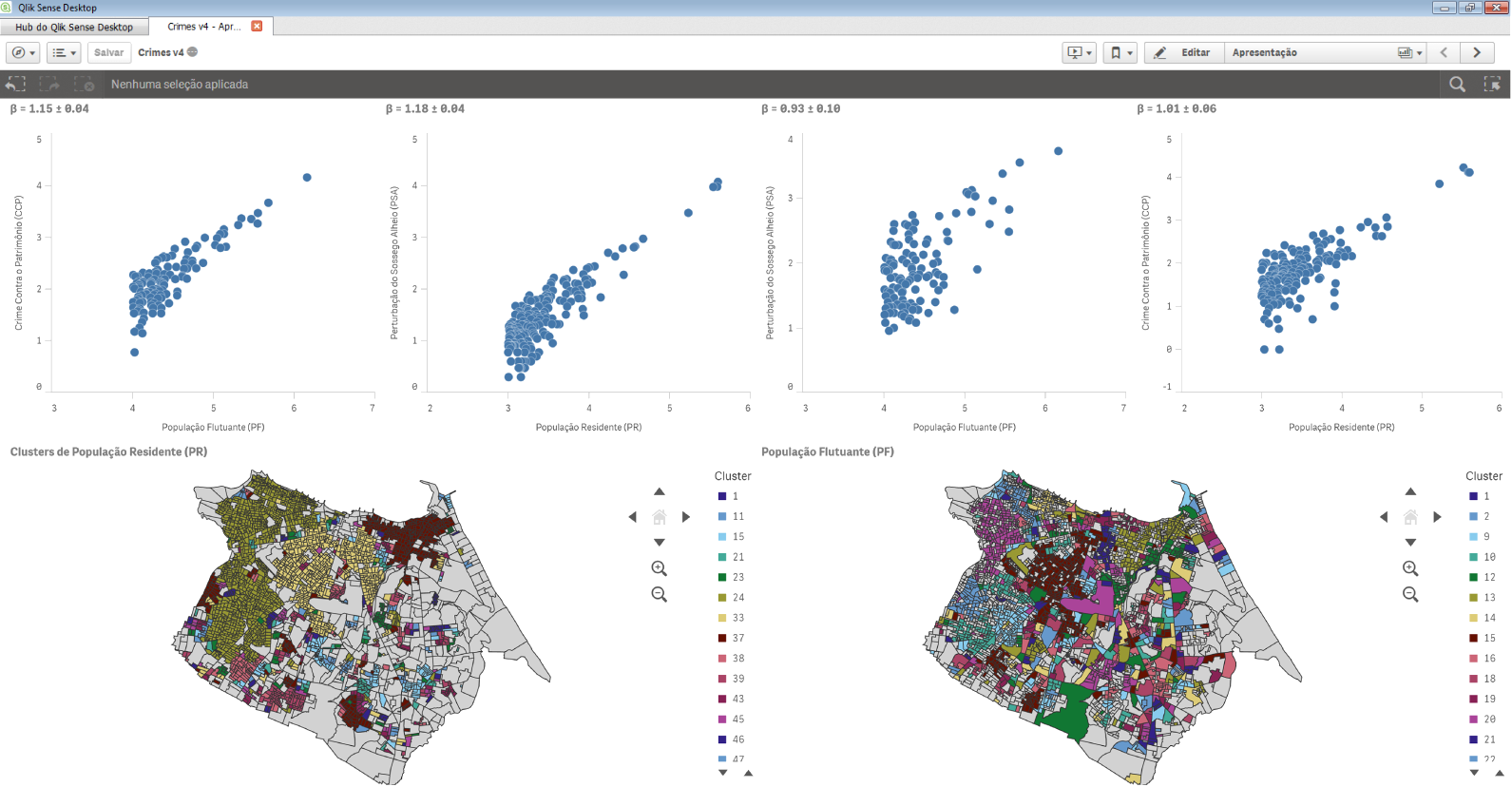}
\caption{{\bf Correlações entre população e crime com seus respectivos expoentes
alométricos encontrados.} A partir da seleção de um ponto (que representa uma
configuração dos parâmetros do CCA) nos gráficos ilustrados no segundo painel de
visualização, é possível verificar a dispersão dos dados de população e crime
para a configuração do CCA escolhida. Esse terceiro painel também ilustra a
divisão espacial dos aglomerados encontrados, tanto para população residente,
quando para população flutuante.}
\label{dash3}
\end{figure}

\end{landscape}

\subsection{Módulo de alocação}

O módulo de alocação policial é um {\it software}, desenvolvido utilizando a
linguagem de programação Java \footnote{http://www.java.com}, atualmente na
versão 1.8. Esse módulo tem como entradas um arquivo KML, com uma divisão de uma
cidade, a quantidade ($T$) de policiais disponíveis para alocação e informações
georreferenciadas de crimes. O {\it software} distribui os recursos policiais
por diferentes locais da cidade de forma proporcional à quantidade de crimes
existentes no local. Desta forma, a parte de um recurso policial, $T_{s_i}$,
alocado em uma sub-região do espaço urbano (seja um aglomerado ou uma unidade
administrativa de terreno), $s_i$ $\in$ S, a partir da quantidade de crimes
ocorridos em $s_i$, $C{s_i}$, pode ser definido formalmente como:
\begin{equation} 
T_{s_i} = \frac{T * C{s_i}}{C}, 
\end{equation}
\noindent onde $T$ corresponde ao total de policiais disponíveis para a alocação
e $C$ ao total de crimes ocorridos em todo o espaço urbano disponível para
alocação.

Foi ainda adotada uma política de alocação interna, mais precisamente a nível de
$s_i$. Tanto aglomerados de população, quanto unidades administrativas de
terreno ({\it e.g.} divisão por bairros ou zonas) são compostos por setores
censitários e internamente é também feito um direcionamento de recursos de forma
proporcional a quantidade de crimes de cada setor dentro de $s_i$. Em outras
palavras, dentro de cada sub-região $s_i$, setores com mais crimes recebem mais
policiais. Essa política de sub-alocação é justificada pela necessidade de
comparar as duas estratégias de alocação. A seguir isso será discutido de
maneira mais profunda.

\begin{figure}[!h] 
\includegraphics[width=1\textwidth]{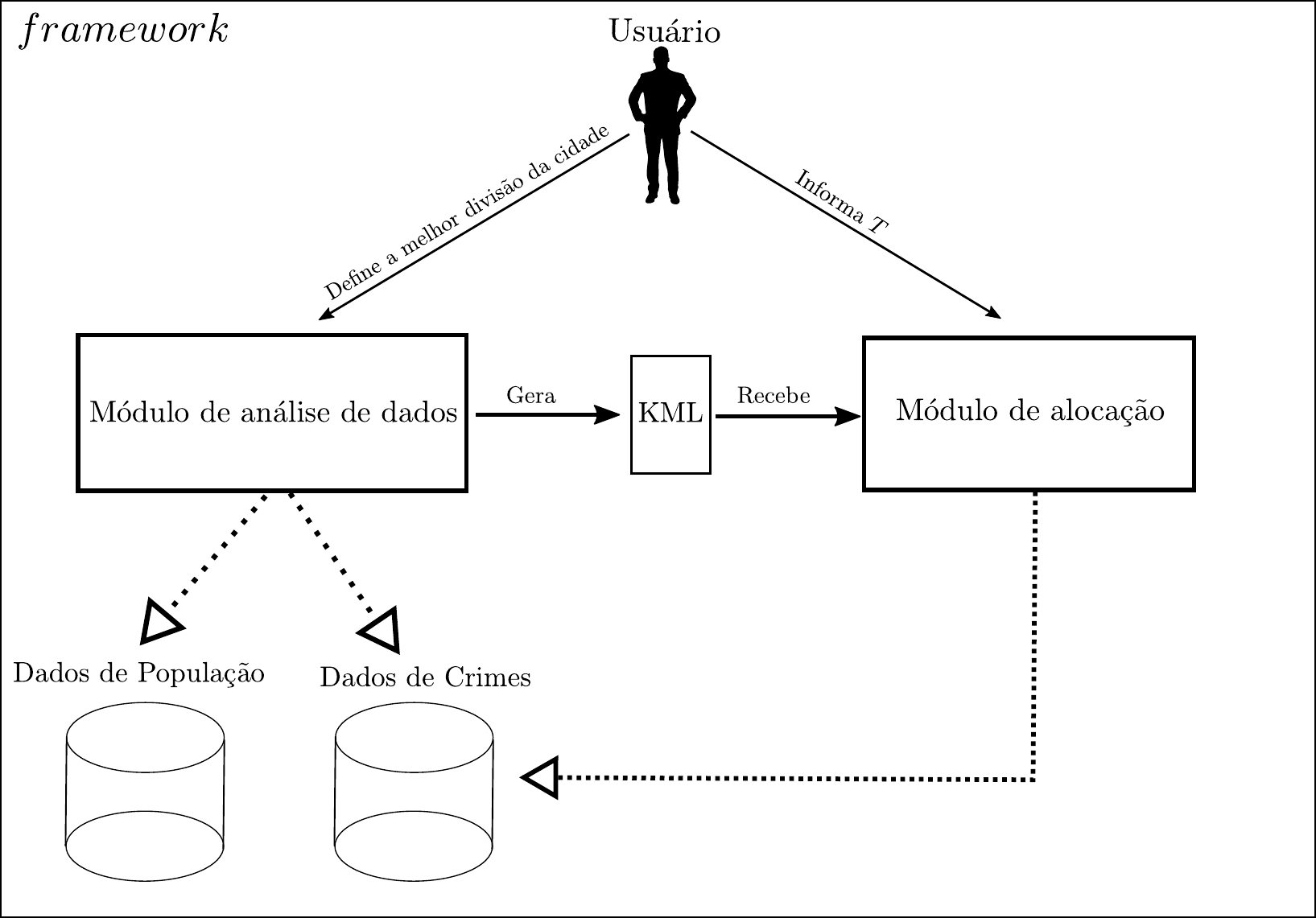}
\caption{{\bf Arquitetura geral do framework desenvolvido.} O usuário utiliza o
módulo de análise de dados para definir a melhor divisão da cidade para a
alocação de uma polícia especializada para combater um determinado tipo de
crime. O módulo de análise de dados gera como saída um arquivo KML com a divisão
definida pelo usuário e essa divisão é utilizada no módulo de alocação policial.
O usuário informa a quantidade de policiais disponíveis para a alocação. O
módulo de análise de dados acessa informações de bancos de dados de crimes e
população de uma cidade, já o módulo de alocação acessa apenas um banco de dados
de crimes.}
\label{comp-framework}
\end{figure}

A arquitetura geral do {\it framework} desenvolvido pode ser visualizada na 
Figura \ref{comp-framework}.

\section{Estratégias}

Serão comparadas aqui duas estratégias de alocação policial, as quais se
inspiram no modelo de alocação mais popular, o modelo baseado em alta densidade
de crimes \cite{sherman1989, sherman1995, weisburd2006, ratcliffe2006,
wortley2016, groff2002, berk2011, kennedy2011}, que aloca o recurso policial
proporcionalmente em função da densidade de crimes nos locais. 

A primeira estratégia, denominada Estratégia de Alocação por População Residente
(APR), uma estratégia cujos recursos são distribuídos de forma proporcional à
quantidade de ocorrências em divisões administrativas de terreno, normalmente
definidas por estatísticas de população residente. Será adotada a divisão por
bairros\footnote{disponível em http://dados.fortaleza.ce.gov.br/}, pois apesar da divisão administrativa por setores censitários também
estar disponível, a mesma é por demais segmentada, com alguns setores sendo
menores que um quarteirão, sendo assim inviável de ser utilizada em uma política
real de alocação.

A segunda estratégia de alocação utilizada, denominada Estratégia de Alocação
por População Flutuante (APF), também distribuirá os recursos policiais de forma
proporcional a quantidade de chamadas à polícia em uma divisão espacial, no
entanto, nessa estratégia serão utilizados os aglomerados de fluxo de pessoas
estimados na seção \ref{cap41} desta tese, ilustrados na Figura \ref{clusters}b.
Foi utilizado o módulo de análise de dados do {\it framework} desenvolvido para
gerar um arquivo KML com a divisão de aglomerados por população flutuante
($\ell=320$ e $D^*=2000$).

\section{Comparação das estratégias}

\begin{figure}[!h] 
\includegraphics[width=1\textwidth]{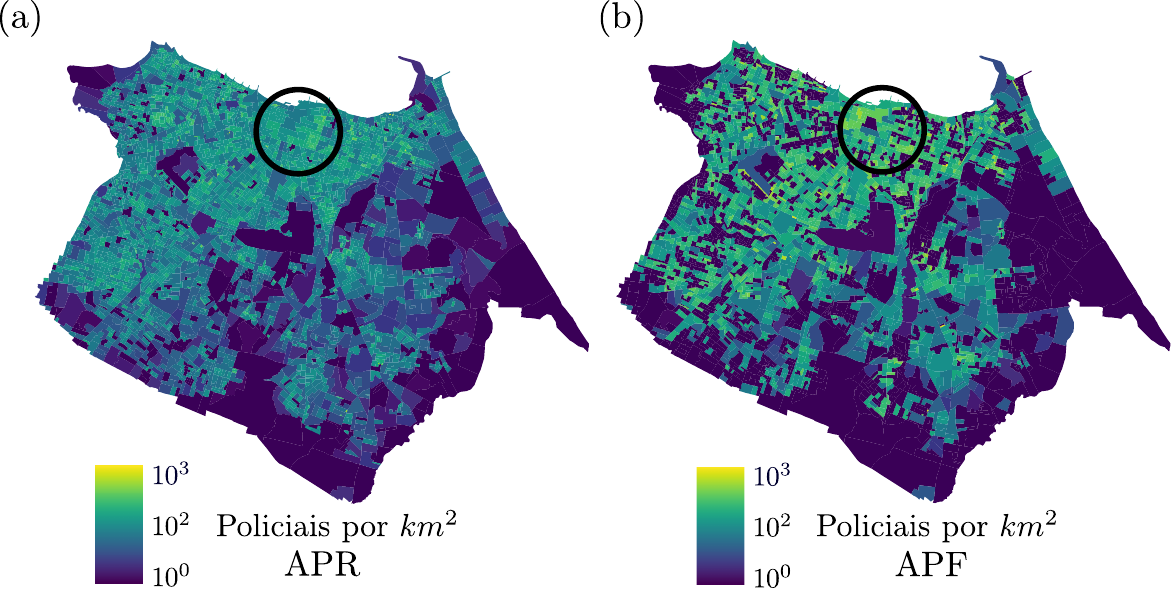}
\caption{{\bf Alocação policial utilizando as duas estratégias estudadas.} Em
(a) é ilustrado mapa de densidade de policiais alocados utilizando APR, em (b) é
ilustrado mapa de densidade de policiais alocados utilizando APF. Os círculos
pretos destacam o centro comercial de Fortaleza, área de grande concentração de
PF e por consequência, grande concentração de CCP.} \label{mapas_aloc}
\end{figure}

Ao aplicar APR e APF em Fortaleza simulando a disponibilidade de um recurso
policial total $T=10000$, obtém-se os mapas de calor ilustrados na Figura
\ref{mapas_aloc}, itens (a) e (b) respectivamente. Observam-se \textit{hot
spots} mais intensos em APF, principalmente no centro comercial da cidade,
destacado pelo circulo preto em ambas as figuras. Isso ocorre devido APF não
alocar recursos policiais em áreas consideradas não populadas $(D_i > D^*)$,
concentrando mais policiais nas regiões mais críticas da cidade.

\begin{figure}[!h] 
\includegraphics[width=1\textwidth]{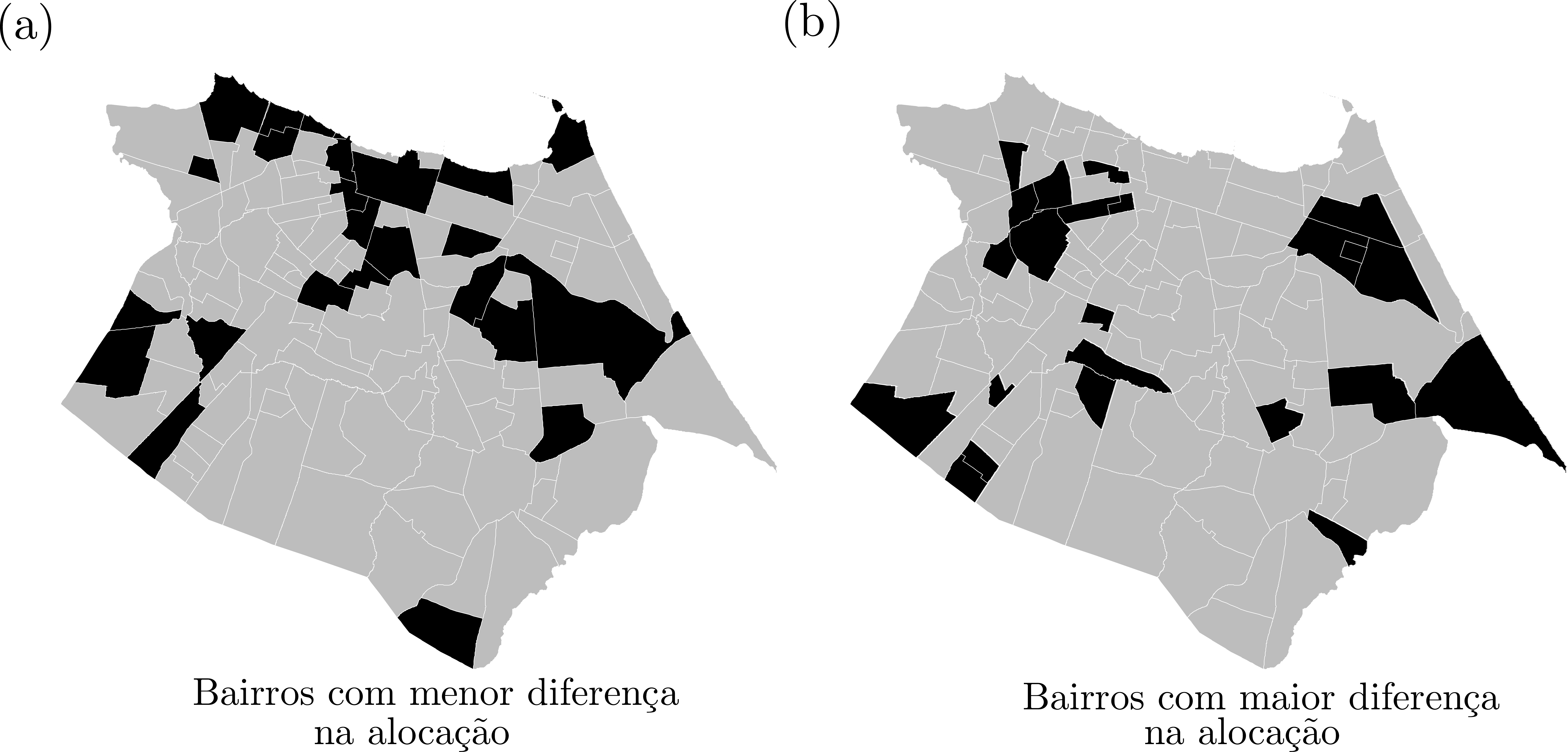}
\caption{{\bf Diferenças e semelhanças entre as alocações estudadas.} Em (a) são
destacados na cor preta os bairros que tiveram alocação mais parecida utilizando
APR e APF. Em (b) são destacados os bairros com maior diferença
na quantidade de policiais alocados. Em ambas as figuras foram destacados 24
 bairros, $20\%$ do total da cidade.}
\label{comp_aloc_bairro}
\end{figure}

Para efeito de comparação, foi calculada a quantidade de policiais alocados por
bairro utilizando APF. Foi somada a quantidade de policiais nos setores
censitários localizados dentro de cada bairro (A Tabela \ref{qtd-pol-alocados}
ilustra a quantidade de policiais por bairro). Após isso, foi calculada a
diferença percentual entre a quantidade de policiais alocados por bairro nas
duas estratégias. Na Figura \ref{comp_aloc_bairro}, itens (a) e (b), são
ilustrados os bairros onde a alocação é mais semelhante e mais diferente
respectivamente. 

\begin{table}[]
\centering
\caption{Quantidade de policiais alocados por bairro pelas duas estratégias 
de alocação policial estudadas.}
\label{qtd-pol-alocados}
\begin{tabular}{ c  c  c  c  c  c  c  c  c}
\hline
{ \textbf{Id Bairro}} & {\textbf{APR}} & {\textbf{APF}} 
& {\textbf{Id Bairro}} & {\textbf{APR}} & {\textbf{APF}}
& {\textbf{Id Bairro}} & {\textbf{APR}} & {\textbf{APF}} \\ 
\hline
\textbf{35}& {67}& {93} & \textbf{36}& {5}& {14}& \textbf{54}& {137}& {99}\\ 
\textbf{33}& {13}& {18} & \textbf{34}& {61}& {71} & \textbf{50}& {131}& {103}  \\ 
\textbf{39}& {50}& {28} & \textbf{37}& {39}& {52} & \textbf{52}& {43}& {111} \\ 
\textbf{38}& {165}& {167} & \textbf{43}& {231}& {242} & \textbf{53}& {100}& {98}  \\ 
\textbf{42}& {68}& {69} & \textbf{41}& {89}& {94} & \textbf{46}& {225}& {215} \\ 
\textbf{40}& {10}& {0} & \textbf{22}& {16}& {18} & \textbf{51}& {37}& {63} \\ 
\textbf{23}& {562}& {571} & \textbf{24}& {17}& {27} & \textbf{44}& {144}& {115} \\ 
\textbf{25}& {131}& {163} & \textbf{26}& {5}& {3} & \textbf{47}& {181}& {120} \\ 
\textbf{27}& {66}& {131} & \textbf{28}& {61}& {60} & \textbf{48}& {19}& {14} \\ 
\textbf{29}& {71}& {58} & \textbf{3}& {385}& {349} & \textbf{45}& {97}& {75}  \\ 
\textbf{2}& {64}& {61} & \textbf{1}& {20}& {28} & \textbf{61}& {167}& {139} \\ 
\textbf{0}& {44}& {62} & \textbf{30}& {40}& {43} & \textbf{49}& {19}& {17} \\ 
\textbf{7}& {29}& {33} & \textbf{6}& {87}& {115} & \textbf{63}& {46}& {59} \\ 
\textbf{32}& {96}& {96} & \textbf{5}& {119}& {120} & \textbf{60}& {53}& {61} \\ 
\textbf{31}& {18}& {20} & \textbf{4}& {95}& {74} & \textbf{65}& {48}& {49} \\ 
\textbf{9}& {50}& {62} & \textbf{8}& {106}& {128} & \textbf{62}& {82}& {111} \\ 
\textbf{19}& {24}& {30} & \textbf{17}& {60}& {60} & \textbf{55}& {69}& {82} \\ 
\textbf{18}& {37}& {48} & \textbf{15}& {199}& {159} & \textbf{64}& {32}& {23} \\ 
\textbf{16}& {125}& {118} & \textbf{13}& {130}& {137} & \textbf{57}& {170}& {118} \\ 
\textbf{14}& {36}& {53} & \textbf{11}& {55}& {64} & \textbf{56}& {25}& {36}  \\ 
\textbf{12}& {85}& {65} & \textbf{21}& {99}& {70} & \textbf{59}& {143}& {88}\\ 
\textbf{20}& {83}& {87} & \textbf{109}& {42}& {34} & \textbf{58}& {59}& {61} \\ 
\textbf{108}& {40}& {63} & \textbf{107}& {109}& {147} & \textbf{75}& {69}& {59} \\ 
\textbf{106}& {103}& {63} & \textbf{105}& {141}& {175} & \textbf{76}& {57}& {36} \\ 
\textbf{104}& {88}& {76} & \textbf{103}& {5}& {5} & \textbf{73}& {148}& {144} \\ 
\textbf{99}& {121}& {72} & \textbf{102}& {7}& {0} & \textbf{74}& {29}& {39} \\ 
\textbf{101}& {110}& {102} & \textbf{100}& {85}& {101} & \textbf{71}& {39}& {43} \\ 
\textbf{98}& {92}& {103} & \textbf{97}& {31}& {22} & \textbf{72}& {63}& {30} \\ 
\textbf{96}& {56}& {77} & \textbf{95}& {35}& {38} & \textbf{68}& {202}& {205} \\ 
\textbf{94}& {61}& {62} & \textbf{93}& {78}& {165} & \textbf{70}& {153}& {137} \\ 
\textbf{92}& {4}& {4} & \textbf{91}& {39}& {28} & \textbf{66}& {42}& {54} \\ 
\textbf{90}& {122}& {111} & \textbf{10}& {271}& {278} & \textbf{69}& {199}& {232} \\ 
\textbf{88}& {125}& {67} & \textbf{89}& {58}& {57} & \textbf{85}& {10}& {3} \\ 
\textbf{79}& {176}& {88} & \textbf{114}& {69}& {96} & \textbf{67}& {26}& {32} \\ 
\textbf{78}& {49}& {42} & \textbf{115}& {138}& {167} & \textbf{87}& {42}& {22} \\ 
\textbf{77}& {123}& {91} & \textbf{112}& {100}& {133} & \textbf{84}& {43}& {58}  \\ 
\textbf{113}& {95}& {117} & \textbf{110}& {173}& {190} & \textbf{81}& {30}& {14} \\ 
\textbf{111}& {31}& {26} & \textbf{82}& {54}& {59} & \textbf{86}& {24}& {6} \\ 
\textbf{83}& {15}& {24} & \textbf{80}& {138}& {181} & \textbf{-} & {-} & {-}\\ 
\hline
\end{tabular}
\end{table}

De uma forma geral, observou-se menor diferença percentual nas alocações dos
bairros com maior presença de PF. Esses bairros estão próximos ao centro
comercial da cidade ou localizados em regiões com grande concentração de
residentes (normalmente locais que são origem de fluxo de pessoas). Observou-se
ainda que os bairros que apresentaram maior diferença percentual entre as
quantidades de policiais alocados usando estratégias de alocação estudadas, são
aqueles que possuem mais setores censitários não populados, ou seja, com
densidade de PF abaixo do limiar $D^*$ estimado na seção \ref{cap41} desta tese.

\begin{figure}[!h] 
\includegraphics[width=1\textwidth]{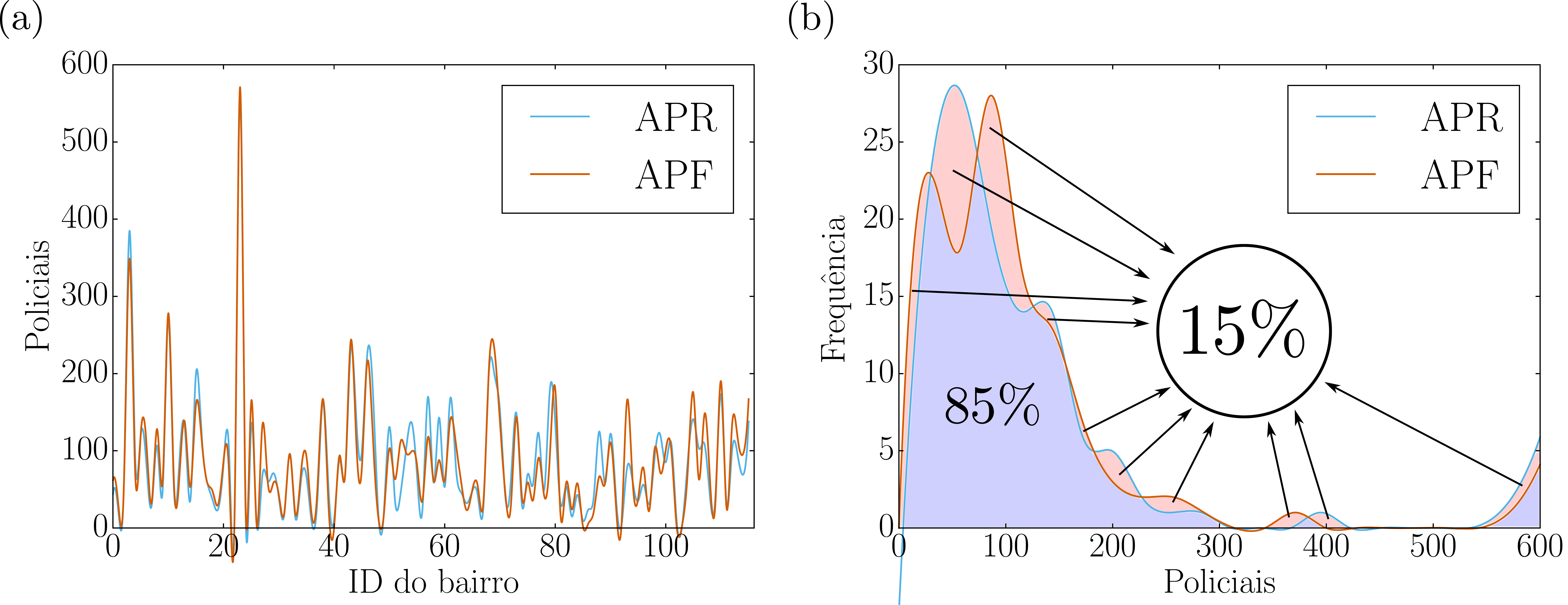}
\caption{{\bf Comparação entre APR e APF.} Em (a) é ilustrada a quantidade de
policiais alocados por bairro. A linha azul representa uma interpolação por
\textit{splines} cúbicos \cite{deboor1978} (mais detalhes no Apêndice I)
aplicada aos valores encontrados para APR. A linha vermelha representa a mesma
interpolação aplicada para as quantidades de policiais alocados utilizando APF.
Em (b) são ilustrados os histogramas das distribuições de alocação nos bairros
da cidade. Para melhor visualização, os histogramas foram gerados em 20
\textit{bins} \cite{wand1997}.}
\label{comp_aloc_distribuicao}
\end{figure}

Na Figura \ref{comp_aloc_distribuicao} pode ser observada uma comparação mais
precisa entre as duas estratégias de alocação. Em (a) são ilustradas as funções
de interpolação \cite{deboor1978} dos bairros pelo número de policiais alocados
pelas duas estratégias estudadas. A interseção das áreas formadas pelas curvas
da interpolação e o eixo $x$ revela aproximadamente 15\% de dissimilaridade
entre as alocações. Essa dissimilaridade pode ser observada de forma mais clara
na Figura \ref{comp_aloc_distribuicao} (b), na qual são ilustradas as funções de
interpolação dos histogramas gerados a partir da quantidade de policiais
alocados por bairros segundo as duas estratégias. A linha azul representa a
função de interpolação dos dados de APR, já a linha vermelha representa a função
estimada para APF. A região na cor azul claro demarca a interseção entre as
distribuições. As regiões na cor vermelho claro representam faixas dos dados
onde não houve interseção. Essas regiões somadas representam 15\% da área
total.

\begin{figure}[!h] 
\includegraphics[width=1\textwidth]{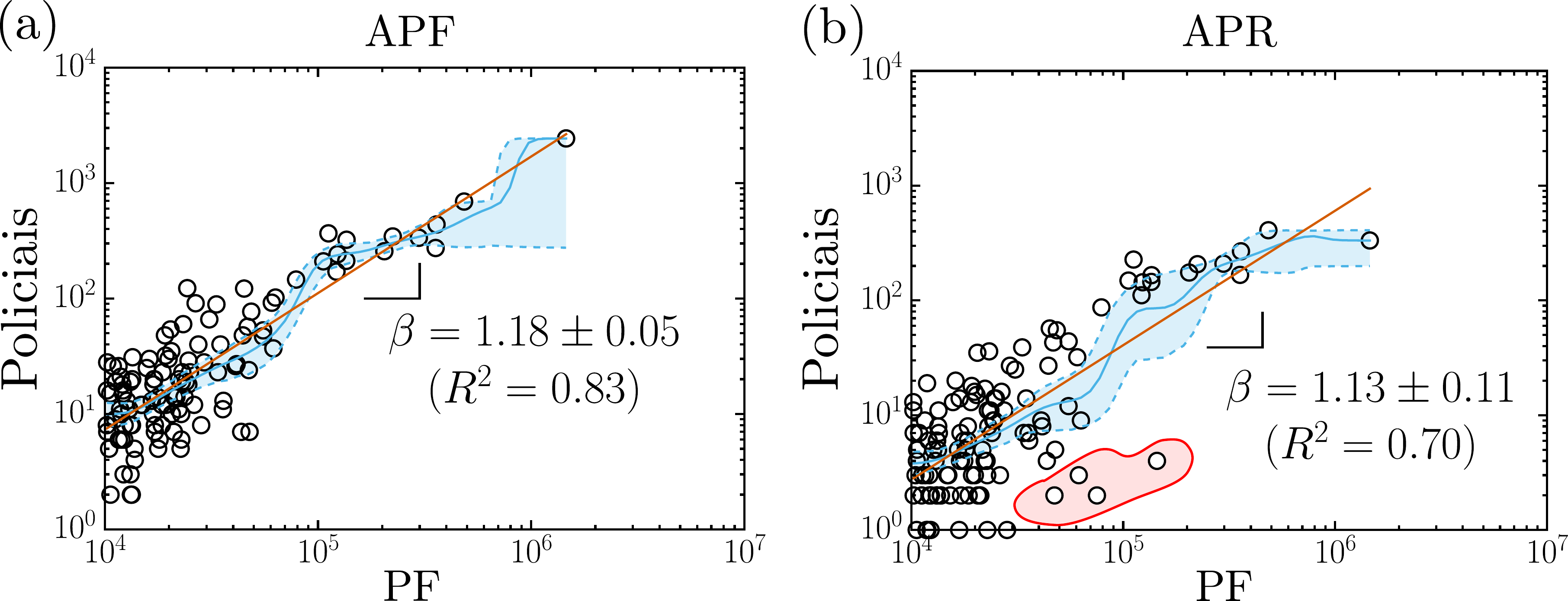}
\caption{{\bf Correlações entre policiais alocados e fluxo de pessoas em APR e
APF.} (a) e (b) ilustram respectivamente as correlações alcançadas para APF e
APR. O eixo $x$ representa a PF e o eixo $y$ a quantidade de policiais alocados.
As linhas vermelhas representam as regressões lineares aplicadas aos dados, as
linhas contínuas azuis representam o método de Nadaraya-Watson
\cite{nadaraya1964,watson1964} e as linhas tracejadas azuis delimitam o
intervalo de confiança 95\% estimado por {\it bootstrap} \cite{racine2004,
li2004}. Em (b) ainda são destacados quatro aglomerados (bolha vermelha) que têm
proporcionalmente um fluxo grande de pessoas para uma pequena presença de
policiais.}
\label{alom-aloc} 
\end{figure}

A diferença constatada entre as estratégias de alocação estudadas pode ser fruto
da ineficiência de APR. Enquanto a alocação produzida por APF é fortemente
correlacionada com o fluxo de pessoas, APR falha ao distribuir a mesma
quantidade de recursos policiais seguindo as leis de escala encontradas na seção
\ref{cap42} desta tese. Na referida seção foi observada relação superlinear,
com expoente $\beta = 1.15 \pm 0.04$, entre CCP e PF. A Figura \ref{alom-aloc}
ilustra as correlações entre a quantidade de policiais alocados, pelas duas
estratégias estudadas, e população flutuante nos aglomerados de fluxo de
Fortaleza. Em (a) é ilustrada a correlação entre as alocações produzidas por APF
e o fluxo de pessoas, observa-se uma evidente relação superlinear, com expoente
$\beta = 1.18 \pm 0.05$ e alto coeficiente de determinação \cite{rawlings2001,
montgomery2015} ($R^2 = 0.83$), similar à encontrada na seção \ref{cap42} desta
tese. Em (b), apesar da aparente relação superlinear, o menor Coeficiente de
Determinação ($R^2 = 0.70$) somado ao maior Erro Padrão do Coeficiente
\cite{rawlings2001, montgomery2015} ($\pm 0.11$) relevam uma possível imprecisão
de APR em alocar recursos policiais de acordo com as necessidades de Fortaleza.
Outro aspecto digno de menção, ainda na Figura \ref{alom-aloc}, é a maior
dispersão dos pontos em APR. Essa dispersão também pode revelar imprecisão dos
métodos de alocação por divisões administrativas. Em (b), são destacados quatro
aglomerados de PF, que mesmo possuindo considerável nível de fluxo de pessoas,
poucos policiais são alocados para atender suas respectivas regiões. Isso ocorre
porque as fronteiras dos bairros por vezes dividem os aglomerados de fluxo, que
são naturalmente formados pela PF, dificultando uma alocação precisa naquela
região.

Existe uma evidente relação de causa e efeito entre PF e CCP. APR é capaz de
produzir uma alocação policial superlinear em função do fluxo de pessoas, porém
de forma imprecisa. Essa imprecisão se deve ao fato de que os métodos
convencionais de alocação policial se baseiam apenas no efeito dessa relação
entre quantidade de transeuntes e crimes. Quando APF considera os aglomerados de
fluxo de pessoas como divisão territorial (causa) e a densidade de crimes para
decidir a proporção de policiais adequada para cada aglomerado (efeito), é
possível também alcançar a relação superlinear evidenciada anteriormente (na
secão \ref{cap42}), no entanto, com maior precisão estatística, mensurada pelo
coeficiente de determinação e erro padrão do coeficiente.

\section{Discussão}

Neste capítulo foi especificado um {\it framework} que apoia o maior
entendimento a respeito da relação entre população e crime dentro de uma cidade.
Esse {\it framework} ainda permite simular a alocação de recursos policiais por
divisões administrativas de terreno e aglomerados de população. A motivação para
a construção desse {\it framework} se deu pelos achados apresentados no capítulo
\ref{cap4} desta tese, onde observou-se que população, residente ou flutuante,
são correlacionadas com crimes, no entanto essa correlação é mais forte ou mais
fraca dependendo do tipo de crime. O módulo de análise de dados desenvolvido
permite manusear as informações e avaliar as correlações entre população e
crime, para a partir daí, extrair a melhor divisão da cidade. Já o módulo de
alocação usa uma divisão da cidade escolhida pelo usuário para simular a
alocação de policiais por essa divisão, sendo essa alocação feita utilizando o
modelo baseado em alta densidade de crimes.

O {\it framework} desenvolvido foi utilizado para gerar os aglomerados de
população flutuante ilustrados na Figura \ref{clusters}b. Essa divisão da cidade
foi utilizada em um estudo comparativo, que revelou que a alocação de recursos
policiais em aglomerados de população flutuante conduz a uma distribuição de
recursos significativamente diferente de estratégias de alocação baseadas em
regiões administrativas, comumente definidas a partir da presença de residentes.
Mais ainda, foi mostrado que a alocação tendo como base os aglomerados de
população flutuante tende a ser mais adequada para combater o CCP, pois a
distribuição de recursos policiais seguirá naturalmente uma lei de potência, o
que é desejável, visto que se espera que o crime cresça desproporcionalmente em
áreas com alta densidade de PF.

Os aspectos aqui discutidos abrem novas linhas de investigação. Em particular, é
importante ressaltar que no capítulo \ref{cap4} desta tese também mostrou-se
que, para PSA, a relação superlinear só é percebida tendo como base aglomerados
definidos a partir de características da população residente. Isso indica que é
necessário elaborar uma estratégia híbrida, em que diferentes políticas de
alocação e diferentes divisões do espaço urbano precisam ser levadas em
consideração para cada tipo de crime.

É digno de destaque ainda que existe a possibilidade de carregar no {\it
framework} dados, de população e crimes, correspondentes a apenas uma parte do
dia. É possível por exemplo carregar dados de PF e CCP apenas para um turno
do dia, permitindo estudo de alocações diferentes por período de um dia. Mesmo
no caso de PR, que não há informação nos conjuntos de dados do horário em que as
pessoas estão de fato em suas residências, é possível fazer diferentes alocações
no decorrer de um dia, pois é sabido que a maior parte das pessoas está em casa
no turno da noite e madrugada. 

\phantompart 

\chapter{Conclusão}\label{cap6}

Este capítulo é dedicado às considerações finais como conclusão desta tese. A
seção \ref{contribuicoes} contém os resultados alcançados com a execução deste
trabalho, onde serão apresentadas também as principais contribuições desta
pesquisa. Na seção \ref{lim-pesq} são descritas as limitações deste trabalho,
será feita uma discussão sobre as principais dificuldades encontradas para
ampliar o escopo da pesquisa. Por fim, na seção \ref{trab-fut}, são sugeridas
proposições para pesquisas futuras.

\section{Resultados alcançados e contribuições} \label{contribuicoes}

Neste trabalho foi identificado que a relação entre população e crime é também
alométrica dentro de uma cidade. Fazendo uso do {\it City Clustering Algorithm}
(CCA) \cite{makse1998, rozenfeld2008, giesen2010, rozenfeld2011, duranton2013,
gallos2012, duranton2015, eeckhout2004} para determinar aglomerados de população
intracidade, foi mensurado volume de influência social com base na presença de
população residente e flutuante nos setores censitários de Fortaleza, uma
grande metrópole brasileira, localizada no estado do Ceará. Ao contrário de
estudos intercidades, onde a influência social foi medida apenas pela presença de
residentes, aqui se propõe que, dentro de uma cidade, a mobilidade humana seja
considerada para entender a dinâmica de crimes contra o patrimônio. Foram
alcançados resultados que mostram que a incidência desse tipo de crime cresce de
forma superlinear, seguindo uma lei de potência em função da população flutuante,
com expoente alométrico $\beta = 1.15 \pm 0.04$. Esse resultado sugere que o
aumento do fluxo de pessoas em uma região da cidade leva a um número
desproporcionalmente maior de crimes contra o patrimônio na mesma região.

A correlação observada entre população flutuante e crimes contra o patrimônio
está de acordo com a Teoria das Atividades Rotineiras \cite{cohen1979},
que afirma que um crime ocorre pela convergência das rotinas de um agressor
motivado, uma vítima desprotegida e a ausência de um guardião capaz de impedir o
delito. Além de fornecer suporte quantitativo para essa teoria, o estudo
apresentado nesta tese de doutorado indica o fato de que a convergência dessas
rotinas produz um crescimento desproporcional da criminalidade. Esse resultado
contrasta claramente com a incidência de chamadas relacionadas à perturbação de
sossego alheio, onde uma relação alométrica também pode ser observada, mas em
função de população residente, com expoente alométrico $\beta = 1.18 \pm 0.04$. 

Sobre esses resultados, é ampla a discussão a respeito de leis de escala em
indicadores urbanos e é sabido que relações superlineares entre população e
esses indicadores não são de fato livres de escala \cite{bettencourt2007,
rozenfeld2008}. Apesar dos efeitos dessas relações crescerem seguindo uma lei de
potência em função das suas causas, não é esperado que esse crescimento ocorra
indefinidamente, tal qual é observado em estudos biológicos. No caso de uma
relação alométrica entre população e homicídios \cite{melo2014}, por exemplo, é
intuitivo pensar que em algum momento essa relação assumirá comportamento
isométrico ou sublinear, pois basicamente existem duas possibilidades: ou a
população crescerá e os homicídios diminuirão, atenuando o comportamento
superlinear, ou ocorrerão tantos homicídios que será impossível que a população
cresça, inviabilizando a percepção da relação superlinear para ordens de
grandeza superiores. Um fato que pode indicar que uma relação superlinear está a
caminho de sofrer uma transição para uma relação isométrica é a presença de
observações do fenômeno abaixo da reta de regressão, principalmente nas suas
maiores ordens de grandeza. Mesmo diante da sua natureza microdinâmica, natureza
ideal para compreensão mais precisa da dinâmica desses indicadores no espaço
urbano, neste estudo, não foi percebido qualquer indício de que essas relações
superlineares entre população e crimes estejam próximas de um ponto
de atenuação.

A compreensão da dinâmica do crime dentro de uma grande metrópole pode ajudar em
eventuais políticas de mitigação da violência, essencialmente na tarefa de
alocação de recursos policiais. A partir dos achados desta pesquisa, foi
conduzido um exercício de aplicação da descoberta de que leis de escala governam
a relação entre população e crime. Foi simulada e comparada a alocação de
policiais em Fortaleza por unidades administrativas de terreno e aglomerados de
população flutuante. Os resultados alcançados revelam que a alocação por
unidades administrativas de terreno falha em distribuir recursos policiais
seguindo as leis de escala observadas anteriormente, transparecendo uma provável
imprecisão de métodos de alocação policial convencionais para prevenir crimes
contra o patrimônio, por exemplo. Neste trabalho se conjectura de que a suposta
imprecisão das estratégias de alocação policial convencionais advêm do fato de
que as autoridades policias distribuem recursos no espaço urbano considerando
apenas o efeito da relação entre população flutuante e crimes contra o
patrimônio, a densidade de crimes.

\section{Limitações da pesquisa} \label{lim-pesq}

Neste trabalho buscou-se compreender um fenômeno, mais precisamente buscou-se
compreender a dinâmica do crime dentro de uma cidade. Embora Fortaleza tenha
sido dividida de inúmeras formas e mais de 80\% dessas divisões tenham
apresentado uma relação alométrica entre população e crime, a generalização
desse achado é condicionada a verificação do fenômeno em outras cidades.
Durante a execução desta tese buscou-se bases de dados de mobilidade humana
para outras grandes metrópoles, no entanto, foi constatada uma grande
dificuldade de se obter acesso a conjuntos de dados desse tipo, essencialmente
por questões de privacidade. Uma alternativa comercial para vencer essa
dificuldade é comprar dados da plataforma {\it LandScan}, que faz uso de um
sistema de informação geográfica e sensoriamento remoto para estimar a população
flutuante em todo o globo terrestre, com aproximadamente 1 {\it km} de
resolução. O {\it LandScan} pode ser definido ainda como uma plataforma de alta
resolução de dados de população ambiente \cite{andresen2011,malleson2015}, onde
é definida a quantidade de população flutuante em cada local por uma média de
cada 24 horas. Recentemente foi adquirido recurso para a compra dos dados de
população ambiente da plataforma para várias cidades americanas, no entanto, por
questões burocráticas, não foi obtido acesso a esses dados antes da conclusão
desta tese.

Outra limitação desta pesquisa está relacionada à validação da estratégia de
Alocação por População Flutuante (APF), proposta no capítulo \ref{cap5} desta
tese. Apesar de ter sido mostrado que a APF tende a ser mais adequada para
combater o crimes contra o patrimônio, essencialmente porque a distribuição de
recursos policiais seguirá naturalmente uma lei de potência, não foi verificado
se APF é de fato mais efetiva na prevenção de crimes que estratégias de alocação
policial convencionais. Para realizar uma validação mais profunda seria
necessário um teste em campo ou a aplicação dessa estratégia em modelos de
simulação \cite{melo2005}. Infelizmente, ambas as soluções se mostraram
inviáveis nesta pesquisa.

\section{Trabalhos futuros} \label{trab-fut}

Os trabalhos futuros identificados nesta tese de doutorado podem ser divididos
em dois grupos:
\begin{enumerate} [label=(\roman*)]

\item Aplicações para a descoberta de que relações alométricas entre população e
crime ocorrem também dentro de cidades.

\item Aplicações para a rede de fluxo de pessoas reconstruída nesta tese a
partir de dados de mobilidade do sistema de ônibus de Fortaleza.
\end{enumerate}

A respeito de trabalhos futuros para a aplicação desse novo conhecimento de que
existe uma relação superlinear entre população e chamadas à polícia dentro de
cidades, pode ser destacado a definição de políticas públicas para propor
estratégias de reorganização do espaço urbano. Sabendo que, a quantidade de
crimes contra o patrimônio, por exemplo, cresce seguindo uma lei de potência em
função da quantidade de transeuntes e assumindo que uma redistribuição
populacional não altera o comportamento superlinear observado, deve ser evitado
que existam lugares que concentrem muitas pessoas. A concentração de muitas
pessoas em poucos lugares produz mais crimes, desta forma, uma redistribuição
populacional que proporcione que parcelas iguais de população flutuante estejam
distribuídas pelo espaço urbano diminuirá a violência em termos quantitativos.
Uma redistribuição do fluxo de pessoas no espaço urbano de uma grande metrópole
pode ser alcançada em médio prazo, a partir de políticas públicas de incentivo
ao uso de certas áreas da cidade, ou em longo prazo, a partir da criação de
zonas autônomas (áreas que atendam por completo pessoas da região, com emprego,
moradia, escolas e lazer, reduzindo o processo de comutação entre as zonas da
cidade) com distribuição populacional uniforme. Essa política pública é de
grande complexidade de execução, especialmente porque pessoas que vivem em
grandes cidades possuem hábitos de mobilidade que dificultam o processo. A
título de exemplo, recentemente foi mostrado que mais de 30\% da população da
cidade passa pelo centro econômico de Fortaleza em um dia de semana
\cite{caminha2016a,caminha2016b}.

Em relação a aplicações para a rede de fluxo de pessoas reconstruída neste
trabalho, atualmente está sendo desenvolvido um simulador de sistemas de
transporte público, projetado a partir de dados reais de mobilidade dos cidadãos
de Fortaleza. Foi estimada a probabilidade de um passageiro surgir em uma parada
de ônibus, escolher uma determinada linha de ônibus e escolher um ponto de
baldeação. Essas probabilidades foram estimadas através da demanda real da rede
de ônibus, representada pela rede de fluxo reconstruída nesta tese. Se
espera que, em breve, seja possível simular o efeito que mudanças no espaço
urbano ou na estrutura da rede de transporte público têm em indicadores sociais
e vice-versa. Trabalhos em andamento, utilizando esse simulador, mostram que os
cidadãos de Fortaleza escolhem rotas ruins \cite{furtado2017} e há indícios de
que muitas pessoas escolhem rotas mais lentas para evitar áreas perigosas
\cite{sullivan2017}. Em curto prazo, se espera que seja possível testar a
criação/alteração de rotas de linhas de ônibus antes das mesmas serem
implementadas. A longo prazo é esperado que seja possível estudar o
comportamento do crime após uma reorganização do espaço urbano, como a criação
das zonas autônomas já mencionadas nesta seção.

Motivado pelos inúmeros trabalhos que buscam compreender a propagação de doenças
em uma macroescala global \cite{colizza2007, hufnagel2004, wang2009}, pelas
recorrentes epidemias de dengue no Brasil e pela carência de estudos que
utilizam dados de mobilidade humana para compreender a dinâmica da propagação de
epidemias em microescala (escala de cidade), é identificado como trabalho futuro
relacionado a esta tese de doutorado a compreensão do espalhamento do vírus da
dengue em Fortaleza. Deseja-se utilizar a rede de mobilidade humana reconstruída
para descobrir focos do mosquito {\it aedes aegypti}, único vetor reconhecido
como transmissor do vírus. A partir de uma base de dados, recém adquirida, de
atendimentos de casos da doença ocorridos entre 2008 e 2016, busca-se estudar as
rotinas de movimentação das pessoas infectadas a fim de encontrar focos
desconhecidos de proliferação do mosquito. Um dos grandes desafios dessa
pesquisa está relacionado ao fato de que não há qualquer informação a respeito
da distribuição espacial dos mosquitos. Deste modo, deseja-se utilizar modelos
estatísticos e de simulação, aliados à utilização de dados reais de mobilidade e
casos da doença, para compreender a propagação da doença em uma grande
metrópole.

\chapter*{Apêndice I} \label{numericos}

Aqui será realizada uma introdução aos métodos numéricos aplicados nesta
tese. Inicialmente será detalhado o método de Regressão Linear, que é a
principal técnica estatística utilizada para estimar relações alométricas. Em
seguida, será formalizado um método de regressão do {\it kernel}, denominado {\it
Nadaraya-Watson}, que frequentemente é utilizado em trabalhos que investigam
relações alométricas. Por fim, serão definidas interpolações por
{\it splines}, que foram utilizadas nesta tese para delimitar a área de
distribuições de dados.

\subsection*{Regressão linear} \label{rls}

Relações alométricas são usualmente estimadas estatisticamente por meio de uma
regressão linear. Esse método é indicado em situações onde é necessário estimar
uma relação de causa e efeito entre duas variáveis. O método estima a reta que
mais se aproxima dos valores observados, considerando um erro possível no eixo
$y$, eixo onde o efeito dessa relação é ilustrado. Seja um conjunto de dados
$\{X_i , Y_i\}^N_i$ e supondo que a função que melhor se ajusta a tais dados
seja uma equação da reta, do tipo,
\begin{equation}\label{eq-ret}
y = a + bx,
\end{equation}
\noindent considera-se $b$ o {\it slope} ou inclinação da reta e $a$ o valor
mínimo assumido por $y$, que naturalmente ocorre quando $x=0$. Uma equação
alométrica, 
\begin{equation}
Y=aX^\beta,
\end{equation}
\noindent pode ser ajustada como uma equação da reta ao aplicar o logaritmo dos
seus dados, 
\begin{equation}
log(Y) = log(a) + \beta log(X),
\end{equation}
\noindent onde o expoente alométrico, $\beta$,
é obtido pela inclinação da reta.

O método dos mínimos quadrados ou {\it ordinary least square} (OLS)
\cite{rawlings2001,montgomery2015,press1987} estima o {\it slope} da função,
formalmente,
\begin{equation}
slope = \sum_{i=1}^{N} \frac{(x_i - \bar{x})(y_i-\bar{y})}{(x_i - \bar{x})^2},
\end{equation}
\noindent onde $\bar{x}$ e $\bar{y}$ correspondem a média aritmética dos
valores de $x$ e $y$ respectivamente, formalmente,
\begin{equation}
\bar{x} = \sum_{i=1}^{N} \frac{x_i}{N}
\end{equation}
e
\begin{equation}
\bar{y} = \sum_{i=1}^{N} \frac{y_i}{N}.
\end{equation}

\begin{figure}[!h] \includegraphics[width=1\textwidth]{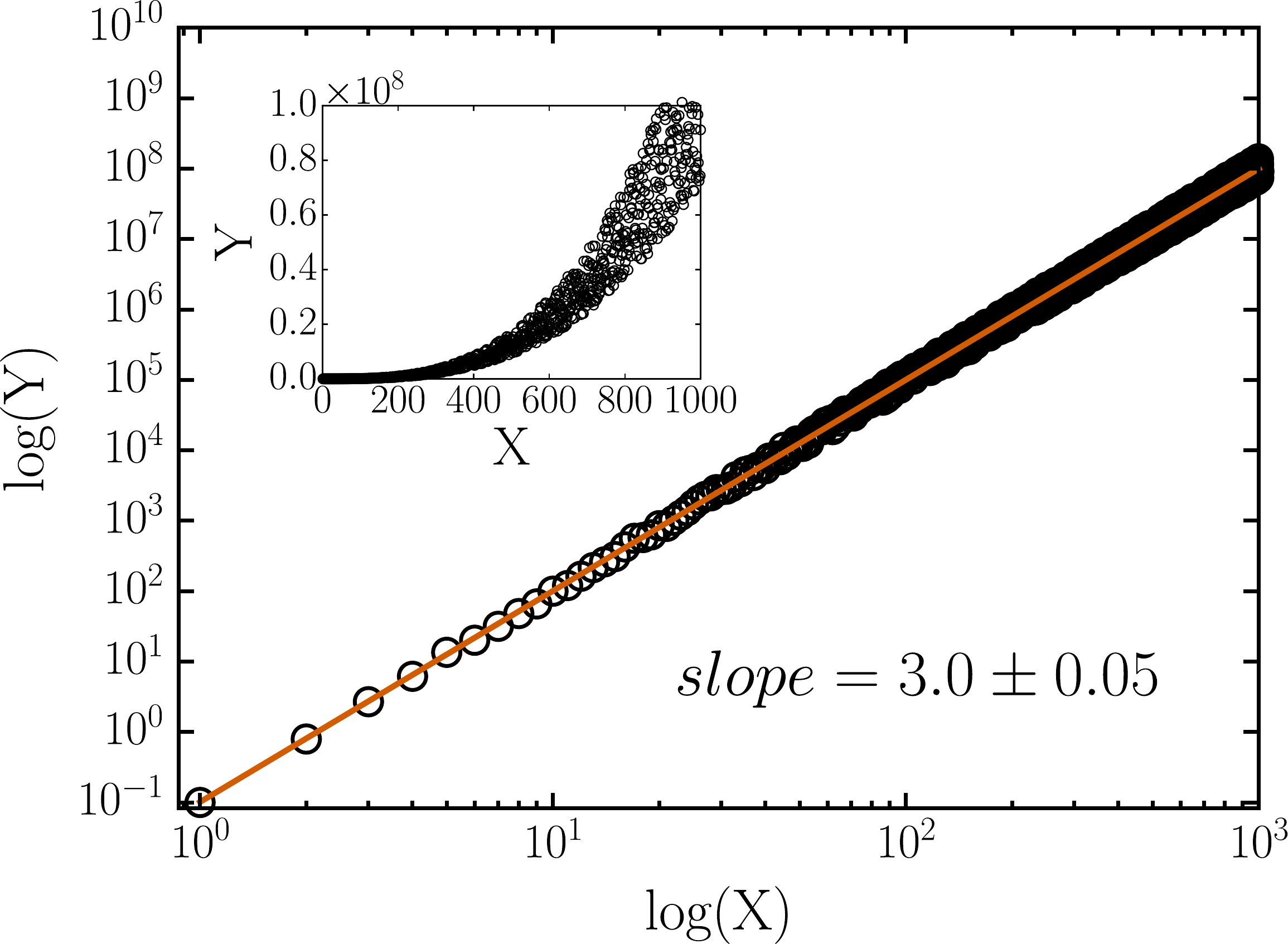}
\caption{{\bf Regressão linear simples aplicada em dados sintéticos.} Cada
círculo representa uma amostra gerada sinteticamente. A reta contínua laranja
ilustra a regressão linear simples aplicada aos dados, com $slope=3.0 \pm 0.05$.
A figura ainda exibe um gráfico embutido, onde é ilustrada a mesma distribuição,
no entanto, em escala linear.}
\label{exemplo-reg}
\end{figure}

A Figura \ref{exemplo-reg} ilustra uma regressão linear aplicada em
dados sintéticos. Foram gerados 1000 valores de $x$, tal que $1 \leq x \leq
1000$, e $y$ foi calculado em função de $x$ para $a=0.1$ e $slope=3.0 \pm 0.05$.

Além disso, é possível calcular o coeficiente de determinação
\cite{rawlings2001,montgomery2015,press1987}, $R^2$, um indicador que mensura a
qualidade do ajuste da função, formalmente: 
\begin{equation}
R^2 = slope \frac{\frac{\sum_{i=1}^{N} (x_i-\bar{x})^2}{N-1}}{\frac{\sum_{i=1}^{N} (y_i-\bar{y})^2}{N-1}}
\end{equation}

$R^2$ tem intervalo entre 0 e 1, onde quanto mais próximo for $R^2$ de 1, melhor
será o ajuste da Eq.\ref{eq-ret} aos dados. 

Outra medida de ajuste associada à técnica de regressão linear simples é o
calculo do erro padrão, $EP$, formalmente,
\begin{equation}
EP = \sqrt { \frac {\sum_{i=1}^{N} (y_i-y'_i)^2}{N} }
\end{equation}

\noindent onde $y'_i$ é o valor esperado de $y_i$ a partir da reta de regressão 
estimada. 

\subsection*{Nadaraya-Watson} \label{nw}

Uma regressão do {\it kernel} ou {\it Kernel Regression} é uma técnica
estatística não-paramétrica para estimar a expectativa condicional de uma
variável aleatória. O objetivo é encontrar uma relação não-linear entre um par
de variáveis aleatórias $x$ e $y$. Para uma distribuição de dados $\{X_i ,
Y_i\}^N_i$, é aplicado o método de {\it Nadaraya-Watson} para construir a função
mais suave do {\it kernel},
\begin{equation}
m_h(x)=\sum\limits_{i=1}^N \frac{K_h(x-X_i)Y_i}{K_h(x-X_i)},
\end{equation}

\noindent onde $N$ é o número de pontos da distribuição e $K_h(x-X_i)$ é 
função gaussiana de {\it kernel} definida formalmente como
\begin{equation}
K_h(x-X_i)=\exp\left[\frac{(x-X_i)^2}{2h^2}\right],
\end{equation}

\noindent onde $h$ é a largura de banda estimada pelo método de validação
cruzada dos mínimos quadrados \cite{nadaraya1964, watson1964}. Nesta tese
foi computado 95\% de IC (intervalo de confiança) em 500 amostras {\it
bootstrapping} aleatórias com substituição para todas as distribuições de dados
analisadas.

\subsection*{Interpolação por splines} \label{inter}

O conceito de {\it spline} é originado de uma técnica de desenho arquitetônico
em que se usa um barbante flexível (chamado {\it spline}) para desenhar uma
curva que conecta um conjunto de pontos em sequência \cite{schoenberg1973}. Uma
curva contínua é usada para intercalar esses pontos afixados. A função de ajuste
dessa curva poder ser representada por uma equação da reta, polinômio de grau
dois ou três \cite{ralston2001}. Mais precisamente, são utilizados polinômios de
pequeno grau para unir dois pontos consecutivos de uma amostragem.

\begin{figure}[!h] \includegraphics[width=1\textwidth]{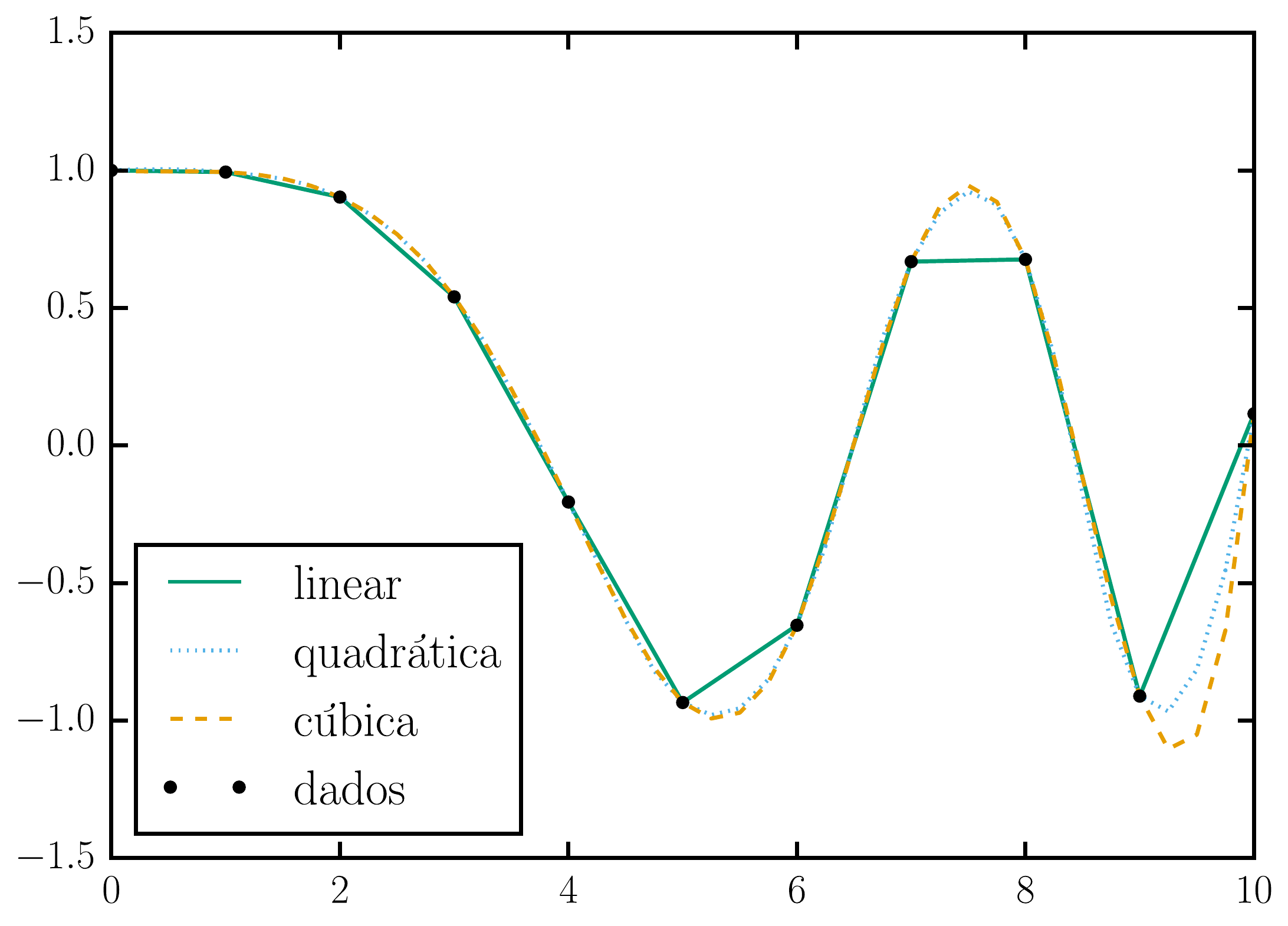}
\caption{{\bf Tipos de interpolação em um conjunto de dados de exemplo.} Os
círculos pretos representam os valores $\{X_i,Y_i\}^N_i$ de um conjunto de dados
hipotético. A linha contínua verde é o resultado de uma interpolação por {\it
splines} lineares nesse conjunto de dados. A linha pontilhada azul ilustra uma
interpolação por {\it splines} quadráticos. Por fim, a linha tracejada amarela
representa uma interpolação por {\it splines} cúbicos.} 
\label{interpol}
\end{figure}

Em uma interpolação linear busca-se desenhar uma reta para cada sequência de
dois pontos de uma distribuição de dados. Para interpolações por {\it splines}
quadráticos e cúbicos o processo é similar, no entanto, são substituídas as
retas por equações do segundo e terceiro grau, respectivamente, na ligação dos
pontos. A Figura \ref{interpol} ilustra interpolações aplicadas a dados
sintéticos.

Nesta tese foram utilizadas interpolações por {\it splines} cúbicos. Esse
tipo de interpolação pode ser definido formalmente a partir de um conjunto de 
pontos $\{X_i , Y_i\}^N_i$, onde não há $x_i=x_{i+1}$ e $a=x_0<x_1<...<x_n=b$,
o {\it spline} $S(x)$ é uma função que satisfaz:
\begin{enumerate} [label=(\roman*)]

\item $S(x) \in C^2[a,b]$ 

\item Em cada subintervalo $x_{i-1},x_i$, $S(x)$ é um polinômio de grau três, 
onde $i=1, ..., N$.

\item $S(x) = y_i$, para todo $i=1, ..., N$.

\end{enumerate} 

A partir disso é assumido que:

$$
S(x)=
\begin{cases}
C_1(x), \quad x_1 \leq x \leq x_1\\

\quad ... \quad \\

C_i(x), \quad x_{i-1} \leq x \leq x_i\\

\quad ... \quad \\

C_n(x), \quad x_{n-1} \leq x \leq x_n\\

\end{cases}
$$

\noindent onde cada $C_i= a_i+b_ix+c_ix^2+d_ix^3$ (para $d \ne 0$) é uma função 
cúbica, $i=1,...,n$. 

\begin{landscape}

\chapter*{Apêndice II} \label{downlods-c-d}

\subsection*{Sumário de dados}

\begin{center}
\begin{tabular}{ p{14em}  p{3.5em}  p{2.5em}  p{36em} } 
\hline
\textbf{Título} & \textbf{Tamanho} & \textbf{Tipo} & \textbf{Disponibilidade} \\
\hline

 Paradas de ônibus &  122k &  CSV &  http://dados.fortaleza.ce.gov.br/catalogo/dataset/dados-de-onibus-11-03-2015  \\ 

 GPS de ônibus &  236,7M &  CSV &  http://dados.fortaleza.ce.gov.br/catalogo/dataset/dados-de-onibus-11-03-2015 \\

 Dados de bilhetagem &  77,8M &  CSV &  http://dados.fortaleza.ce.gov.br/catalogo/dataset/dados-de-onibus-11-03-2015 \\

 Crimes: CCP e PSA &  14,9M &  CSV &  http://wikicrimes.org/dados-oficiais/crimes.csv \\

 Mapa de setores censitários &  7,9M &  KMZ &  http://www.ibge.gov.br/english/ \\
 
 Mapa de bairros &  504,2 K &  KML &  http://mapas.fortaleza.ce.gov.br/ \\

 População residente &  7,8M &  CSV &  http://www.ibge.gov.br/ \\
\hline
\end{tabular}
\end{center}

\end{landscape}

\chapter*{Publicações}

\section*{Artigos publicados em periódicos}

\noindent CAMINHA, C. et al. Human mobility in large cities as a proxy for crime. {\it
PLoS ONE}, v. 2, n. 12, p. e0171609, 2017.

\noindent CAMINHA, C.; FURTADO, V. Modeling user reports in crowdmaps as a complex
network. In: {\it Journal of Information and Data Management} 3 (3), 179, 2013.

\noindent FURTADO, V. et al. Open government and citizen participation in law enforcement
via crowd mapping. {\it IEEE Intelligent Systems}, IEEE, v. 27, n. 4, p. 63–69,
2012.

\section*{Artigos publicados em anais de congresso}

\noindent CAMINHA, C. et al. Detecção de comunidades em redes complexas para
identificar gargalos e desperdício de recursos em sistemas de ônibus. In: {\it
IV Brazilian Workshop on Social Network Analysis and Mining}, 2017. p. 1-7.

\noindent CAMINHA, C. et al. Micro-interventions in urban transportation from pattern
discovery on the flow of passengers and on the bus network. In: IEEE. {\it Smart
Cities Conference (ISC2), 2016 IEEE International}. [S.l.], 2016. p. 1–6.

\noindent PONTE, C.; CAMINHA, C.; FURTADO, V. Busca de melhor caminho entre dois pontos
quando múltiplas origens e múltiplos destinos são possíveis. In: {\it ENIAC.}
[S.l.: s.n.], 2016.

\noindent CAMINHA, C.; FURTADO, V. Impact of Human Mobility on Police Allocation. In: {\it
Conference of Intelligence and Security Informatics (ISI)}, 2017.

\noindent SULLIVAN, D.; et al.Towards Understanding the Impact of Crime
in a Choice of a Route by a Bus Passenger. In: {\it Agent-Based Modelling for
Criminological Research}, 2017.


\section*{Artigos sob avaliação}

\noindent FURTADO, V. et al. Increasing the Likelihood of Finding Public
Transport Riders that Face Problems Through a Data-Driven approach. {\it
In. Journal of Transportation Research}, 2017.

\postextual

\providecommand{\abntreprintinfo}[1]{%
 \citeonline{#1}}
\setlength{\labelsep}{0pt}

\phantompart
\printindex

\end{document}